\documentclass{aastex62}

\shorttitle{Evolution of high-energy particle distribution in SNRs}
\shortauthors{Zeng et al.}

\begin{document}

\title{Evolution of high-energy particle distribution in Supernova Remnants}

\correspondingauthor{Houdun Zeng}
\email{zhd@pmo.ac.cn, liusm@pmo.ac.cn}

\author[0000-0001-8500-0541]{Houdun Zeng}
\affil{Key Laboratory of Dark Matter and Space Astronomy \\
Purple Mountain Observatory, Chinese Academy of Sciences \\
Nanjing 210008, China}

\author{Yuliang Xin}
\affiliation{Key Laboratory of Dark Matter and Space Astronomy \\
Purple Mountain Observatory, Chinese Academy of Sciences \\
Nanjing 210008, China}

\author{Siming Liu}
\affiliation{Key Laboratory of Dark Matter and Space Astronomy \\
Purple Mountain Observatory, Chinese Academy of Sciences \\
Nanjing 210008, China}



\begin{abstract}

Supernova remnants (SNRs) have been considered as the dominant contributors to Galactic cosmic rays. However, the relation between high-energy particles trapped in SNRs and cosmic rays observed at the Earth remains obscure. In this paper, we fit the spectral energy distributions of 35 SNRs with a simple one-zone emission model and analyze correlations of model parameters to uncover the evolution of high-energy particle distribution in SNRs. We find that (1) the particle distribution in general can be described by a broken power-law function with a high-energy cutoff for all SNRs; (2) the low-energy spectrum becomes harder and the break energy decreases with aging of SNRs, (3) for most middle-age SNRs, the energy loss timescale of electrons at the high-energy cutoff is approximately equal to the age of the corresponding remnant implying quenching of very high-energy electron acceleration; for young SNRs, this energy loss timescale is shorter than the age of SNRs implying continuous electon acceleration at the cutoff energy; and for a few old age SNRs, the energy loss timescale is longer than the corresponding age which may suggest escaping of higher energy particles from SNRs. Finally, we comment on the implications of these results on the SNR origin of Galactic cosmic rays.

\end{abstract}

\keywords{cosmic rays --- gamma rays: ISM --- ISM: supernova remnants ---
radiation mechanisms: non-thermal}


\section{Introduction} \label{sec:intro}

As early as in the 1930s, supernova remnants (SNRs),
have been considered as possible contributors to the galactic cosmic rays \citep{1934PNAS...20..259B}. Shocks of typical SNRs have a total energy of
about $10^{51}$ erg. If 10\% of this energy can be converted into cosmic rays,
three supernovae per century in the Milky Way Galaxy can maintain the observed
cosmic ray flux.
The development of diffusive shock acceleration theory further strengths this scenario \citep{1983SSRv...36...57D}. 
Observations also reveal presence of high-energy electrons \citep{2015AARv..23....3D,1995Natur.378..255K} and protons in SNRs \citep{2013Sci...339..807A}.

The radio and non-thermal X-ray from SNRs are produced by relativistic electrons
via the synchrotron process, and gamma rays can be produced via inverse Compton scattering (IC) and bremsstrahlung of electrons, and/or decay of $\pi^0$ produced via hadronic processes.
Radio observations of SNRs show that the distribution of the radio spectral index peaks around 0.5 \citep{2011ApSS.336..257R,2015AARv..23....3D}, which is consistent with the diffusive particle acceleration by strong shocks. However a consensus has not been reached on the gradual hardening mechanism of the
radio spectra with aging of SNRs. Gamma-ray observations of SNRs reveal a variety of spectral shape \citep{2012ApJ...761..133Y,2013Sci...339..807A,2016ApJS..224....8A}.
The nature of gamma-ray emission from some SNRs is still a matter of debate. In general, the gamma-ray spectra can be fitted with a broken-power model with the break energy ranging from 1 GeV to 1 TeV \citep{2009ApJ...706L...1A,2010ApJ...718..348A,2015AA...574A.100H,2016ApJ...817...64X}.
A high-energy spectral cutoff is also obvious for some sources.

In the scenario of SNR origin of Galactic cosmic rays, \cite{2012ApJ...761..133Y}
proposed a unified model for gamma-ray emission from SNRs and found that the hardness
of gamma-ray emission (and the emission mechanism) is determined by density of the
emission region. They conclude that hard gamma-ray spectra originate from low density
regions via the IC process, while softer gamma-ray spectra come from high-density
regions via the hadronic process.
\cite{2017ApJ...834..153Z} carried out a comparative study of the spectral energy
distribution of nonthermal emission from three shell-type SNRs (RX J1713.7-3946,
CTB 37B, and CTB 37A) located within $2^{\circ}$ on the sky and  found that high-energy
particle distribution in SNRs is generally characterized by a double power law distribution and the particle distribution
becomes harder with aging of SNRs. Here we expand this study to a sample of 35 SNRs.
In section 2, we discuss the sample selection and have a brief review of our one-zone
emission model. Section 3 shows results of the spectral fitting and their implications.
Discussion and conclusions are drawn in section 4.

\section{Sample and Model}
\subsection{Sample}
More than 300 SNRs have been identified via radio observations \citep{2014BASI...42...47G,2017Green}. Fermi collaboration reported detection of 30
SNRs and 14 gamma-ray sources marginal associated with SNRs and 14 of these radio remnants have
synchrotron X-ray emission \citep{2016ApJS..224....8A}. \cite{2012AdSpR..49.1313F}
presented a Census of High-Energy Observations of Galactic SNRs. We use their latest
database (http://www.physics.umanitoba.ca/snr/SNRcat/), which summarizes the basic
physical properties and observational status of all SNRs \citep{2017Green} and
find 35 SNRs with both radio and GeV spectra. Complex SNRs, such as Cas A, is not
considered since its emission is not consistent with a simple one-zone emission model.
Among these 35 sources, 6 are detected with non-thermal X-ray emission
and 16 have TeV spectral measurements.
The distance, age, shock velocity, medium density, and the related references
are shown in Table. \ref{tab:sample}.

The GeV spectra of SNR G166.0+4.3 and SNR G296.5+10 are obtained
by analyzing the Pass 8 data from the \textit{Fermi}-LAT and the GeV data of other
SNRs are obtained from published literatures. The left panel of Figure 1
shows the multi-wavelength spectral data of those 35 SNRs. To demonstrate
the spectral evolution of SNRs, the right panel shows the spectra normalized
at 100 GeV. To obtain the 100 GeV flux of individual SNRs, their spectra from
1 GeV to 300 GeV are fitted with a single power-law. It can be seen that the
gamma-ray spectra always have convex shape which may be fitted with a broken-
power law with a high-energy cut-off. There is also evidence that the break energy
shifts from a few TeV to a few GeV as SNR evolves. Younger remnants have relatively
harder GeV spectra with a high-energy cutoff in the tens of TeV range. Older remnants
interacting with molecular clouds have soft gamma-ray spectra with a spectral break in
the GeV range. Most of the radio spectra can be fitted with a single power-law but
spectral curvature is evident in some radio spectra. The X-ray spectra are soft
implying a cutoff. Next we will use simple emission models to fit these spectra to
uncover evolution of high-energy particle distribution in SNRs.

\subsection{Model description and spectral fitting strategy}
Given the quality of the spectral data, we consider a simple one-zone emission
model with high-energy particle distributions given by
\begin{eqnarray}
\label{eq:Np}
N(P_i)= N_{0,i} \textrm{exp} (-\frac{P_i}{P_{i,cut}})
 \left\{
\begin{array}{lcl}
P_i^{-\alpha}&
& { \textrm{if}~~~ P_i < P_{\rm br} } \\
 P_{i,\rm br} P_i^{-(\alpha+1)}  &
& { \textrm{if}~~~P_i \geq P_{\rm br} }\;,
\end{array}
\right.
\end{eqnarray}
which $"i"$ represents different particle species, $P_{\rm e,cut} < P_{\rm p,cut}$
are the high-energy cutoffs of electron and proton distributions, respectively,
and we have assumed an identical spectral break for electron and protons. Considering
the SNR origin of Galactic cosmic rays, $N_{\rm 0,e}/N_{\rm 0,p}$ is fixed at 0.01 \citep{2012ApJ...761..133Y,2017ApJ...834..153Z}.
For emission of electrons, the synchrotron, bremsstrahlung, and IC process are
considered. For protons, we only consider $\gamma$-ray emission via decay of neutral
pion.
For better comparison, besides considering the cosmic microwave background radiation,
we introduce an infrared field with a temperature of 30 K and an energy density of 1
eV cm$^{-3}$ \citep{2006ApJ...648L..29P} for IC. The density of the emission region
can be estimated with IR or X-ray observations. In cases that the density is not well constrained, two models with density differing by more than one order of magnitude
are considered. The magnetic field is treated as a free parameter. There are therefore
six free parameters: five of which, two cutoff energies, one break energy, spectral
index $\alpha$ and normalization, come from the particle distribution, and the other
one is the magnetic field for synchrotron radiation. The other parameters (e.g.
the distance $D$, the shock speed $V_s$) can be obtained and/or
estimated with observations and/or from other literatures. Our model belongs to the hybrid
model containing leptonic (Brem and IC) and hadronic ($\pi^0$ decay) $\gamma$-ray
emission processes, which are the same as in the previous paper \citep{2017ApJ...834..153Z}.
We also use the Markov Chain Monte Carlo (MCMC)
algorithm \footnote{The MCMC code is CosRayMC \citep{2012PhRvD..85d3507L}
adapted from the COSMOMC package \citep{2002PhRvD..66j3511L}} to carry out the spectral fit.
The MCMC method is widely used for high-dimensional parameter space investigation in which
the Metropolis-Hastings algorithm is used.
A brief introduction of the basic procedure of the MCMC sampling can be found in \cite{2010AA...517L...4F} and \cite{2017ApJ...834..153Z}.
 For cases
where $P_{\rm e,cut}$ is poorly constrained, we fix $P_{\rm e,cut}$ by requiring
the radiative energy loss timescale being equal to the age of the corresponding
SNR. Then there will be 5 free parameters. If $P_{\rm e,cut}$ is
smaller than $P_{\rm br}$, the electrons then have a single power-law distribution.
For some sources, simpler spectral models are allowed, and we adopt the same single
power-law spectrum for both electrons and protons with their normalization the same
as the above broken power-law model.
Revision of the
above model is only considering for spectral fits with a reduced $\chi^2$ greater
than 2 and systematic deviations in residuals of the spectral fit.

\section{The results}
Figure 2 shows the fitting results and the one-dimensional (1D) probability
distributions of model parameters. The best fit model parameters are indicated
by the dashed lines in the 1D probability distributions. For SNRs in the inner
Galaxy, free-free absorption of low-frequency radio emission can be significant.
The corresponding data are not considered in our fits and shown as open boxes in
Figure 2. For radio data obtained from literatures without errors, a ten percent
error is assumed. The TeV data of W30 are likely associated with a pulsar wind nebula
and treated as upper limits for the SNR. For SNR 1006, the TeV spectral data of both
limbs are multiplied by a factor of 2 and used as the TeV spectrum for the whole remnant.
The very hard X-ray spectrum of G78.2+02.1 is not considered in our spectral fit.

Table 2 lists the model parameters: $\alpha$, $E_{\rm br}$,
$E_{\rm e,cut}$, $E_{\rm p,cut}$, the magnetic field for synchrotron emission $B$,
the total energy content of protons $W_{p}$ with $E >1$ GeV for a given density and
distance, and the ratio $W_{B}/W_{e}$, the adopted density, and the
reduced $\chi^2$ for the
best-fit of these 35 SNRs, where $W_{e}$ is the total energy content
in electrons with $E > 1$ GeV and $W_{B}$ is the energy content in the magnetic
field assuming a volume filling factor of unity and a uniform emission sphere
with a radius $R$. Note that the normalization $N_{i,0}$ are given through the
total energy of protons above 1 GeV $W_{p}$ and the electron-to-proton number
ratio $k_{ep}=0.01$ at 1 GeV.

It can be seen that all the spectral fits have a reduced $\chi^2 <  2.0$ except
for W28, Kes 79, CTB 109. The poor quality of their gamma-ray or radio data, as
can be seen from the corresponding residuals of the SED fits, prevents them from
improved spectral fits with smaller values of the reduced $\chi^2$. For RX J1713.7-3946, given the high
quality of X-ray and gamma-ray data, a super-exponential high-energy cut off is
needed to obtain a reduced $\chi^2$ of 1.85. An exponential cutoff of the electron
distribution will lead to a $\chi^2$ greater than 2 \citep{2008ApJ...685..988T,2010AA...517L...4F,2017ApJ...834..153Z}.
For W51C, the reduced $\chi^2$ is slightly greater than 2 and there is a gradual
increase with frequency in the residuals of the radio data (Figure 2: W51C (a)).
To improve the spectral fit, one needs to increase the magnetic field since the
break energy is fixed by the gamma-ray data. This implies a density much larger
than 100 cm$^{-3}$ adopted here \citep{2012AA...541A..13A}. To improve the spectral
fit, we assume a single power-law with a high-energy cutoff for the electron
distribution and a broken power-law model for the protons. The spectral index
of electrons is the same as the low-energy proton spectral index. Figure 2:
W51C ($a^{'}$) shows the corresponding best fit. For Tycho, G166.0+4.3, S147, MSH 15-56,
and CTB 37B, there are significantly uncertainties on the gas density and
two values are adopted for comparison. For Tycho, our results are more in favor of a
scenario where that the gamma-ray emission is dominated by $\pi^0$ decay, which is
consistent with the result of a number of researchers \citep[e.g.][]{2013ApJ...763...14B,
2013MNRAS.429L..25Z,2014ApJ...783...33S}. SNR G166.0+4.3 is interacting with the
interstellar medium as revealed with HI observations\citep{1989MNRAS.237..277L}. Here the hadronic interpretation for the gamma-ray emission is favored since a lower ambient
density ($n=0.01$ cm$^{-3}$) implied by X-ray observations \citep{1994ApJ...421L..19B}
results in an unrealistically large total proton energy and a low magnetic filed.
The total SED of S147 may come from the joint contributions of a
low density diffuse zone and a high density filament \citep{2012ApJ...752..135K},
our results show that the gamma-ray emission is dominated by $\pi^0$ decay with $n=1$
cm$^{-3}$. For RX J0852-4622, the emission is dominated by electrons \citep[e.g.][]{2007ApJ...661..236A,2011ApJ...740L..51T,2018AA...612A...7H}. We considered
both single and double power-law models. For Kes 79, G73.9+0.9, Cygnus Loop, HB 21,
HB 3, HB 9, G166.0+4.3, the data quality
is poor, we adopt a single power-law distribution for electrons and protons with the
same high energy cutoff (their normalization is still different by two orders of magnitude). For G296.5+10.0, Kes 17, RCW 103, the data quality
is also poor, the spectral can be fitted with a single power-law
energetic particle distribution and the cutoff energy of the electron distribution is
obtained by setting
the synchrotron energy loss time at this energy to be equal to the age. We note that the
recently identified SNR G150.3+4.5 produces very strong GeV emission \citep{2014A&A...567A..59G,2017ApJ...843..139A}. According to our spectral fit, it can be
readily detected by future experiments such as LHAASO and CTA. Moreover, the weak radio
emission of this source implies a magnetic field as low as 3 $\mu$G. The electron
distribution then cuts off at $\sim$ 1 PeV if the synchrotron energy loss timescale at
this energy is equal to the age of the remnant. The cutoff energy of proton is even higher!
Molecular clouds near this source may be illuminated by escaping high-energy proton from
this remnant.

Figure 3 (a) shows correlation between the electron synchrotron cooling time
at the cutoff energy and the age of SNRs. The evolution of high-energy electron
distribution in SNR appears to go through three stages. In very young SNRs with
ages less than 1000 years, the synchrotron cooling time is shorter than the age of
the corresponding SNR implying very efficient high-energy electron acceleration so
that the cutoff results from a balance between the acceleration and the loss process
\citep{2017JHEAp..13...17O}. For intermediate age SNRs, the electron synchrotron
energy loss at the cutoff
energy is comparable to the age. Although for many sources, this close correlation is
a result of model selection\footnote{If the high-energy cutoff of electron distribution
is not well constrained by the data, we fix this parameter by requiring the synchrotron
energy loss time being equal to the age.}, these results do suggest that acceleration of
high-energy electrons might have stopped so that their spectral evolution is dominated
by the energy
loss and/or escape processes. For Kes 79, G73.9+0.9, Cygnus Loop, HB 21,
HB 3, HB 9, G166.0+4.3, the synchrotron energy loss
time scale at the cutoff energy is much longer than their age, this implies very efficient
escape of high energy particles so that only particles with relatively low energies are
still trapped within the remnants \citep{2012MNRAS.427...91O}.

Figure 3 (b) shows the correlation of low-energy spectral index and the age of SNRs. This result is
consistent with our previous work \citep{2017ApJ...834..153Z}, and also agrees with the
radio spectral hardening of older SNRs. The softer spectra of younger SNRs has been attributed
to amplification of magnetic field in the shock upstream so that the effective compression
ratio is reduced \citep{2011JCAP...05..026C}. Considering the turbulent amplification of particle diffusion in the shock
downstream, \cite{2017ApJ...844L...3Z} suggests that such a soft spectrum may be caused by
time-dependent effect of the acceleration process. For older SNRs, the spectral hardening
can be attributed to Coulomb collision energy loss of low energy electrons, re-acceleration
of cosmic ray electrons, and stochastic particle acceleration in shock downstream. It may
also suggest very efficient injection into the acceleration process so that the adiabatic
index of the downstream plasma becomes less than 5/3 giving rise to a high shock compression
ratio. Energy losses at the shock front can also lead to a high shock compression ratio.
It should be noted that the SED here is for
the spatially integrated distribution, and the detail spatial structure of SNRs is not
considered. More detailed study is needed to clarify this issue.

Figure 3 (c) shows that the break energy of particle distribution decreases with
the age of the corresponding SNRs, which may be related to the gradual weakening
of shock waves with aging of SNRs \citep{2017JHEAp..13...17O,2017ApJ...844L...3Z}.
For a few old SNRs without a spectral break, we use the corresponding high-energy
cutoff instead. This result is compatible to Figures 3 (a) and (b). Combining these
results together, we conclude that the acceleration of highest energy particles
occurs in young SNRs presumably in the free expansion phase of the shock wave. The
maximum energy that the shock can accelerate particles to decreases quickly in the
Sedov phase so that the evolution of very high-energy particle distribution is
dominated by the energy loss and/or escape processes \citep{2012MNRAS.427...91O,
2012SSRv..173..369H}. However the particle injection into
the acceleration may be very efficient in this stage, which leads to a broken
power-law spectrum for the spatially integrated particle distributions and a gradual
hardening of the low-energy particle distribution. When SNRs further evolve in
a high density environment, high-energy particles may escape from SNR efficiently
so that only particles with relatively low-energies are still trapped.

Figure 3 (d) shows that $W_{p}$ varies
from $10^{48}$ erg to $10^{51}$ erg. They should be regarded as a lower limit of
their contribution to Galactic cosmic rays since some particles have
escaped earlier.
Figure 3(e) shows the correlation between the high-energy cutoffs of electrons and
protons. The $E_{\rm p,cut}$ of protons for almost all sources are higher than that of
electrons except for some escape sources for which the high energy cutoffs for electrons
and protons are assumed to be equal. For most SNRs, $E_{\rm p,cut}$ is not well constrained
and a lower limit can be obtained. Only 4 of SNRs have a $E_{\rm p,cut} > E_{\rm e,cut}$
constrained by the SED. The corresponding $E_{\rm p,cut}$ is always lower than 100 TeV.
If SNRs can indeed accelerate protons to PeV energy, this result implies that protons
above $E_{\rm p,cut}$ have already escaped from these SNRs. Figure 3 (f) shows the
correlation of the high-energy cutoffs of protons and the energy of protons with a gyro
radius being equal to the radius of the corresponding remnant ($9.25 \times 10^{5}\times
\frac{B}{\rm \mu G}\times \frac{R}{\rm pc}$ GeV) (Hillas criterion).
This figure shows that the lower limits of $E_{\rm p,cut}$ below the solid line
satisfy the Hillas condition.
Figure 3 (g) and (h) respectively show  the evolution of
the magnetic field and the gas density. There is no obvious
pattern except for a few young SNRs when both the magnetic
field and the density decrease with the age of SNRs. This is
consistent with shock evolution in a stellar wind bubble
\citep[e.g.][]{2008AIPC.1085..169V,2012ApJ...759...89H}.

Based on the results above, we also find a good correlation between
the low-energy spectral index $\alpha$ and the electron synchrotron cooling time at the
cutoff energy $\tau_{syn}$ (Figure 3 (i)). Due to escape of high energy particles in old SNRs,
these sources occupy the upper portion of the Figure.  
Figure 3 (j) shows the correlation
between the spectral index $\alpha$ and the break energy $E_{br}$.
It also shows a good correlation, which implies young SNRs having high
break energies and soft spectra, and old SNRs having low break energies
and harder spectra.

Figure 3 (k) shows the dependence of the ratio of the mean magnetic-field energy
density to the ambient gas density $B^2/(8 \pi n m_p)$ on the shock velocity $V_s$.
The thin lines show a $V_s^2$ dependency for several ratios of the shock
kinetic energy density to the magnetic energy density, the thick dotted line shows
a $V_s^3$ dependency and the thick solid line represents the best fit in log-log
space. Note that the magnetic energy density is always lower than the kinetic energy
density and $\eta=\frac{B^2}{8 \pi}/(0.5 n m_p V_s^2)$ decreases with increasing $V_s$
along the line of the best fit. Our sample shows the correlation
coefficient between $B^2/(8 \pi n m_p)$ and $V_s$ is $0.69$ with a chance
probability of $1.2 \times 10^{-6}$, which suggests that the magnetic field is
indeed amplified near the shocks of SNRs. The fitting slope
of 1.49 seems more close to the dependency of $ B^2 \propto n V_s^2$
\citep{2005A&A...433..229V}. \cite{2008AIPC.1085..169V} instead of
$ B^2 \propto n V_s^3$ expected by the other magnetic field amplification
mechanisms \citep{2004MNRAS.353..550B}. However, different mechanisms may play
the dominant role in different stages of SNR evolution \citep{2012AARv..20...49V}.

\section{Discussion and Conclusions}
Multi-wavelength observations of SNRs provide a unique opportunity to study evolution
of high-energy particle distribution trapped with SNRs, which is closely related
to the particle acceleration and escape processes. In this paper, we extend our
earlier comparative spectral study of three SNRs to a simple of 35 SNRs \citep{2017ApJ...834..153Z}.
In general, our results are compatible to the scenario of SNR origin of Galactic
cosmic rays. Young SNRs have relatively soft spectra, however their break energies
are high so that they may dominate the flux of high-energy cosmic rays. Although the
low-energy spectral of old SNR are hard, their break energies are low. They likely
dominate fluxes of cosmic rays with relatively lower energies \citep{2017ApJ...844L...3Z}. Previously,
\cite{2012ApJ...761..133Y} show that the averaged injection spectrum of cosmic rays
needs to be a broken power-law with a break energy of a few GeV, which is consistent
with parameters of old SNRs. The simple unified emission model proposed by
\cite{2012ApJ...761..133Y} attribute the gamma-ray spectral hardness to density of the
emission region and the related emission process. Our modeling not only improves the
gamma-ray spectral fit of individual SNRs, in combination with radio and X-ray observations,
parameters of the emission model are well constrained.

We find that in general a simple one-zone emission model with a broken power distribution
with a high-energy cutoff can fit non-thermal spectra of SNRs. The low-energy spectral
index decreases with aging of SNRs, which is consistent with the observed radio spectral
harding. The break energy also decreases with the increase of the age, which is consistent
with gamma-ray spectral evolution of SNRs. The evolution of high-energy electron
distribution in SNRs goes through three stages. Very young SNRs with an age of of
a few hundreds of years produce strong synchrotron X-ray emission. The electron synchrotron
energy loss time at the cutoff energy is shorter than the age of the SNR so that there is
very efficient high-energy electron acceleration. For intermediate age SNRs, the synchrotron
energy loss at the cutoff energy is comparable to the age, which implies that
the high-energy spectral particle evolution is dominated by the energy
loss process and the acceleration may already
stop at very high-energies. However, the continuous spectral hardening at low energies and
the decrease of the break energy with aging of SNRs show that low energy particle
acceleration should be very efficient. For a few old SNRs, the electron synchrotron energy
loss time at the cutoff energy is much longer than the age of the SNRs, implying escape
of particles at even high-energies. There are 4 SNRs whose proton cutoff energy is well
constrained by the SED and is higher than the electron cutoff energy. Since these proton
cutoff energies are lower than 100 TeV, higher energy protons may have already escaped
from these remnants before high energy electrons start to escape significantly.
There are many old radio SNRs without gamma-ray
counterparts. They are likely similar to SNRs with most of their high-energy particles
already escaped. Similarly, the evolution of high-energy ion distribution also
experiences three stages. Since the effect of radiation loss of high-energy ions
on spectral evolution is negligible, ions can be accelerated to much higher energies.
The highest energy ions are likely accelerated near the end of the free expansion phase
when the shock speed is highest. As the shock slows down, high-energy ion acceleration
is suppressed. However, the acceleration of ions with relatively low energies can still
be very efficient leading to gradual spectral hardening and a broken power-law energy
distribution of the spatially integrated spectrum. In old SNRs, the escape process
dominates at high-energies. At the beginning of significant escape, electrons and protons
share the same high-energy cutoff.

\cite{2015APh....65...80M} fitted multi-wavelength data of 24 SNRs by using a simpler
one-zone hybrid model, and found that 21 of the 24 SNR gamma-ray spectra can be attributed
to a hadronic origin. In order to obtain
the corresponding neutrino spectra, they adopted the proton
distribution with a single power-law form, completely independent of the electron
distribution, and the ratio $k_{ep}$ was allowed to vary between $10^{-4}$ and 50.
Most importantly, they appeared to attribute thermal X-ray emission from many sources
to electron synchrotron process. Our model have much less free parameters and the data
are selected carefully, which leads to many results not seen via their approach.
Several sources in our sample can be detected with future gamma-ray telescopes.
In particular, we expect strong emission above 100 TeV from G150.3+04.5 that can
be readily detected with the LHAASO \citep{2014APh....54...86C}.

\acknowledgments
This work is partially supported by National Key R\&D Program of
China: 2018YFA0404203, NSFC grants: U1738122, 11761131007 
and by the International Partnership Program of Chinese Academy
of Sciences, Grant No. 114332KYSB20170008.


\floattable
\begin{deluxetable}{l|c|c|c|c|c|c|c|c|c|c|c}
\tabletypesize{\tiny}
\tablewidth{0.99\textwidth}
\tablecaption{The sample of SNRs and related physical information \label{tab:sample}}
\tablehead{
\colhead{SNR name} & \colhead{Other name} & \colhead{Radius} & \colhead{Distance} & \colhead{Age} & \colhead{Density} &
\colhead{Shock speed}& \multicolumn{4}{c}{} & \colhead{References for related}  \\
\colhead{} & \colhead{} & \colhead{(pc)} & \colhead{(Kpc)}& \colhead{(kyr)} & \colhead{(cm$^{-3}$)} &
\colhead{(km/s)}& \colhead{Radio} & \colhead{X-ray}& \colhead{GeV}& \colhead{TeV}& \colhead{physical information }
}
\startdata
G006.4$-$00.1 & W28 & $\sim$ 13 & $\sim$ 2.0 &  40(33-150) & $\sim$ 100 &
60-80 &$\checkmark$ & &$\checkmark$ & $\checkmark$ & [1-4] \\
G008.7$-$00.1 & W30 & $\sim$ 26 & $\sim$4.0 & 25(15-28) & $\sim$ 100 &
530-750 &$\checkmark$ & &$\checkmark$ &  & [5][6] \\
G031.9$+$00.0 & 3C 391 & $\sim$ 7 & $\sim$ 7.2 & $\sim$ 4 & $\sim$ 300 &
620-730 &$\checkmark$ & &$\checkmark$ & &[7-10]  \\
G033.6$+$00.1 & Kes 79 & $\sim$ 9.6 & $\sim$ 7.0 & $\sim$ 4.4-6.7 & $\sim$ 3(1-5) &
$400\pm5$ &$\checkmark$ & &$\checkmark$ & &  [11-13]\\
G034.7$-$00.4 & W44 & $\sim$ 12.5 & $\sim$ 3.0 & $\sim$ 20 & $\sim$ 200 &
100-150 &$\checkmark$ & &$\checkmark$ & &  [14-16]\\
G043.3$-$00.2 & W49B & $\sim$ 5 & $\sim$ 10 & $\sim$ 5.7(5-6) & $\sim$ 700 &
$\sim$ 400 &$\checkmark$ & &$\checkmark$ & $\checkmark$ &  [17][18]\\
G049.2$-$00.7 & W51C & $\sim$ 18 & $\sim$ 4.3 & $\sim$ 30 & $\sim$ 10 &
$\sim$ 100 &$\checkmark$ &$\top$ &$\checkmark$ & $\checkmark$ & [19-22] \\
G073.9$+$00.9 &  & $\sim$ 16/5.2 & $\sim$ 4.0/1.3 & $\sim$ 11-12 & $\sim$ 10 &
$\sim$ 200-300 &$\checkmark$ &$\top$ &$\checkmark$ &  & [23][24] \\
G074.0$-$08.5 & Cygnus loop & $\sim$ 16 & $\sim$ 0.54 & $\sim$ 14 & $\sim$ 5.0 &
240-330 & $\checkmark$ & & $\checkmark$ &  & [25-28] \\
G078.2$+$02.1 & $\gamma$ Cygni & $\sim$ 17 & $\sim$ 2.0 & $\sim$ 8.25(6.8-10)
& $\sim$ 2.5(0.1-20) &
700-1100 &$\checkmark$ &$\top$ &$\checkmark$ & $\checkmark$ & [29][30] \\
G089.0$+$04.7 & HB21 & $\sim$ 26 & $\sim$ 1.7 & $\sim$ 40(36 or 45) & $\sim$ 15 &
$\sim$ 125 &$\checkmark$ &$\top$ &$\checkmark$ & &  [31-35]\\
G109.1$-$1.00 & CTB109 & $\sim$ 16 & $\sim$ 3.1 & $\sim$ 9.0(9.0-9.2) & $\sim$ 1.1 &
$\sim 230\pm 5$ &$\checkmark$ &$\top$ &$\checkmark$ & & [36][37] \\
G120.1$+$01.4 & Tycho & $\sim$ 3.3 & $\sim$ 3.0 & $\sim$ 0.44 & $\sim$ 10/0.3 &
4600-4800 &$\checkmark$ &$\checkmark$ &$\checkmark$ &$\checkmark$ &  [38][39]\\
G132.7$+$01.3 & HB3 & $\sim$ 26.4 & $\sim$ 2.2 & $\sim$ 30.0 & $\sim$ 2.0 &
303-377 &$\checkmark$ & &$\checkmark$ & &  [40-42]\\
G150.3$+$04.5 &  & $\sim$ 9.4 & $\sim$ 0.40 & $\sim$ 1.5(0.5-5) & $\sim$ 1.0 &
$< 2500$ &$\checkmark$ & $\top$&$\checkmark$ & & [43] \\
G160.9$+$02.6 & HB9 & $\sim$ 15 & $\sim$ 0.8 & 5.3(4-7) & $\sim$ 0.1 &
$\sim$ 740 &$\checkmark$ &$\top$ &$\checkmark$ & &  [44][45]\\
G166.0$+$04.3 &  & $\sim$ 26 & $\sim$ 4.5 & $24.0$ & $\sim$ 0.01 &
$\sim$ 680 &$\checkmark$ & &$\checkmark$ & & [46][47] \\
G180.0$-$01.7 & S147 & $\sim$ 38 & $\sim 1.3$ & 30(20-100) & $\sim$ 250(100-500) &
$\sim$ 500 &$\checkmark$ & &$\checkmark$ & &  [48][49]\\
G189.1$+$03.0 & IC 443 & $\sim$ 11 & $\sim$ 1.5 & $\sim$ 30 & $\sim$ 140 &
60-100 &$\checkmark$ & &$\checkmark$ &$\checkmark$ &  [50-52]\\
G205.5$+$0.50 & Monoceros & $\sim$ 63.36 & $\sim$ 1.98 & $\sim$ 30 & $\sim$ 3.6 &
$\sim$ 50 &$\checkmark$ & &$\checkmark$ & &  [53-55]\\
G260.4$-$03.4 & Puppis A & $\sim$ 15 & $\sim$ 2.2 & 4.45(3.75-5.20) & $\sim$ 4.0 &
700-2500 &$\checkmark$ & &$\checkmark$ & $\top$&  [56-59]\\
G266.2$-$01.2 & RX J0852-4622 & $\sim$ 13 & $\sim$ 0.75 & 2.7(1.7-4.3) & $\sim$ 3.8 &
$\sim$ 3000 &$\checkmark$ & $\checkmark$&$\checkmark$ & $\checkmark$& [60][61] \\
G296.5$+$10.0 &   & $\sim$ 26 & $\sim$ 2.1 & $\sim$ 10.0 & $\sim$ 13.0 &
$< 1000$  &$\checkmark$ & &$\checkmark$ & &  [62][63]\\
G304.6$+$00.1 & Kes 17 & $\sim$ 10 & $\sim$ 10 &  4.2(2-5.2) & $\sim$ 10 &
150-200 &$\checkmark$ & $\top$ &$\checkmark$ & &  [64][65]\\
G315.4$-$02.3 & RCW 86 & $\sim$ 15 & $\sim$ 2.5 & $\sim$ 1.8 & $\sim$ 0.1-2.0 &
700-2000 &$\checkmark$ & $\checkmark$&$\checkmark$ & $\checkmark$& [66-68] \\
G326.3$-$01.8 & MSH 15-56 & $\sim$ 22.2 & $\sim 4.1$ & $\sim$ 10.0(10-16.5) & $\sim$ 0.1/1.0 &
500-860 &$\checkmark$ &  &$\checkmark$ &  & [67][69][70] \\
G327.6$+$14.6 & SN 1006 & $\sim$ 9.0 & $\sim$ 2.2 & $\sim$ 1.0 & $\sim$ 0.085 &
3200-5800 &$\checkmark$ & $\checkmark$ &$\checkmark$ & $\checkmark$ &  [71][72]\\
G332.4$-$00.4 & RCW 103 & $\sim$ 5 & $\sim$ 3.3 & $\sim$ 2.0 & $\sim$ 10 &
$\sim$ 1100 &$\checkmark$ & &$\checkmark$ & &  [73-75]\\
G337.0$-$00.1 &CTB 33 & $\sim$ 2.55 & $\sim$ 11.0 & $\sim$ 5.0 & $\sim$ 60 &
$< 200$ &$\checkmark$ & &$\checkmark$ & &  [76-78]\\
G347.3$-$00.5 & RX 1713.7-3946 & $\sim$ 10 & $\sim$ 1.0 & $\sim$ 1.6 & $\sim$ 0.01 &
$\sim$ 5000 &$\checkmark$ & $\checkmark$&$\checkmark$ &$\checkmark$ & [78-81] \\
G348.5$+$00.1 & CTB 37A & $\sim$ 10 & $\sim$ 7.9 & $\sim$ 30 & $\sim$ 100 &
75-100 &$\checkmark$ & $\top$&$\checkmark$ &$\checkmark$ & [82-86] \\
G348.7$+$00.3 & CTB 37B & $\sim$ 20 & $\sim$ 13.2 & $\sim$ 5 & $\sim$ 10/0.5 &
$\sim$ 800 &$\checkmark$ & $\top$&$\checkmark$ &$\checkmark$ & [85-88] \\
G349.7$+$00.2 &  & $\sim$ 3.3 & $\sim$ 11.5 & $\sim$ 2.8 & $\sim$ 35.0 &
700-900 &$\checkmark$ & $\top$&$\checkmark$ &$\checkmark$ &  [89-92]\\
G353.6$-$00.7 &Hess J1731-347  &$\sim  14.0$ & $\sim 3.2$ & $\sim$ 2-6 & $\sim$ 0.01 &
$\sim$ 2100 &$\checkmark$ &$\checkmark$ & $\checkmark$&$\checkmark$ &  [93][94]\\
G359.1$-$00.5 &Hess J1745-303  &$ \sim 16.0$ & $\sim 4.6$ & $\sim$ 70 & $\sim$ 100 &
$\sim$ 300 & $\checkmark$ & $\top$ &$\checkmark$ &$\checkmark$ &  [95-98]\\
\enddata
\tablecomments{The $\checkmark$ means that the flux has been measured, while the $\top$ means that an upper limit is available. References for the related physical information: [1]\citet{1993ApJ...409L..57K}, [2]\citet{2010ApJ...718..348A}, [3]\citet{1983RMxAA...8..155B}, [4]\citet{2002AJ....124.2145V},
[5]\citet{1994ApJ...434L..25F}, [6]\citet{2012ApJ...744...80A},
[7]\citet{2004ApJ...616..885C}, [8]\citet{1972ApJS...24...49R},
[9]\citet{2014IAUS..296..372S}, [10]\citet{1998AJ....115..247W},
[11]\citet{2009AA...507..841G}, [12]\citet{2016ApJ...831..192Z},
[13]\citet{2014ApJ...783...32A}, [14]\citet{1991ApJ...372L..99W},
[15]\citet{2013ApJ...768..179Y}, [16]\citet{2000ApJ...544..843R},
[17]\citet{2017arXiv170705107Z}, [18]\citet{2001ApJ...550..799B},
[19]\citet{1995ApJ...447..211K}, [20]\citet{2012AA...541A..13A},
[21]\citet{2013ApJ...769L..17T}, [22]\citet{1997ApJ...475..194K},
[23]\citet{1993ARep...37..240L}, [24]\citet{2013ApJS..204....4P},
[25]\citet{1998ApJS..118..541L}, [26]\citet{2005AJ....129.2268B},
[27]\citet{1994ApJ...420..721H}, [28]\citet{2009ApJ...702..327S},
[29]\citet{2013MNRAS.436..968L}, [30]\citet{2001AstL...27..233G},    
[31]\citet{2007AA...461..991M},[32]\citet{1991ApJ...382..204K},
[33]\citet{2006ApJ...637..283B}, [34]\citet{2013ApJ...779..179P},
[35]\citet{1991ApJ...382..204K}, [36]\citet{2012ApJ...756...88C},
[37]\citet{2018MNRAS.473.1705S}, [38]\citet{2010ApJ...725..894H},
[39]\citet{2007ApJ...665..315C}, [40]\citet{2006ApJ...647..350L},
[41]\citet{1991AA...247..529R}, [42]\citet{2005AstL...31..179G},
[43]\citet{2016PhDT.......190C}, [44]\citet{1995AA...293..853L},
[45]\citet{2007AA...461.1013L}, [46]\citet{1994ApJ...421L..19B},
[47]\citet{1989MNRAS.237..277L}, [48]\citet{2009ApJ...698..250C},
[49]\citet{2012ApJ...752..135K}, [50]\citet{2001ApJ...554L.205O},
[51]\citet{2003AA...408..545W}, [52]\citet{2014ApJ...788..122S},
[53]\citet{1986MNRAS.220..501L}, [54]\citet{2018arXiv180201069Z},
[55]\citet{2012AA...545A..86X}, [56]\citet{2012ApJ...755..141B},
[57]\citet{2003MNRAS.345..671R}, [58]\citet{2010ApJ...725..585A},
[59]\citet{2015AA...575A..81H}, [60]\citet{2008ApJ...678L..35K},
[61]\citet{2001ApJ...548..814S}, [62]\citet{1997ApJ...476L..43V},
[63]\citet{2000AJ....119..281G}, [64]\citet{2013ApJ...777..148G},
[65]\citet{2009ApJ...694.1266H}, [66]\citet{2013MNRAS.435..910H},
[67]\citet{1996AA...315..243R}, [68]\citet{2000AA...360..671B},
[69]\citet{2013ApJ...768...61T}, [70]\citet{2013ApJ...773...25Y},
[71]\citet{2003ApJ...585..324W},
[72]\citet{2009ApJ...692L.105K}, [73]\citet{1997PASP..109..990C},
[74]\citet{1975AA....45..239C}, [75]\citet{2015ApJ...810..113F},
[76]\citet{1997ApJ...483..335S}, [77]\citet{1999ApJ...526L..29C},
[78]\citet{2013ApJ...774...36C} 
[79]\citet{1997AA...318L..59W}, [80]\citet{2011ApJ...735..120Y},
[81]\citet{2003PASJ...55L..61F}, [82]\citet{2011MNRAS.417.1387S},
[83]\citet{2011ApJ...742....7A}, [84]\citet{2000ApJ...545..874R},
[85]\citet{2012MNRAS.421.2593T}, [86]\citet{2017ApJ...834..153Z},
[87]\citet{2008AA...486..829A},[88]\citet{2009PASJ...61S.197N},
[89]\citet{2002ApJ...580..904S}, [90]\citet{2014ApJ...783L...2T},
[91]\citet{2015ApJ...804..124E}, [92]\citet{2005ApJ...618..733L},
[93]\citet{2008ApJ...679L..85T}, [94]\citet{2011AA...531A..81H},
[95]\citet{2011PASJ...63..527O}, [96]\citet{2012PASJ...64....8H},
[97]\citet{2007IAUS..242..366Y}, [98]\citet{2008ApJ...683..189H}. \\
References for observation data: W28: \citet{2010ApJ...718..348A,2008AA...481..401A,2000AJ....120.1933D},
W30: \citet{1990Natur.343..146K,2012ApJ...744...80A},
W41: \citet{1970AAS....1..319A,1992AJ....103..943K,
2007ApJ...657L..25T,2009ApJ...691.1707M,2015AA...574A..27H},
3C391: \citet{1989ApJS...71..799K,2010ApJ...717..372C,2014ApJ...790...65E},
Kes 79: \citet{2014ApJ...783...32A},
W44: \citet{2007AA...471..537C,2013Sci...339..807A,2011ApJ...742L..30G},
W49B: \citet{1994ApJ...437..705M,2018AA...612A...5H},
G73.9+0.9: \citet{2006AA...457.1081K,2016MNRAS.455.1451Z,2015ApJ...813....3A},
W51C: \citet{1994JKAS...27...81M,2016ApJ...816..100J,2012AA...541A..13A},
Cygnus Loop: \citet{2004AA...426..909U,2015arXiv150203053R},
$\gamma$ Cygni: \citet{2018ApJ...861..134A,2013ApJ...770...93A,
2002ApJ...571..866U,1991AA...241..551W,1977AJ.....82..329H,1997AA...324..641Z,
2006AA...457.1081K,2011AA...529A.159G},
HB21: \citet{2006AA...457.1081K,2013ApJ...779..179P,2010AJ....140.1787P},
Tycho: \citet{2011AA...536A..83S,2012ApJ...744L...2G,2009PASJ...61S.167T,2015ICRC...34..769P},
CTB 109: \citet{2006AA...457.1081K,2012ApJ...756...88C},
HB3: \citet{2005AA...436..187T,2016ApJ...818..114K},
G150.3+4.5: \citet{2014AA...566A..76G,2016PhDT.......190C},
HB9: \citet{2014MNRAS.444..860A,1982JApA....3..207D,1995AA...293..853L},
G166.0+4.3: \citet{2005AA...440..929L,2006AA...451..991T},
S147: \citet{2008AA...482..783X,2012ApJ...752..135K},
MSH 15-56: \citet{2000ApJ...543..840D,2018arXiv180511168D},
IC443: \citet{2011AA...534A..21C,2013Sci...339..807A,
2009ApJ...698L.133A,2007ApJ...664L..87A,2010ApJ...710L.151T},
G205.5+0.5: \citet{2012AA...545A..86X,2016ApJ...831..106K,2015arXiv150805489M},
Puppis A: \citet{2006AA...459..535C,2013AA...555A...9D,
2012ApJ...759...89H,2017ApJ...843...90X,2015AA...575A..81H},
RX J0852-4662: \citet{2000AA...364..732D,2007ApJ...661..236A,
2011ApJ...740L..51T,2018AA...612A...7H},
G296.5+10.0: \citet{1994MNRAS.270..106M},
Kes 17: \citet{1970AuJPA..14..133S,1996AAS..118..329W,2013ApJ...777..148G},
RCW 86: \citet{1975AuJPA..37....1C,2012AA...545A..28L,
2016ApJ...819...98A,2018AA...612A...4H},
SN 1006: \citet{2009AJ....137.2956D,2008PASJ...60S.153B,
2010AA...516A..62A,2016ApJ...823...44X,2017ApJ...851..100C},
RCW 103: \citet{1996AJ....111..340D,2014ApJ...781...64X},
CTB 33: \citet{1997ApJ...483..335S,2017Green,2013ApJ...774...36C},
RX J1713.7-3946: \citet{2009AA...505..157A,2008ApJ...685..988T,2018AA...612A...6H},
CTB 37A: \citet{1991ApJ...374..212K,2008AA...490..685A,2017ApJ...834..153Z},
CTB 37B: \citet{1991ApJ...374..212K,2006ApJ...636..777A,
2008AA...490..685A,2016ApJ...817...64X},
G349.7+0.2: \citet{2017Green,1976MNRAS.174..267C,1996AAS..118..329W,2015AA...574A.100H},
Hess J1731-347: \citet{2008ApJ...679L..85T,2017AA...608A..23D,2017ApJ...851..100C,
2018ApJ...853....2G},
Hess J1745-303: \citet{2008AA...483..509A,2011ApJ...735..115H,1984AAS...57..165R,1991MNRAS.249..262A}
}
\end{deluxetable}

\clearpage
\floattable
\begin{table}
\begin{center}
\caption{Spectral fitting parameters. \label{tbl-1}}
\tiny
\begin{tabular}{l|c|c|c|c|c|c|c|c|c}
\tableline\tableline
Source Name
& $\alpha$ &log$_{10} {E_{\rm br}\over{\rm GeV}}$
& log$_{10} {E_{\rm e, cut}\over{\rm GeV}}$
& log$_{10} {E_{\rm p, cut}\over{\rm GeV}}$
& log$_{10} {B\over\mu{\rm G}}$
& log$_{10} {W_{p}\over{\rm erg}}$
& ${W_{B}\over W_{e}}$
& $n \over \rm{cm}^-3$
& $\chi^2 \over \rm {NDF}$
\\

\tableline
{W28}
& {$1.76_{-0.03}^{+0.03}$}& {$0.18_{-0.11}^{+0.11}$}
& {$1.63$} & {$>5.72$}
& {$1.94_{-0.04}^{+0.04}$}
& {$49.36_{-0.02}^{+0.02}$}
&{$3000$}
&100
&{$\frac{24.3}{10}=2.43$}\\

\tableline
{W30}
& {$1.63_{-0.11}^{+0.10}$}& {$0.24_{-0.38}^{+0.33}$}
& {$2.06$} & {$>4.29$}
& {$1.86_{-0.13}^{+0.12}$}
& {$49.69_{-0.07}^{+0.07}$}
&{$736$}
&100
&{$\frac{6.0}{7}=0.86$}\\

\tableline
{3C391}
& {$1.99_{-0.05}^{+0.05}$}& {$1.15_{-0.14}^{+0.14}$}
& {$1.86$} & {$>3.81$}
& {$2.31_{-0.04}^{+0.04}$}
& {$49.03_{-0.03}^{+0.03}$}
&{$619$}
&300
&{$\frac{37.1}{19}=1.95$}\\

\tableline
Kes79$^{c}$
& $2.00_{-0.08}^{+0.08}$& $\rm NA$
& $E_{\rm p,cut}$ & $1.07_{-0.16}^{+0.16}$
& $1.70_{-0.05}^{+0.05}$
& $49.47_{-0.04}^{+0.04}$
&$21.3$
&100.0
&$\frac{64.2}{24}=2.68$\\

\tableline
{W44}
& {$1.60_{-0.04}^{+0.04}$}& {$0.73_{-0.09}^{+0.09}$}
& {$1.23$} & {$1.87_{-0.09}^{+0.11}$}
& {$2.28_{-0.03}^{+0.03}$}
& {$49.43_{-0.01}^{+0.01}$}
&{$1480$}
&200
&{$\frac{44.4}{48}=0.93$}\\

\tableline
{W49B}
& {$1.47_{-0.04}^{+0.04}$}& {$-0.21_{-0.24}^{+0.23}$}
& {$1.55$} & {$3.70_{-0.13}^{+0.13}$}
& {$2.40_{-0.06}^{+0.06}$}
& {$49.43_{-0.02}^{+0.02}$}
&{235}
&700
&{$\frac{18.9}{20}=1.00$} \\

\tableline
{W51C}
& {$1.56_{-0.02}^{+0.02}$}& {$0.31_{-0.08}^{+0.08}$}
& {$1.64$} & {$4.39_{-0.29}^{+0.30}$}
& {$2.08_{-0.03}^{+0.03}$}
& {$49.83_{-0.01}^{+0.01}$}
&{$708$}
&100
&{$\frac{59.7}{29}=2.06$} \\
\tableline
{W51C$^{b}$}
& {$1.64_{-0.02}^{+0.02}$}& {$0.32_{-0.05}^{+0.05}$}
& {$1.57$} & {$>5.78$}
& {$2.02_{-0.03}^{+0.03}$}
& {$49.79_{-0.01}^{+0.01}$}
&{$201$}
&100
&{$\frac{34.9}{29}=1.20$} \\

\tableline

{G73.9+0.9$^{c}$}
& {$0.78_{-0.19}^{+0.19}$}& $\rm NA$
& $E_{\rm p,cut}$ & {$0.96_{-0.09}^{+0.09}$}
& {$1.57_{-0.05}^{+0.05}$}
& {$49.34_{-0.04}^{+0.04}$}
&{$393$}
&{10} 
&{$\frac{22.4}{13}=1.72$}
\\

\tableline
{Cygnus Loop$^{c}$}
& {$1.86_{-0.08}^{+0.08}$}& {$\rm NA$}
& {$E_{\rm p,cut}$}
& {$1.09_{-0.10}^{+0.10}$}
& {$1.46_{-0.03}^{+0.03}$}
& {$48.72_{-0.02}^{+0.02}$}
& {$197$}
& 5.0
&{$\frac{20.5}{20}=1.03$}  \\

\tableline
{$\gamma$ Cygni}
& {$2.00_{-0.05}^{+0.05}$}& {$3.17_{-0.18}^{+0.17}$}
& {$2.63$}
& {$>4.97$}
& {$1.78_{-0.05}^{+0.05}$}
& {$50.25_{-0.02}^{+0.02}$}
& {$61.5$}
& 2.5
&{$\frac{23.6}{18}=1.31$}  \\
\tableline

HB21$^{c}$
& $1.20_{-0.12}^{+0.12}$ & $\rm NA$
& $E_{\rm p,cut}$ & $0.69_{-0.05}^{+0.04}$
& $1.71_{-0.01}^{+0.01}$
& $49.42_{-0.01}^{+0.01}$
&$564$
&15
&$\frac{36.2}{23}=1.57$ \\

\tableline
CTB109
& $1.94_{-0.09}^{+0.09}$ & $2.66_{-0.81}^{+0.38}$
& $3.28$ & $>4.82$
& $1.47_{-0.20}^{+0.17}$
& $49.84_{-0.12}^{+0.12}$
&$19.6$
&1.1
&$\frac{20.9}{8}=2.61$ \\

\tableline
Tycho 
& $2.15_{-0.02}^{+0.02}$& $3.37_{-0.12}^{+0.12}$
& $4.14_{-0.09}^{+0.08}$ & $ > 5.04$
& $2.15_{-0.05}^{+0.04}$
& $49.01_{-0.08}^{+0.08}$
&$23.5$
&0.3
&$\frac{55}{35}=1.57$  \\

\tableline
Tycho
& $2.16_{-0.02}^{+0.02}$& $3.36_{-0.11}^{+0.11}$
& $4.06_{-0.07}^{+0.07}$ & $ > 4.93$
& $2.29_{-0.05}^{+0.04}$
& $48.78_{-0.07}^{+0.07}$
&$92.2$
&10.0
&$\frac{44}{35}=1.26$  \\

\tableline

HB3$^{c}$
& $1.84_{-0.16}^{+0.17}$& $\rm NA$
& $E_{\rm p,cut}$ & $1.08_{-0.17}^{+0.17}$
& $1.04_{-0.04}^{+0.04}$
& $50.04_{-0.03}^{+0.03}$
&$6.4$
&2.0
&$\frac{16.4}{19}=0.86$\\

\tableline
G150.3+4.5
& $1.73_{-0.22}^{+0.22}$& $2.65_{-0.42}^{+0.36}$
& $6.04$ & $>6.37$
& $0.45_{-0.13}^{+0.13}$
& $48.33_{-0.05}^{+0.05}$
&$1.42$
&1.0
&$\frac{11.5}{12}=0.96$\\

\tableline

{HB9$^{c}$}
& ${ 1.75_{-0.34}^{+0.32}}$ & $\rm NA$
& $E_{\rm p,cut}$ & ${1.11_{-0.40}^{+0.43}}$
& ${ 0.61_{-0.08}^{+0.07}}$
& ${ 50.25_{-0.08}^{+0.08}}$
&{ 0.07}
&0.1
&${ \frac{7.3}{12}=0.61} $ \\

\tableline
{G166.0+4.3$^{c}$}
& ${ 1.32_{-0.18}^{+0.17}}$ & $\rm NA$
& $E_{\rm p,cut}$ & ${ 1.87_{-0.15}^{+0.14}}$
& ${ 0.57_{-0.24}^{+0.24}}$
& ${ 50.92_{-0.25}^{+0.26}}$
&{ 0.12}
&0.01
&${ \frac{6.92}{5}=1.38} $ \\

\tableline
{G166.0+4.3$^{c}$}
& ${ 1.26_{-0.18}^{+0.17}}$ & $\rm NA$
& $E_{\rm p,cut}$ & ${ 1.18_{-0.16}^{+0.16}}$
& ${ 1.62_{-0.10}^{+0.10}}$
& ${ 49.18_{-0.07}^{+0.07}}$
&{717}
&10.0
&${ \frac{7.70}{5}=1.54} $ \\

\tableline
{S147}
& ${ 1.36_{-0.06}^{+0.06}}$ & ${ -0.14_{-0.13}^{+0.12}}$
& ${ 0.09} $ & ${ >3.86}$
& ${ 2.77_{-0.09}^{+0.09}}$
& ${ 47.71_{-0.05}^{+0.05}}$
&{ $2.7 \times 10^8$}
&250
&${ \frac{17.3}{17}=1.02} $ \\

\tableline
{S147}
& ${ 1.53_{-0.11}^{+0.11}}$ & ${ 0.51_{-0.12}^{+0.12}}$
& ${ 3.57} $ & ${ >4.65}$
& ${ 1.03_{-0.05}^{+0.05}}$
& ${ 49.94_{-0.04}^{+0.04}}$
&{ 31.6}
&1.0
&${ \frac{19.8}{17}=1.16} $ \\

\tableline

{IC 443}
& ${ 1.38_{-0.03}^{+0.03}}$
& ${ 0.12_{-0.07}^{+0.07}}$
& ${ 1.35} $ & ${ 3.22_{-0.10}^{+0.10}}$
& ${ 2.14_{-0.02}^{+0.02}}$
& ${ 48.96_{-0.01}^{+0.01}}$
&{ $2280$}
&140
&${ \frac{92.0}{64}=1.44} $\\

\tableline
Monoceros Loop
& $1.63_{-0.02}^{+0.02}$& $0.74_{-0.11}^{+0.11}$
& $2.97$ & $>5.77$
& $1.31_{-0.03}^{+0.03}$
& $50.29_{-0.03}^{+0.03}$
&$224$
&3.6
&$\frac{42.5}{16}=1.63$ \\

\tableline
{Puppis A}
& ${ 2.08_{-0.02}^{+0.02}}$ & ${ 3.23_{-0.56}^{+0.48}}$
& ${ 2.50} $ & ${ >4.57}$
& ${ 1.97_{-0.02}^{+0.02}}$
& ${ 49.53_{-0.04}^{+0.04}}$
&{ 500}
&4.0
&${ \frac{48.8}{30}=1.46} $ \\

\tableline
{RX J0852-4622$^{d}$}
& ${ 2.21_{-0.04}^{+0.04}}$ & $\rm NA$
& ${ 4.30_{-0.06}^{+0.06}} $ & ${ >5.15}$
& ${ 1.03_{-0.04}^{+0.04}}$
& ${ 49.61_{-0.05}^{+0.05}}$
&{ 2.79}
&{ 0.01}
&${ \frac{26.0}{15}=1.73} $\\

\tableline
{RX J0852-4622}
& ${ 1.33_{-0.05}^{+0.05}}$ & ${ 1.13_{-0.16}^{+0.18}}$
& ${ 4.38_{-0.06}^{+0.06}} $ & ${ >5.15}$
& ${ 1.04_{-0.04}^{+0.04}}$
& ${ 49.70_{-0.04}^{+0.04}}$
&{ 2.6}
&{ 0.01}
&${ \frac{16.6}{14}=1.19} $\\

\tableline

{G296.5+10.0$^{d}$}
& ${ 1.82_{-0.10}^{+0.10}}$ & $\rm NA$
& ${ 2.11} $ & ${ >4.26}$
& ${ 2.01_{-0.11}^{+0.10}}$
& ${ 49.74_{-0.22}^{+0.23}}$
&{ 6770}
&1.0
&${ \frac{3.48}{6}=0.58} $\\

\tableline
{Kes 17$^{d}$}
& ${ 2.04_{-0.16}^{+0.17}}$ & $\rm NA$
& ${ 3.01} $ & ${ >4.20}$
& ${ 1.77_{-0.16}^{+0.16}}$
& ${ 50.39_{-0.14}^{+0.13}}$
&{ $7.6$}
&10.0
&${ \frac{1.06}{3}=0.35} $\\
\tableline

{RCW 86}
& ${ 2.26_{-0.02}^{+0.02}}$ & ${ 3.92_{-0.09}^{+0.08}}$
& ${ 4.42_{-0.03}^{+0.04}} $ & ${ >5.23}$
& ${ 1.44_{-0.02}^{+0.02}}$
& ${ 49.82_{-0.03}^{+0.03}}$
&{ 15.3}
&{ 0.01}
&${ \frac{31.5}{22}=1.43} $\\

\tableline
{MSH 15-56}
& ${ 1.43_{-0.14}^{+0.14}}$ & ${ 2.13_{-0.20}^{+0.17}}$
& ${ 2.40}$& ${ >3.06}$
& ${ 1.81_{-0.08}^{+0.09}}$
& ${ 51.05_{-0.13}^{+0.13}}$
&{ 34.3}
&{ 0.1}
&${ \frac{10.1}{7}=1.44} $\\

\tableline
{MSH 15-56}
& ${ 1.61_{-0.12}^{+0.12}}$ & ${ 1.17_{-0.11}^{+0.08}}$
& ${ 3.02}$& ${ >3.83}$
& ${ 1.60_{-0.10}^{+0.10}}$
& ${ 50.75_{-0.06}^{+0.06}}$
&{ 9.7}
&{ 1.0}
&${ \frac{8.0}{7}=1.14} $\\

\tableline
{SN 1006$^{d}$}
& ${ 2.12_{-0.02}^{+0.02}}$ & $\rm NA$
& ${ 3.91_{-0.04}^{+0.04}} $ & ${ >4.91}$
& ${ 1.78_{-0.03}^{+0.03}}$
& ${ 48.94_{-0.05}^{+0.05}}$
&{ 150}
&{ 0.085}
&${ \frac{66.0}{39}=1.69} $\\
\tableline

{RCW 103$^{d}$}
& ${ 2.12_{-0.07}^{+0.07}}$ & $\rm NA$
& ${ 3.92} $ & ${ >4.77}$
& ${ 1.44_{-0.08}^{+0.08}}$
& ${ 50.04_{-0.06}^{+0.06}}$
&{ 0.41}
&{ 10}
&${ \frac{1.1}{7}=0.16} $\\

\tableline

{CTB 33}
& ${ 1.89_{-0.26}^{+0.29}}$ & ${1.02_{-0.45}^{+0.44}}$
& ${ 3.46} $ & ${ >4.66}$
& ${ 1.50_{-0.08}^{+0.09}}$
& ${ 49.50_{-0.07}^{+0.07}}$
&{ 0.17}
&{ 600}
&${ \frac{6.92}{5}=1.38} $\\

\tableline
RX J1713.7-3946$^{a}$
& $1.81_{-0.02}^{+0.02}$& $3.10_{-0.05}^{+0.05}$
& $4.89_{-0.004}^{+0.004}$ & $>5.57$
& $1.29_{-0.004}^{+0.004}$
& $49.46_{-0.03}^{+0.03}$
& 6.0 &0.01
& $\frac{445}{240}=1.85$
\\

\tableline

{CTB 37A}
& ${ 1.47_{-0.02}^{+0.02}}$ & ${ 0.36_{-0.17}^{+0.19}}$
& ${ 1.0} $ & ${ >5.96}$
& ${ 2.40_{-0.10}^{+0.12}}$
& ${ 49.82_{-0.02}^{+0.02}}$
&{ 607}
&{ 100}
&${ \frac{23.4}{16}=1.46} $
\\

\tableline

{CTB 37B}
& ${ 1.49_{-0.11}^{+0.11}}$ & ${ 2.40_{-0.34}^{+0.33}}$
& ${ 0.81} $ & ${ >5.34}$
& ${ 2.84_{-0.15}^{+0.15}}$
& ${ 50.51_{-0.04}^{+0.04}}$
&{ $1.04 \times 10^5$}
&{ 10}
&${ \frac{15.6}{14}=1.11} $
\\

\tableline

{CTB 37B}
& ${ 1.58_{-0.07}^{+0.07}}$ & ${ 3.06_{-0.20}^{+0.19}}$
& ${ 2.47} $ & ${ >5.32}$
& ${ 1.97_{-0.06}^{+0.06}}$
& ${ 51.60_{-0.04}^{+0.04}}$
&{ 28.3}
&{ 0.5}
&${ \frac{14.1}{14}=1.00} $
\\

\tableline

{G349.7+0.2}
& ${ 2.06_{-0.12}^{+0.13}}$ & ${ 2.82_{-0.38}^{+0.30}}$
& ${ 2.70} $ & ${ >5.00}$
& ${ 2.00_{-0.12}^{+0.12}}$
& ${ 50.09_{-0.04}^{+0.04}}$
&{ 1.30}
&{ 35}
&${ \frac{5.2}{10}=0.52} $
\\

\tableline
Hess J1731-347
& $1.86_{-0.04}^{+0.04}$& $3.65_{-0.10}^{+0.10}$
& $4.27_{-0.02}^{+0.02}$ & $>5.19$
& $1.46_{-0.02}^{+0.02}$
& $49.42_{-0.04}^{+0.04}$
&45.1
&0.01
&$\frac{283.9}{322}=0.88$
\\

\tableline

{Hess J1745-303}
& ${ 1.64_{-0.04}^{+0.04}}$ & ${ 0.52_{-0.18}^{+0.20}}$
& ${ 2.03} $ & ${ >5.37}$
& ${ 1.66_{-0.08}^{+0.08}}$
& ${ 49.53_{-0.08}^{+0.08}}$
&{167}
&{ 100}
&${ \frac{3.62}{8}=0.45} $
\\

\tableline\tableline
\end{tabular}
\end{center}
\tablecomments{Note: $^{a}$, the cutoff shape parameter have been changed to 2.0--super-exponential high energy cutoff; $^{b}$, the electron distribution of a single
power-law have been adopted in our modified model; $^{c}$, $E_{e,cut}=E_{e,cut}$ have been used in modified model; $^{d}$,
the particle distribution have been modified to that with a power-law form plus an exponential cutoff .\\
}
\end{table}

\begin{figure}
\gridline{\fig{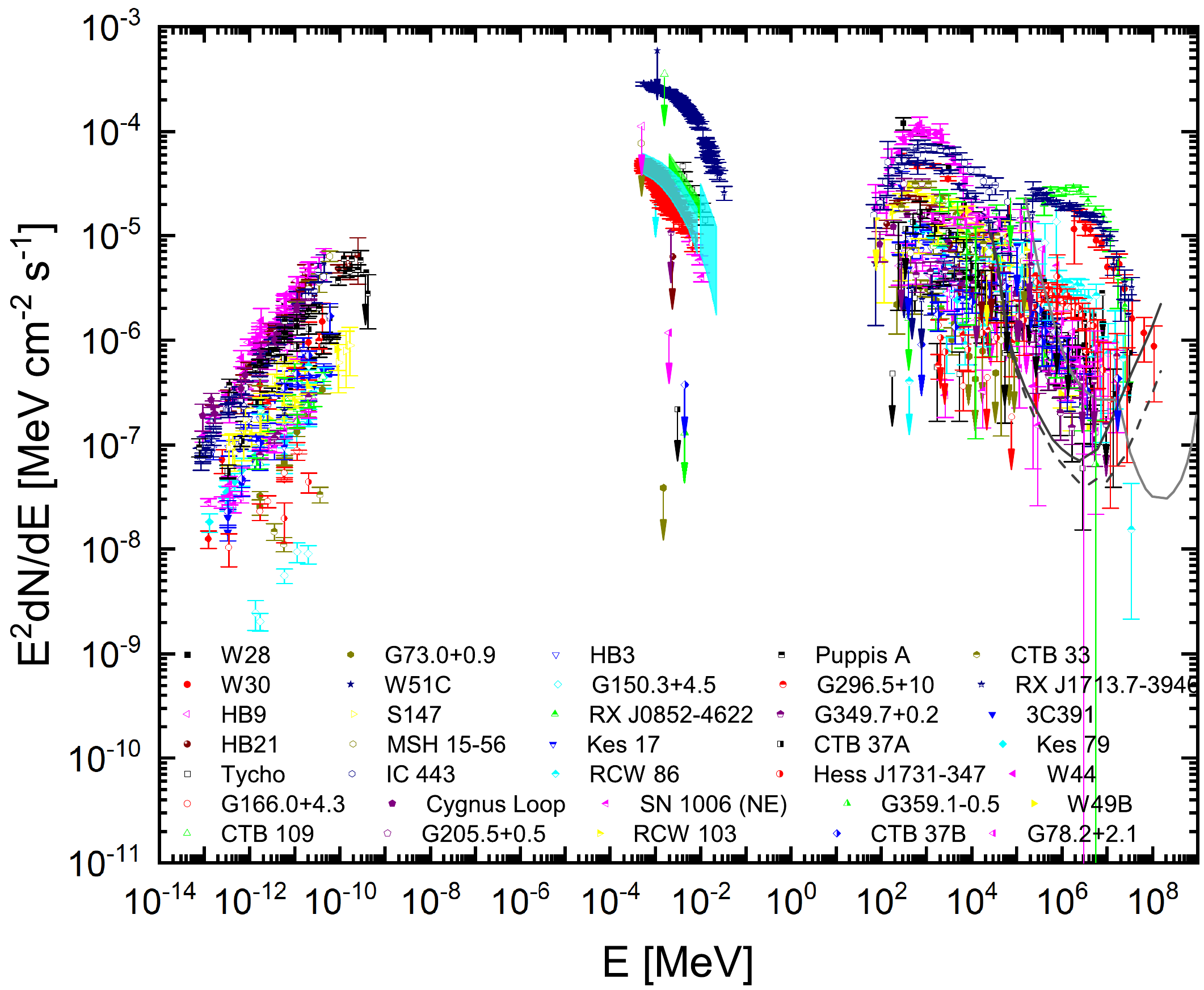}{0.4\textwidth}{ }
          \fig{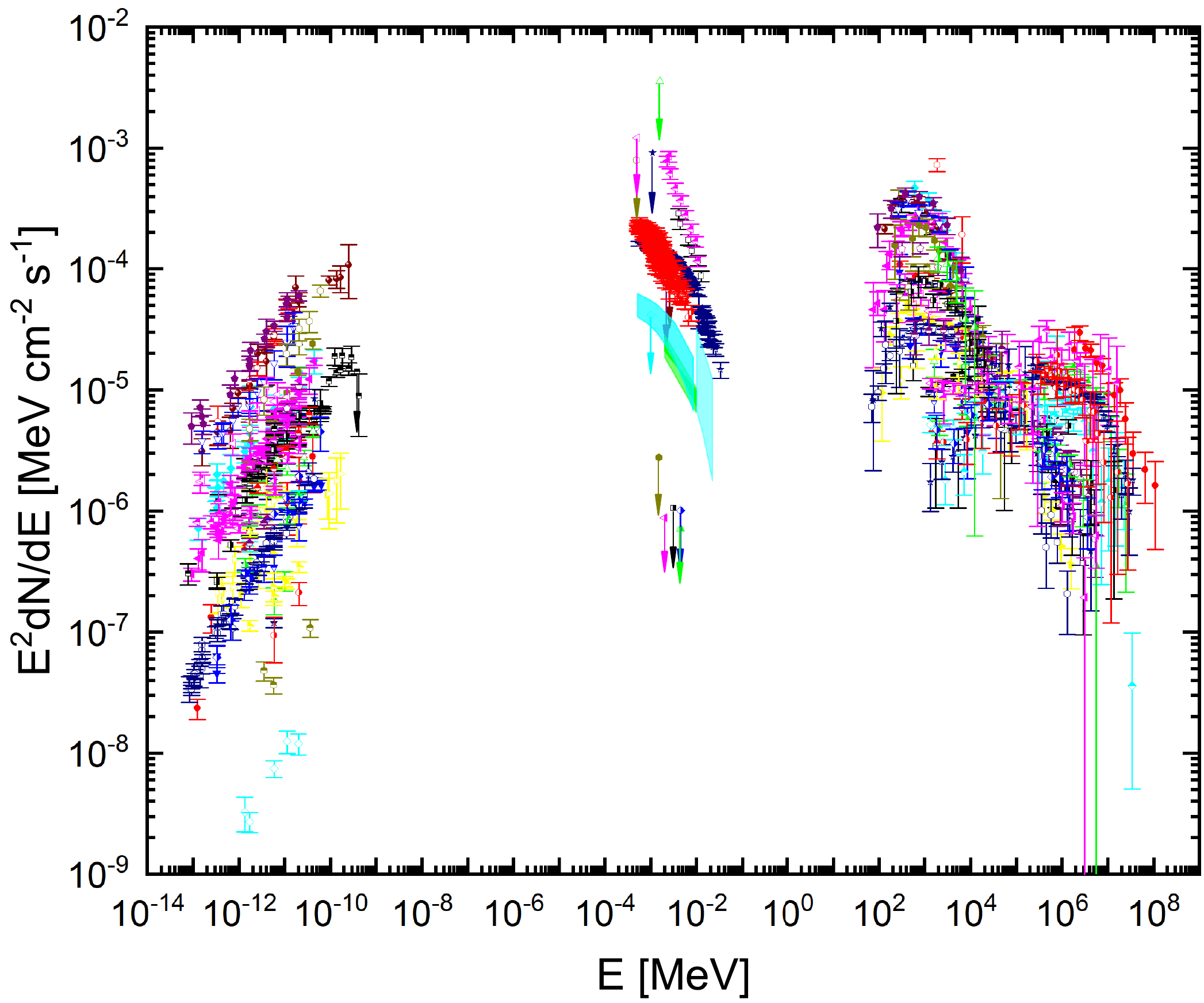}{0.4\textwidth}{ }
          }
\caption{The left panel shows
the multi-wavelength spectral data of 35 SNRs. The right panel
shows the spectra normalized at 100 GeV to $10^{-5}$ MeV cm$^{-2}$ s$^{-1}$.
The gray line is the differential sensitivity (one year) of LHAASO (Decl. from $-10^{0}$ to $70^{0}$)
\citep{2016NPPP..279..166D}, and the dark gray and dashed dark gray lines represent
the differential sensitivities of North and South (50 hour) of CTA, respectively
(https://www.cta-observatory.org/). Noted that all sources have radio data and GeV emission detected by \textit{Fermi}-LAT, but only six sources have non-thermal X-ray data, 12 sources have the upper limits of non-thermal X-ray emission and 16 sources have TeV data.}
\end{figure}

\figsetstart
\figsetnum{2}
\figsettitle{The best fit to the spectral energy distribution (SED) and 1D probability distribution of the parameters for individual SNRs in our sample}

\figsetgrpstart
\figsetgrpnum{2.1}
\figsetgrptitle{W28 (a)
}
\figsetplot{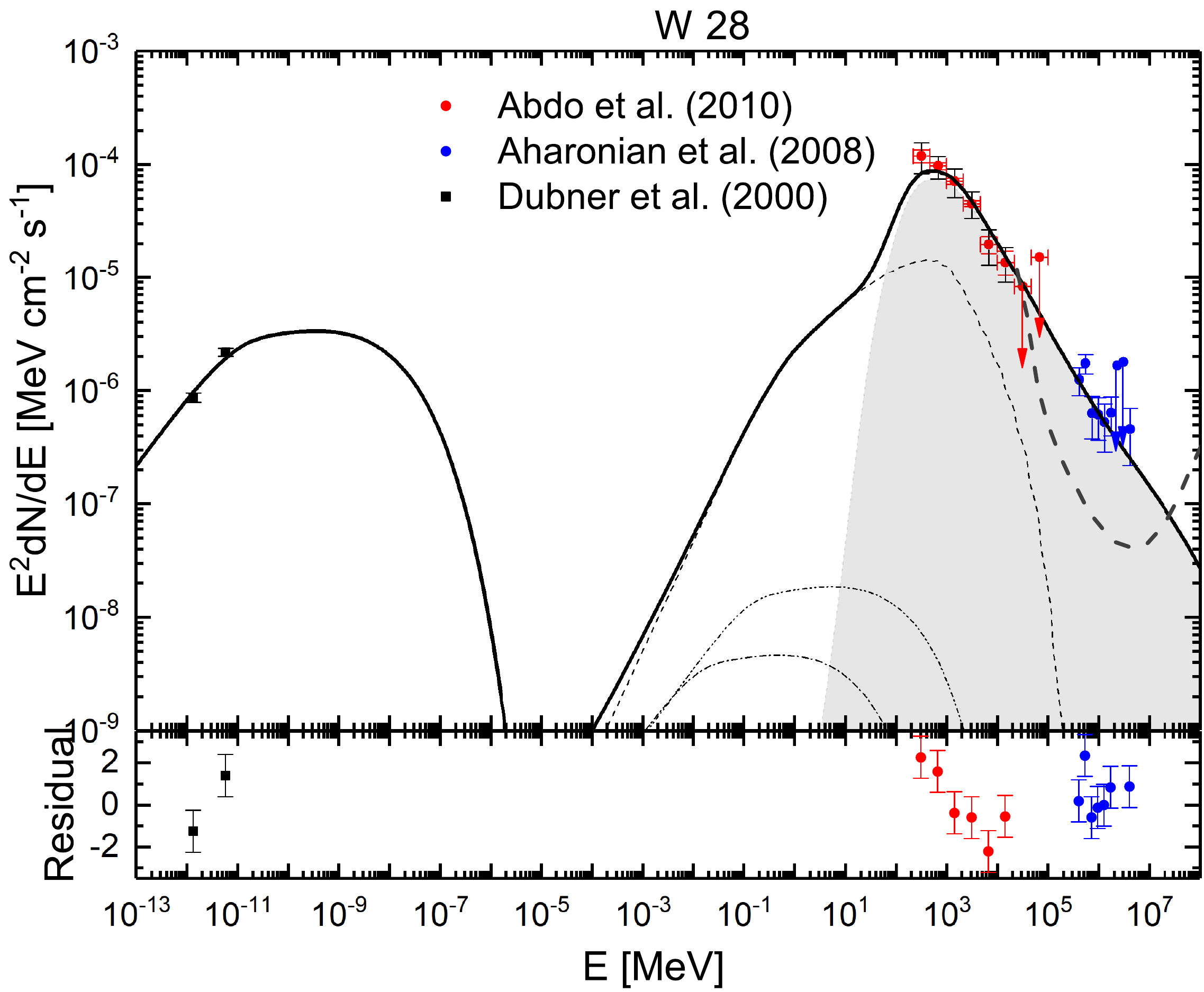}
\figsetgrpnote{The best fit to the spectral energy distribution (SED) for W28 }
\figsetgrpend

\figsetgrpstart
\figsetgrpnum{2.2}
\figsetgrptitle{W28 (b)
}
\figsetplot{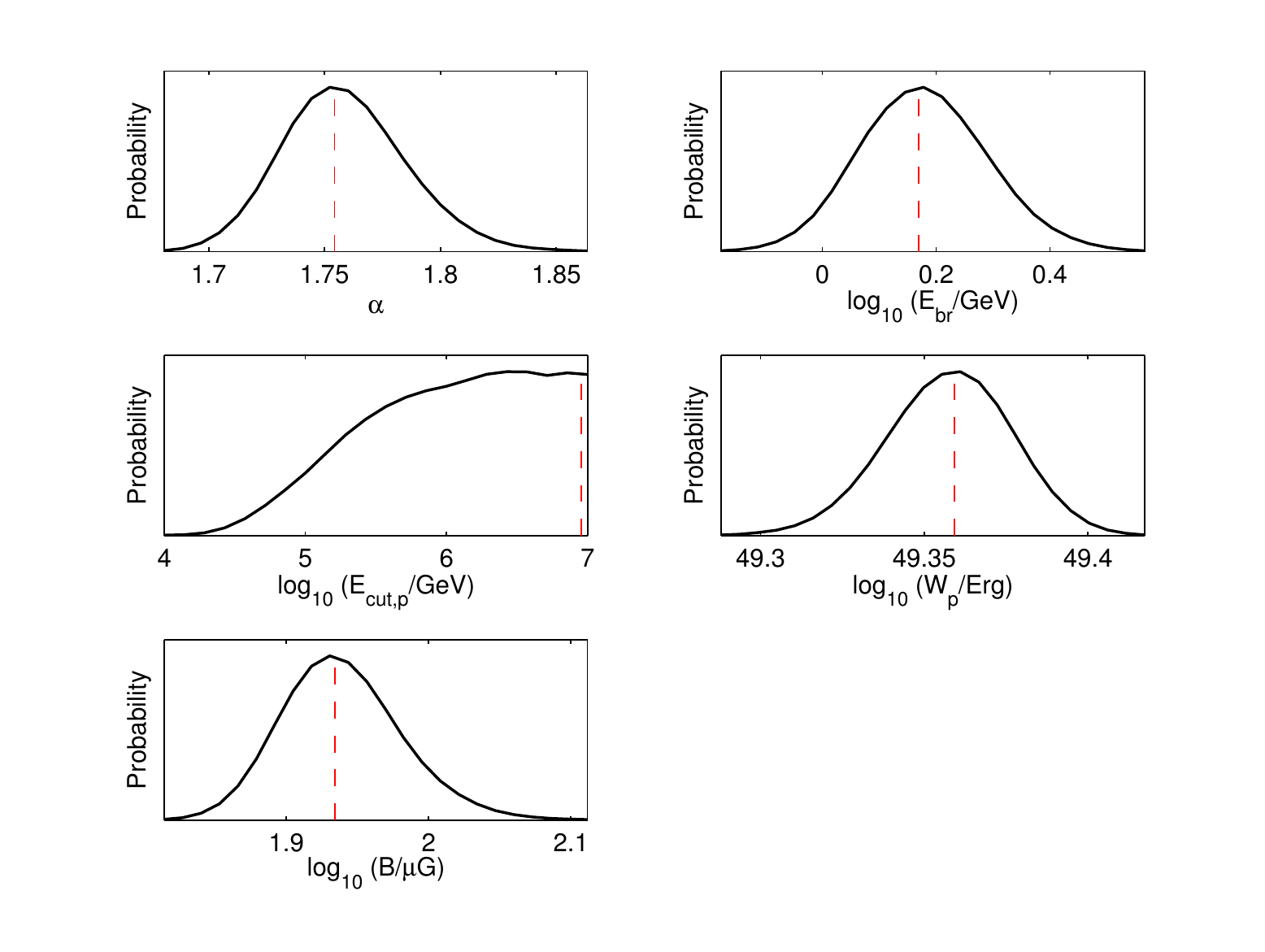}
\figsetgrpnote{1D probability distribution of the parameters for W28 }
\figsetgrpend

\figsetgrpstart
\figsetgrpnum{2.3}
\figsetgrptitle{W30 (a)
}
\figsetplot{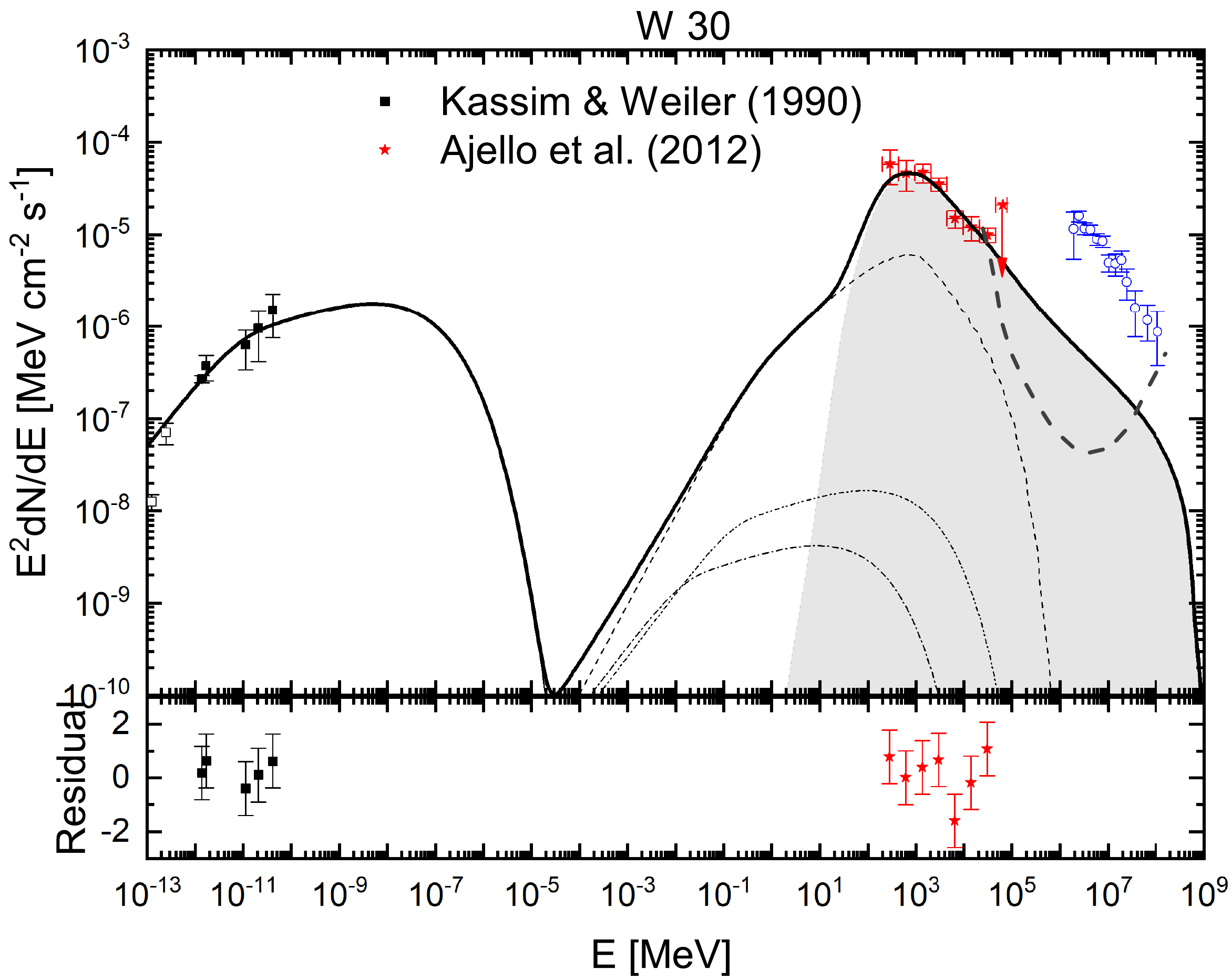}
\figsetgrpnote{The best fit to the spectral energy distribution (SED) for W30 }
\figsetgrpend

\figsetgrpstart
\figsetgrpnum{2.4}
\figsetgrptitle{W30 (b)
}
\figsetplot{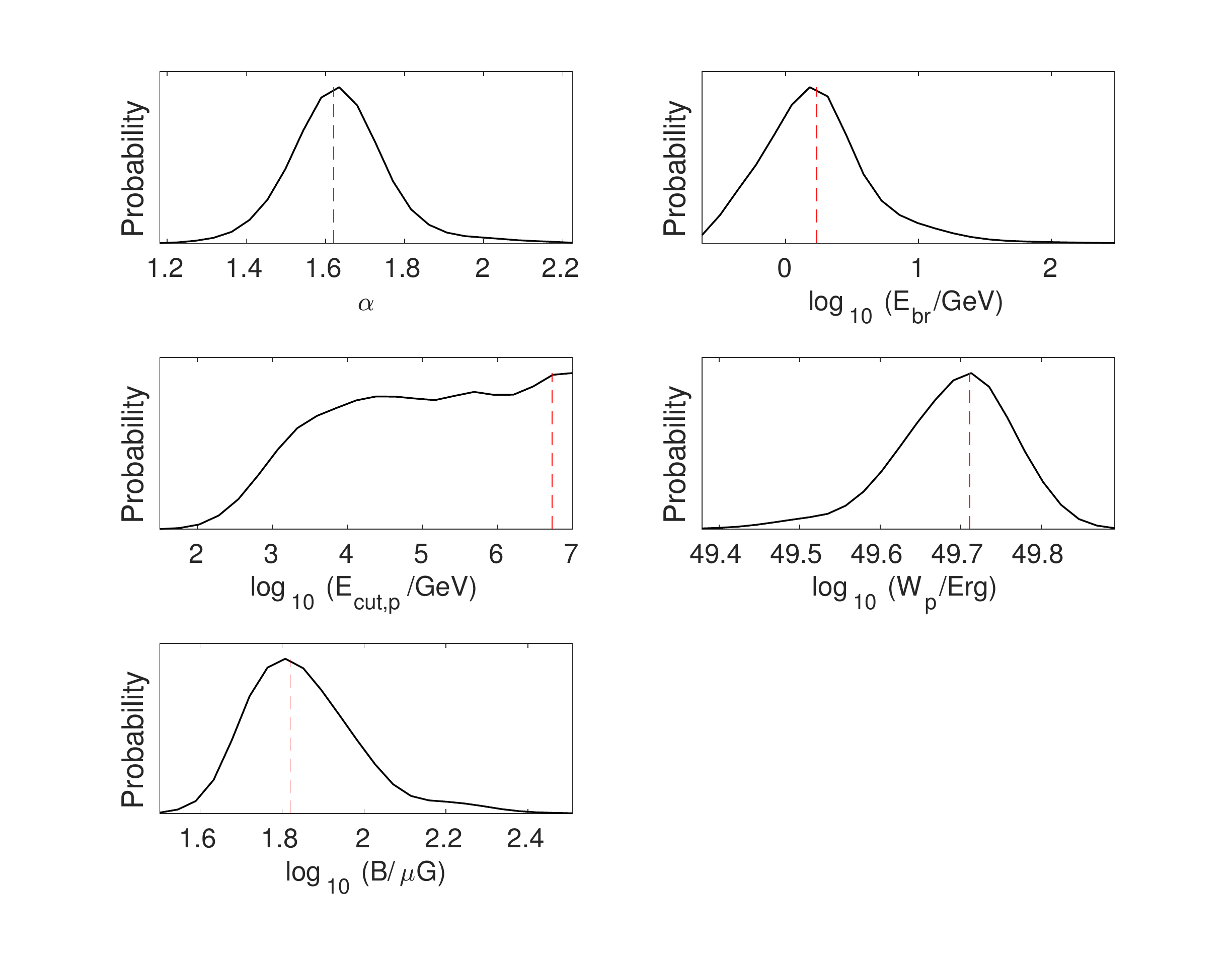}
\figsetgrpnote{1D probability distribution of the parameters for W30 }
\figsetgrpend

\figsetgrpstart
\figsetgrpnum{2.5}
\figsetgrptitle{3C 391 (a)
}
\figsetplot{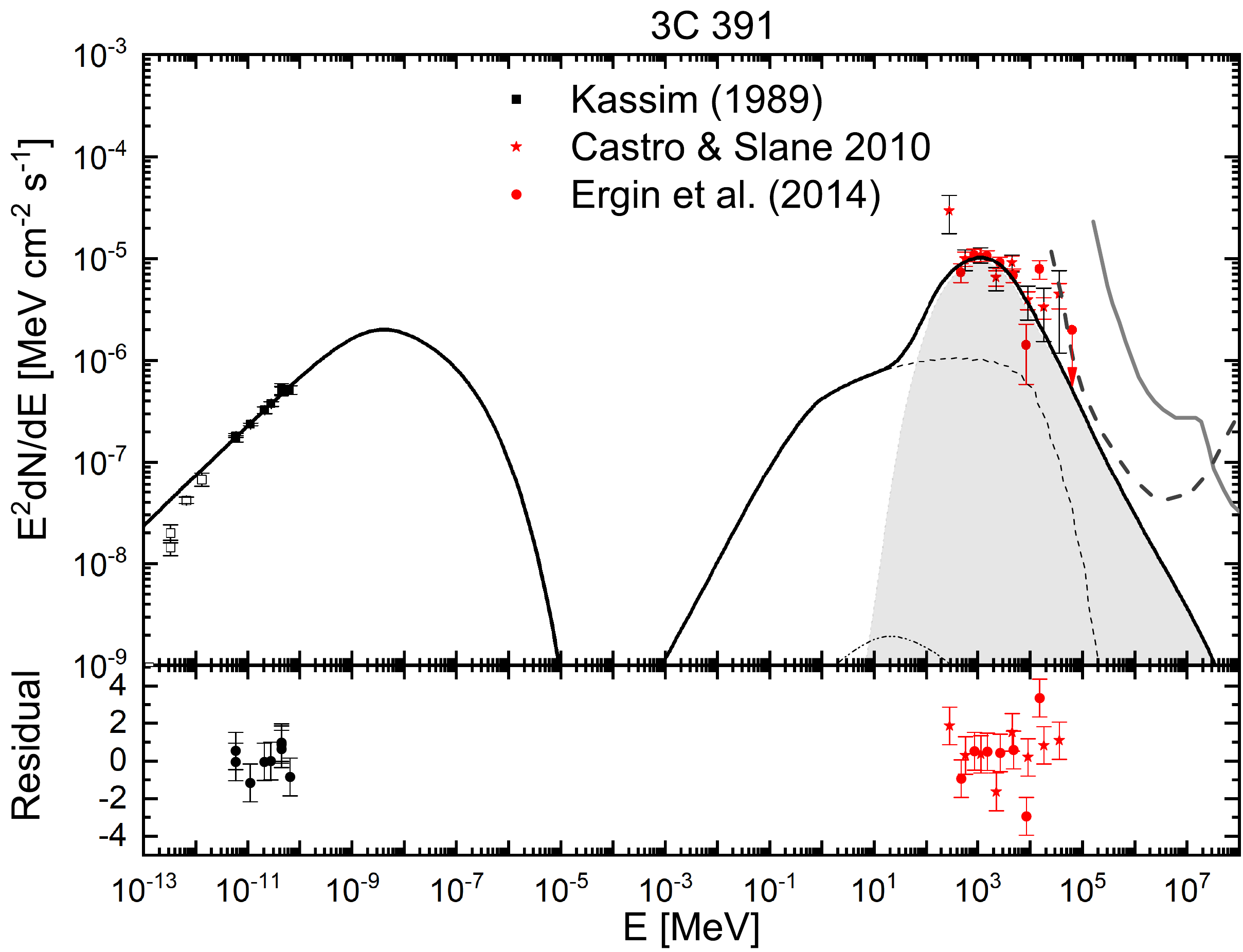}
\figsetgrpnote{The best fit to the spectral energy distribution (SED) for 3C 391}
\figsetgrpend

\figsetgrpstart
\figsetgrpnum{2.6}
\figsetgrptitle{3C 391 (b)
}
\figsetplot{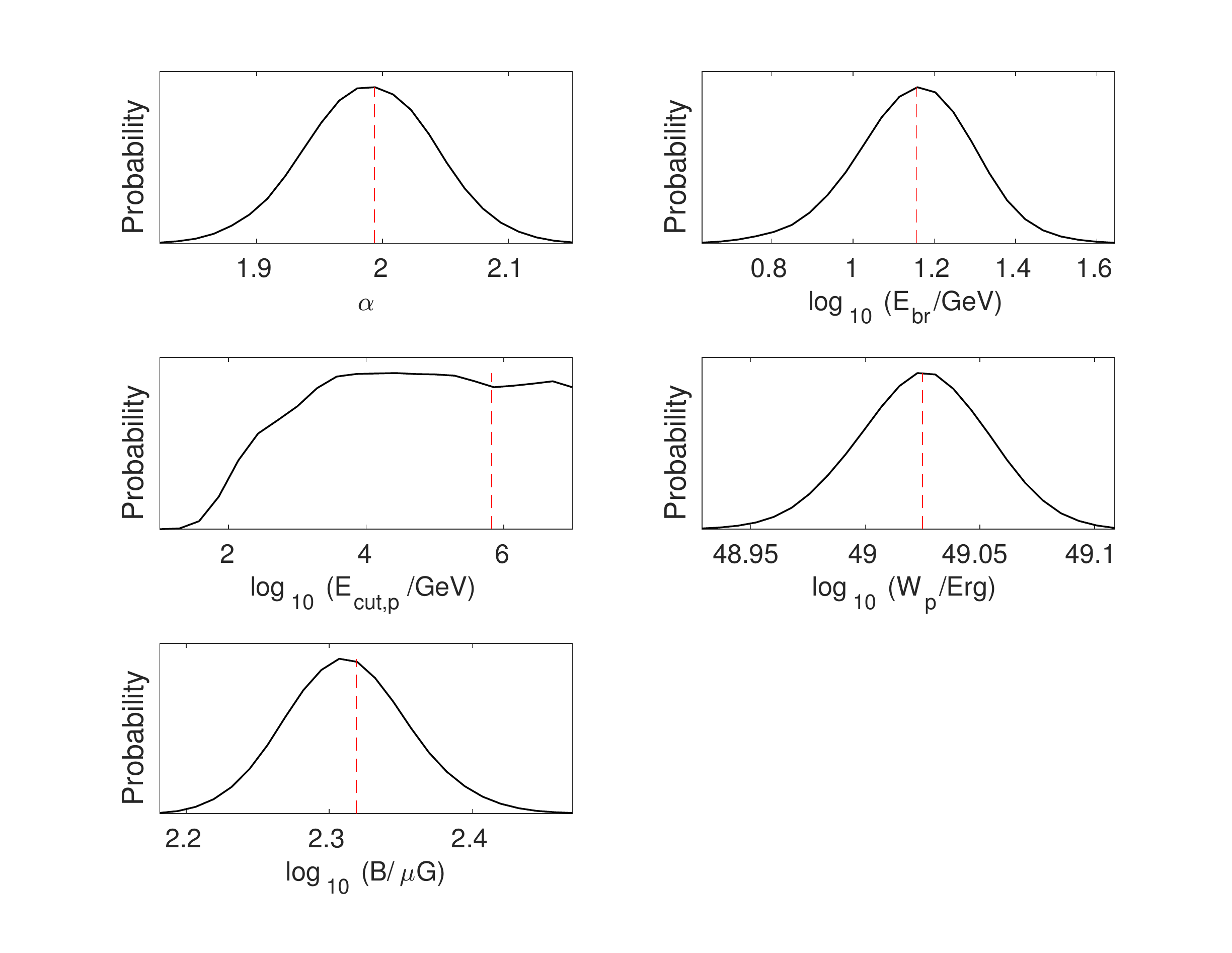}
\figsetgrpnote{1D probability distribution of the parameters for 3C 391}
\figsetgrpend

\figsetgrpstart
\figsetgrpnum{2.7}
\figsetgrptitle{Kes 79 (a)
}
\figsetplot{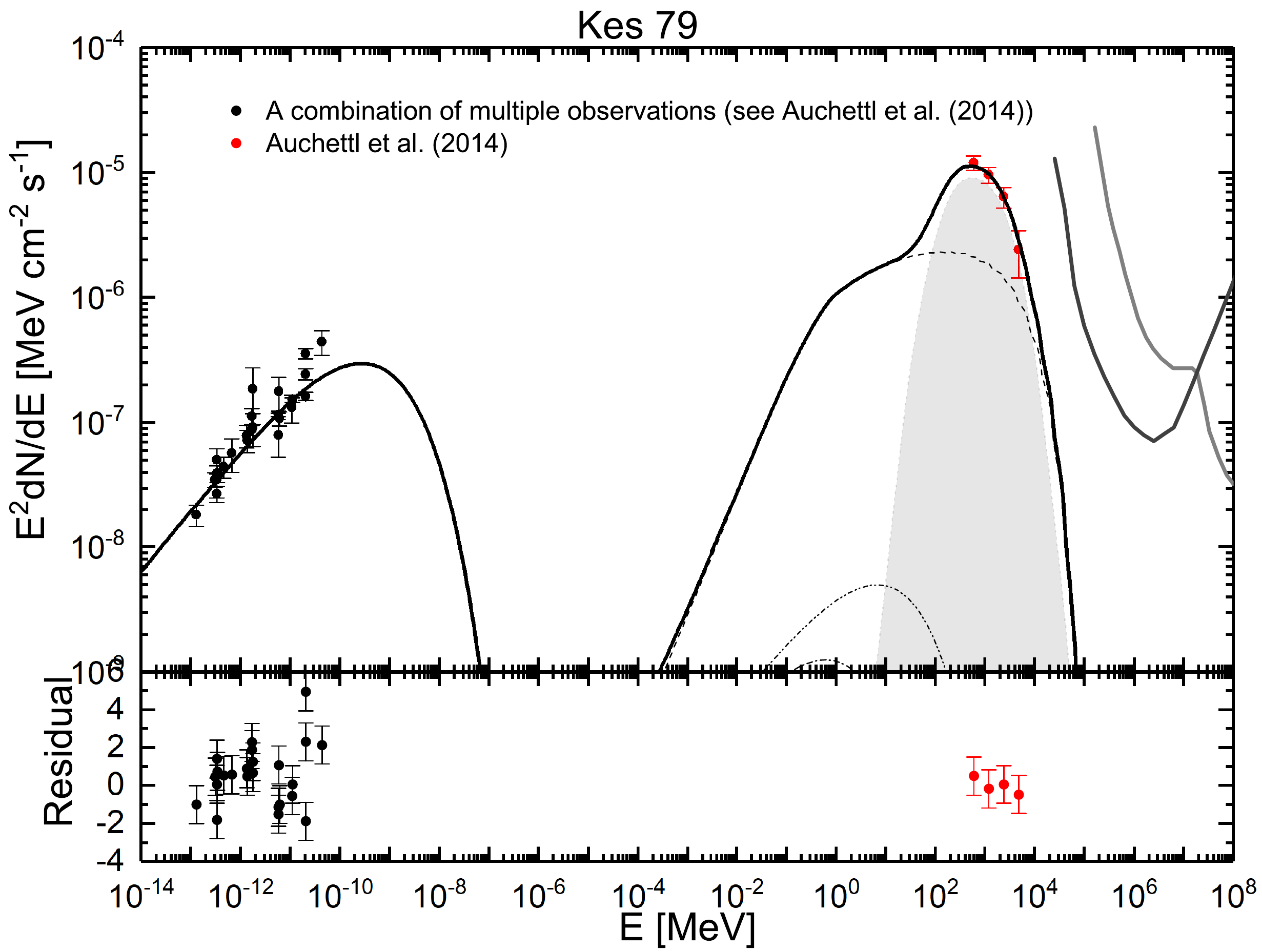}
\figsetgrpnote{The best fit to the spectral energy distribution (SED) for Kes 79 }
\figsetgrpend

\figsetgrpstart
\figsetgrpnum{2.8}
\figsetgrptitle{Kes 79 (b)
}
\figsetplot{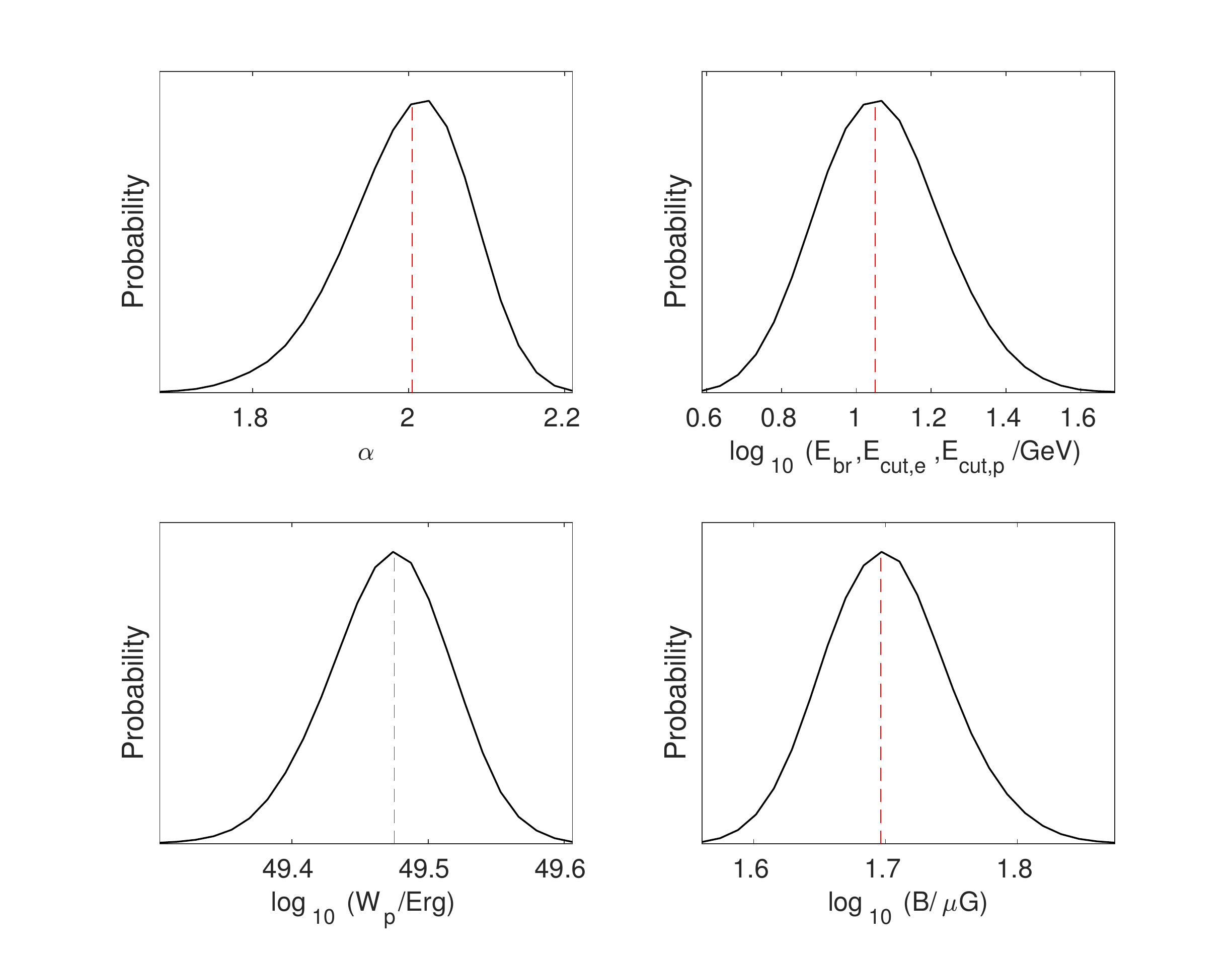}
\figsetgrpnote{1D probability distribution of the parameters  for Kes 79}
\figsetgrpend

\figsetgrpstart
\figsetgrpnum{2.9}
\figsetgrptitle{W44 (a)
}
\figsetplot{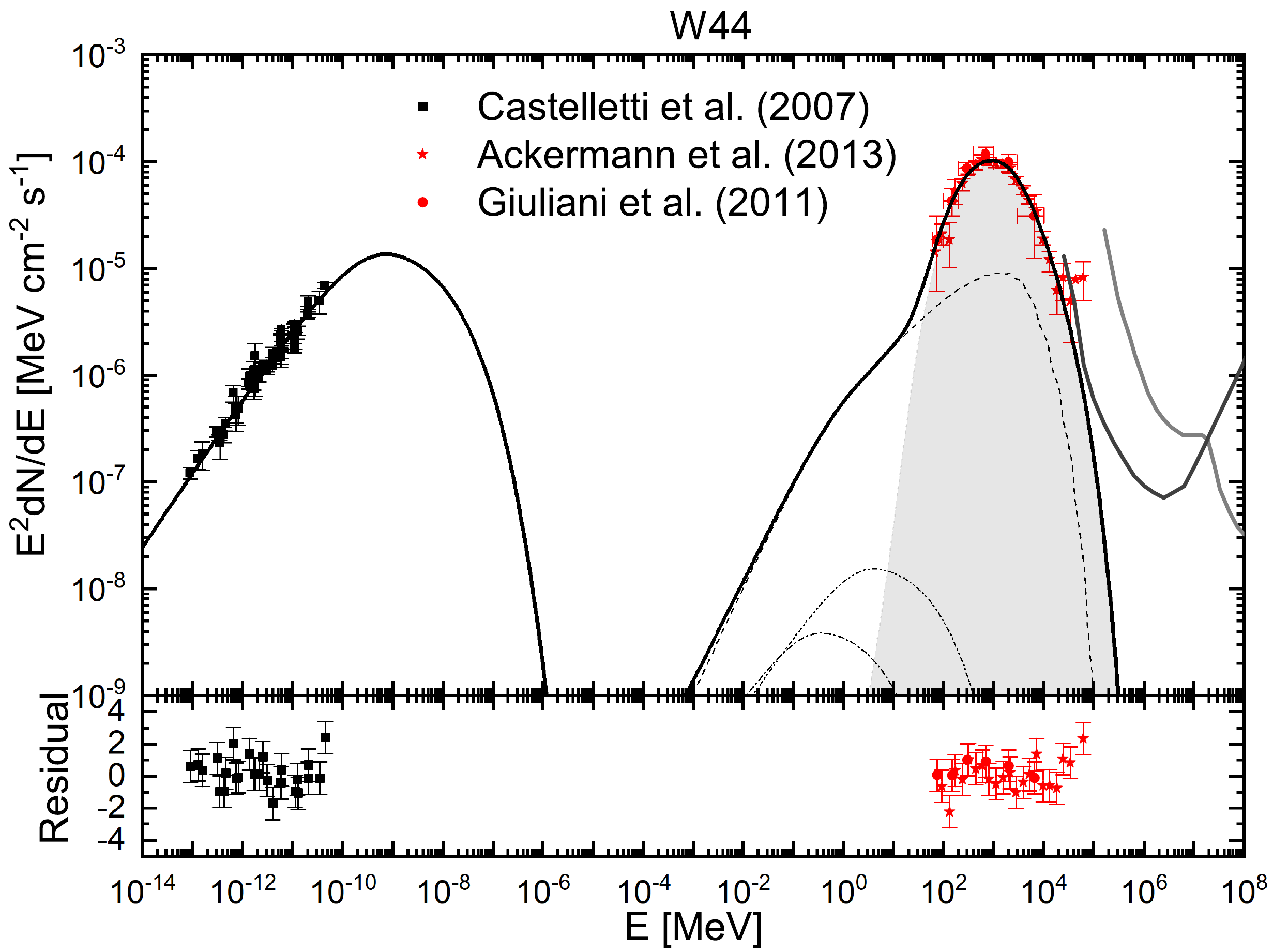}
\figsetgrpnote{The best fit to the spectral energy distribution (SED) for W44 }
\figsetgrpend

\figsetgrpstart
\figsetgrpnum{2.10}
\figsetgrptitle{W44 (b)
}
\figsetplot{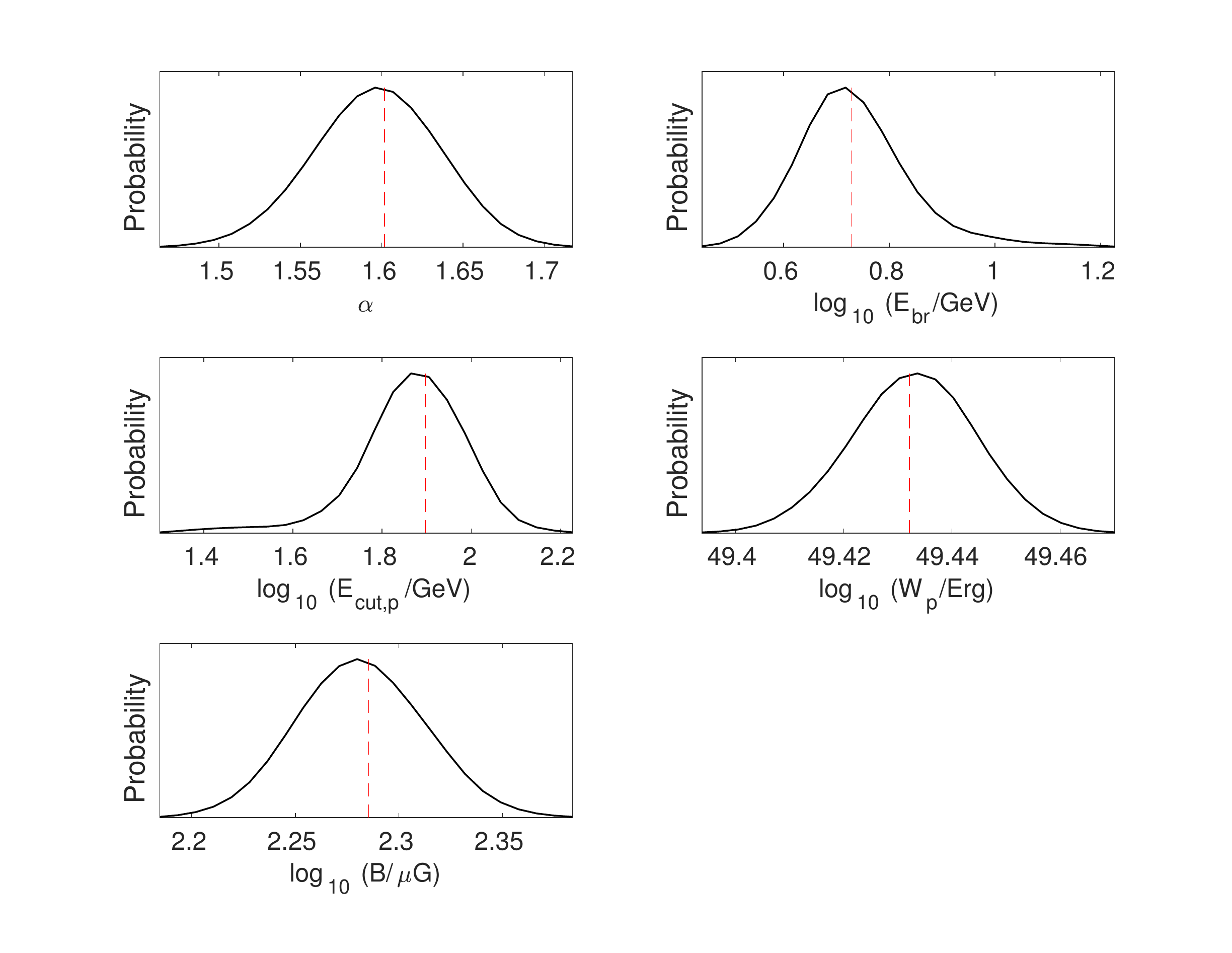}
\figsetgrpnote{ 1D probability distribution of the parameters for W44 }
\figsetgrpend

\figsetgrpstart
\figsetgrpnum{2.11}
\figsetgrptitle{W49B (a)
}
\figsetplot{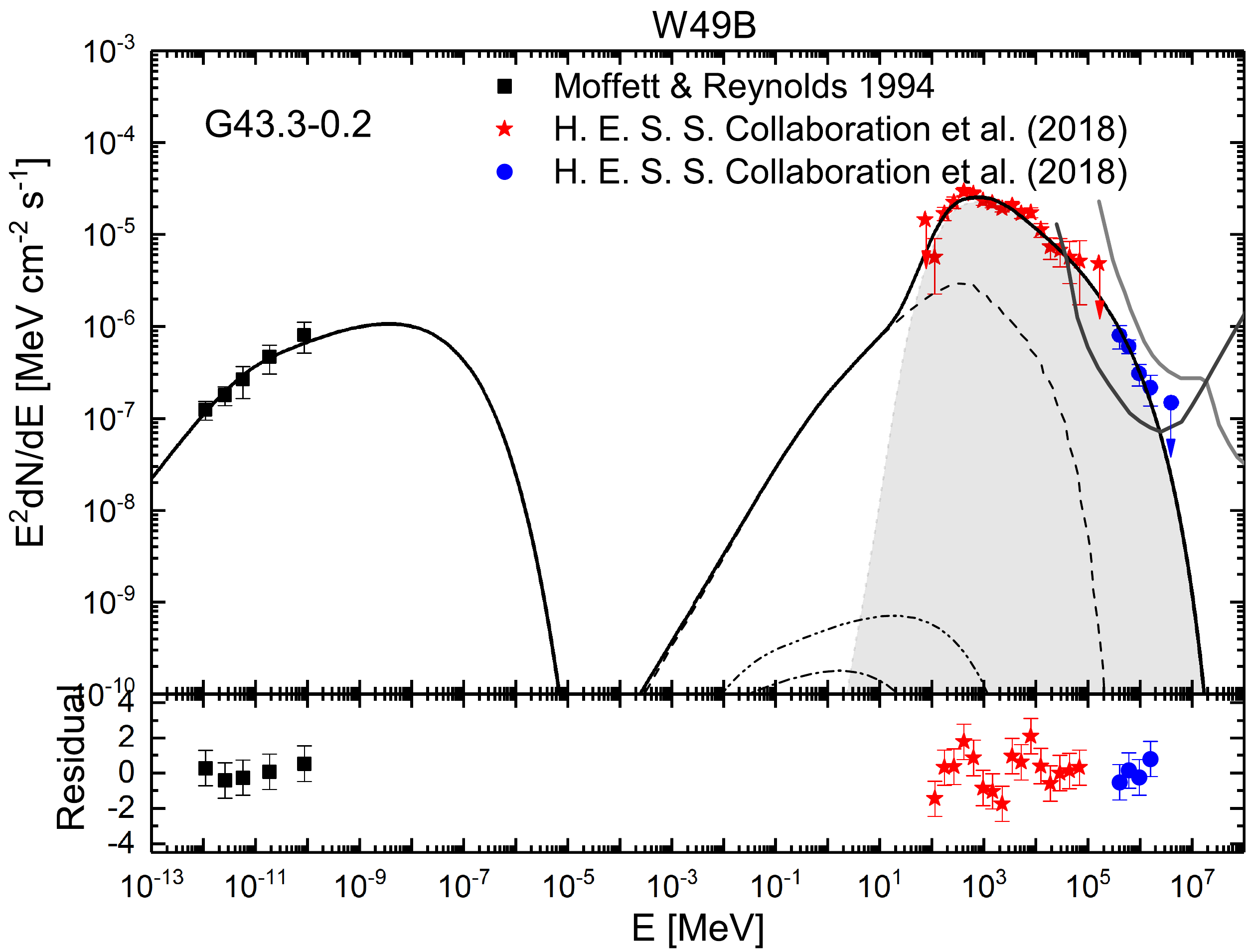}
\figsetgrpnote{The best fit to the spectral energy distribution (SED) for W49B }
\figsetgrpend

\figsetgrpstart
\figsetgrpnum{2.12}
\figsetgrptitle{W49B (b)
}
\figsetplot{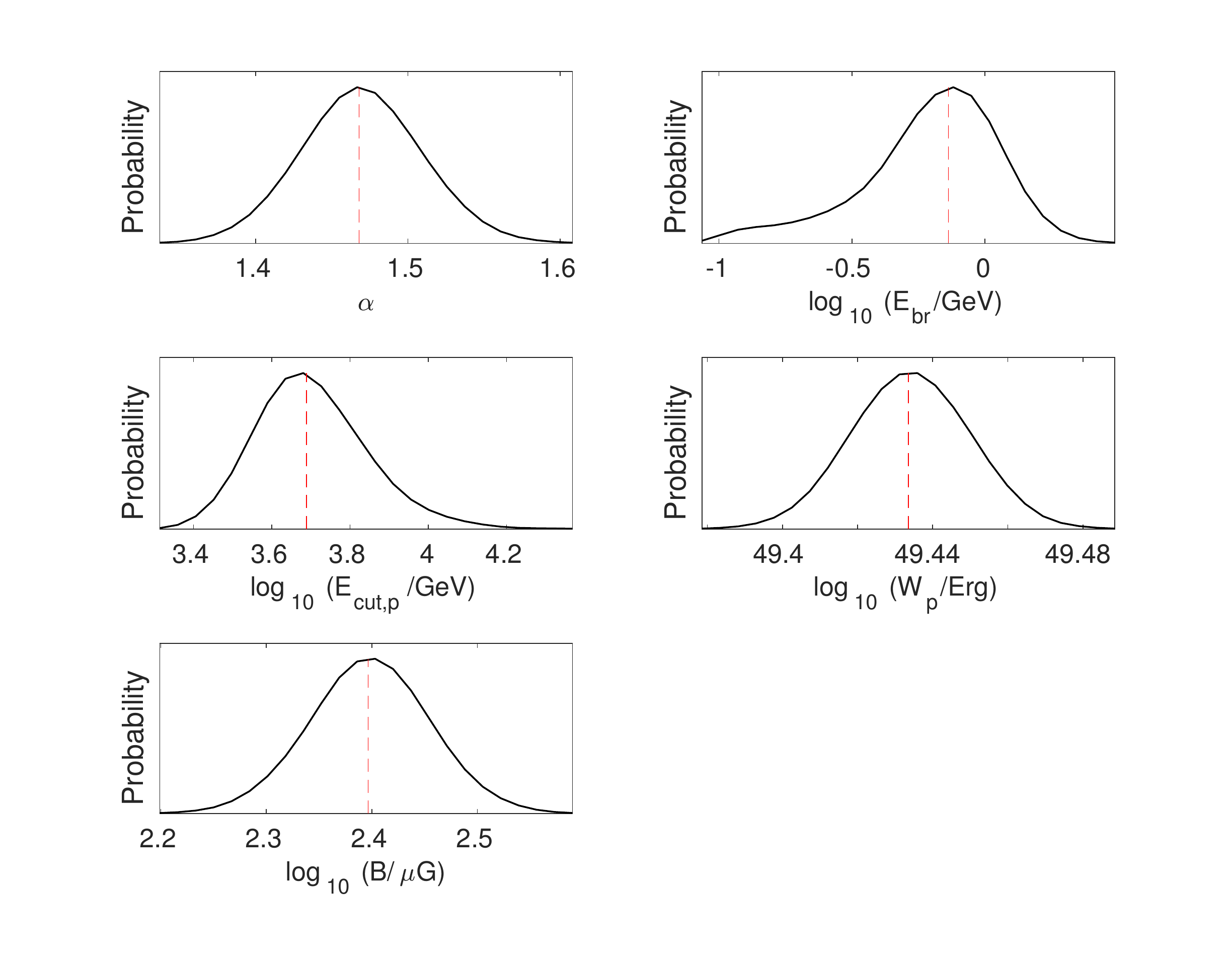}
\figsetgrpnote{1D probability distribution of the parameters for W49B}
\figsetgrpend

\figsetgrpstart
\figsetgrpnum{2.13}
\figsetgrptitle{W51C (a)
}
\figsetplot{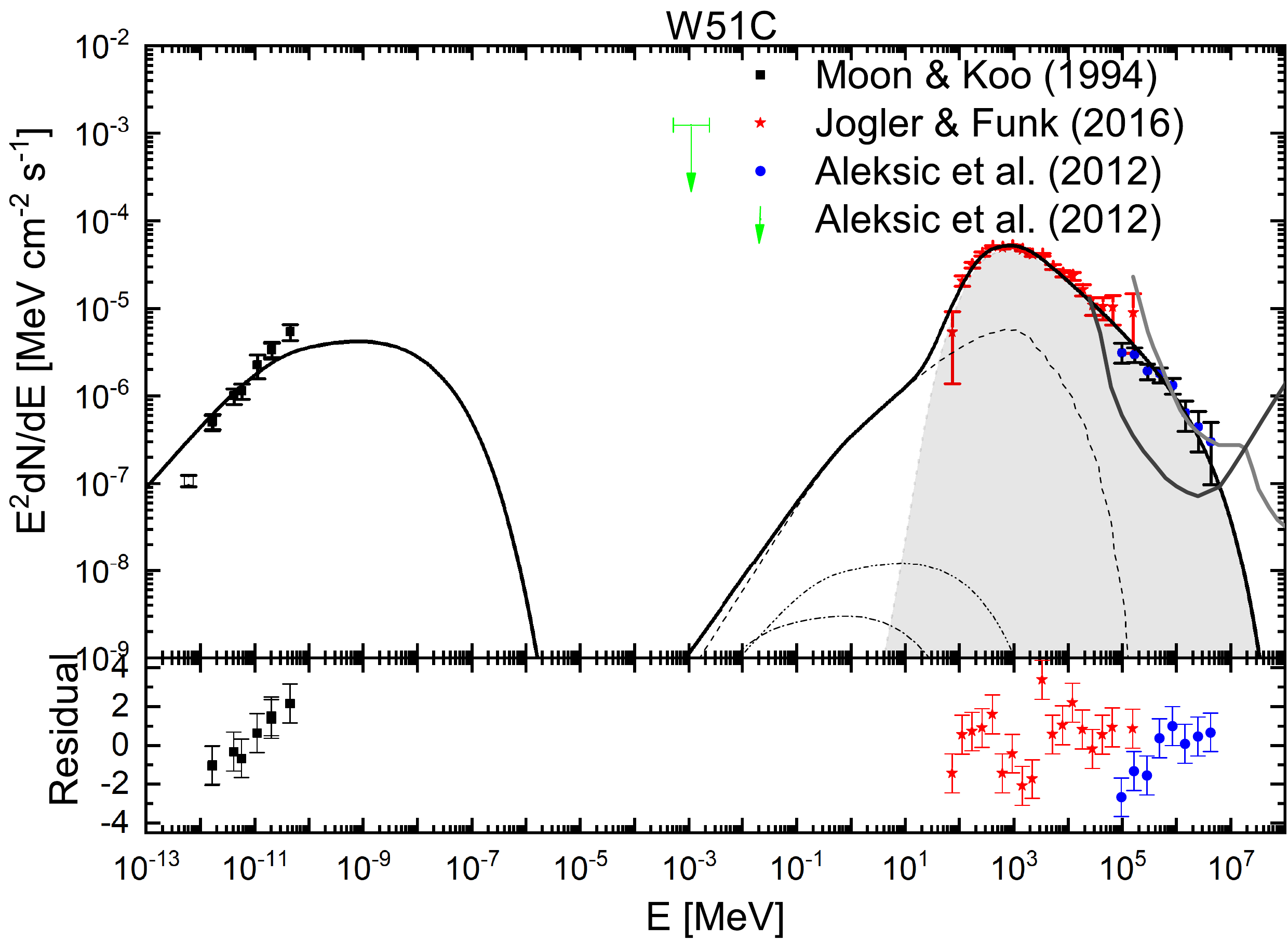}
\figsetgrpnote{The best fit to the spectral energy distribution (SED) for W51C }
\figsetgrpend

\figsetgrpstart
\figsetgrpnum{2.14}
\figsetgrptitle{W51C (b)
}
\figsetplot{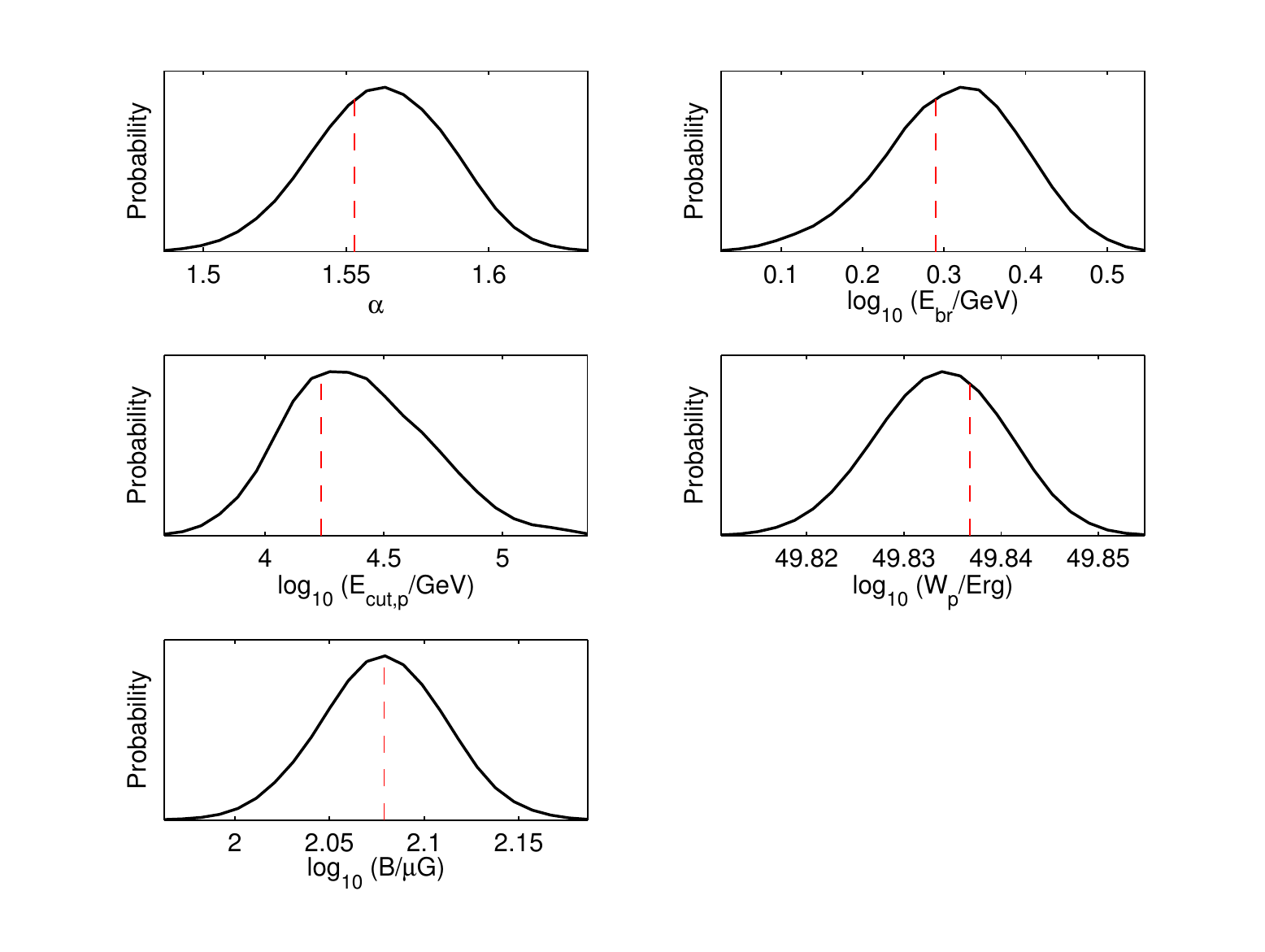}
\figsetgrpnote{1D probability distribution of the parameters for W51C }
\figsetgrpend

\figsetgrpstart
\figsetgrpnum{2.15}
\figsetgrptitle{W51C ($\rm a^{'}$)
}
\figsetplot{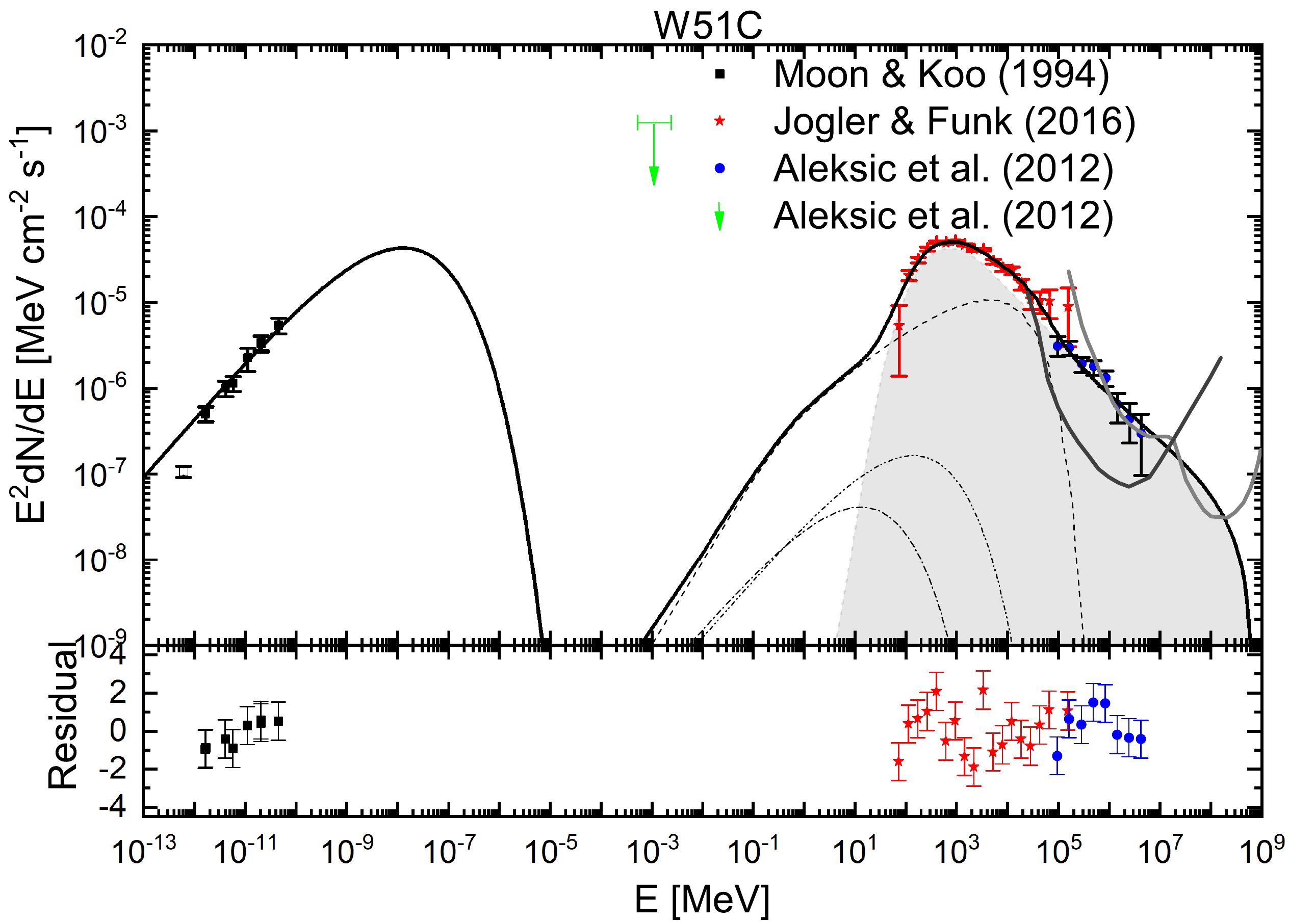}
\figsetgrpnote{The best fit to the spectral energy distribution (SED) for W51C }
\figsetgrpend

\figsetgrpstart
\figsetgrpnum{2.16}
\figsetgrptitle{W51C ($\rm b^{'}$)
}
\figsetplot{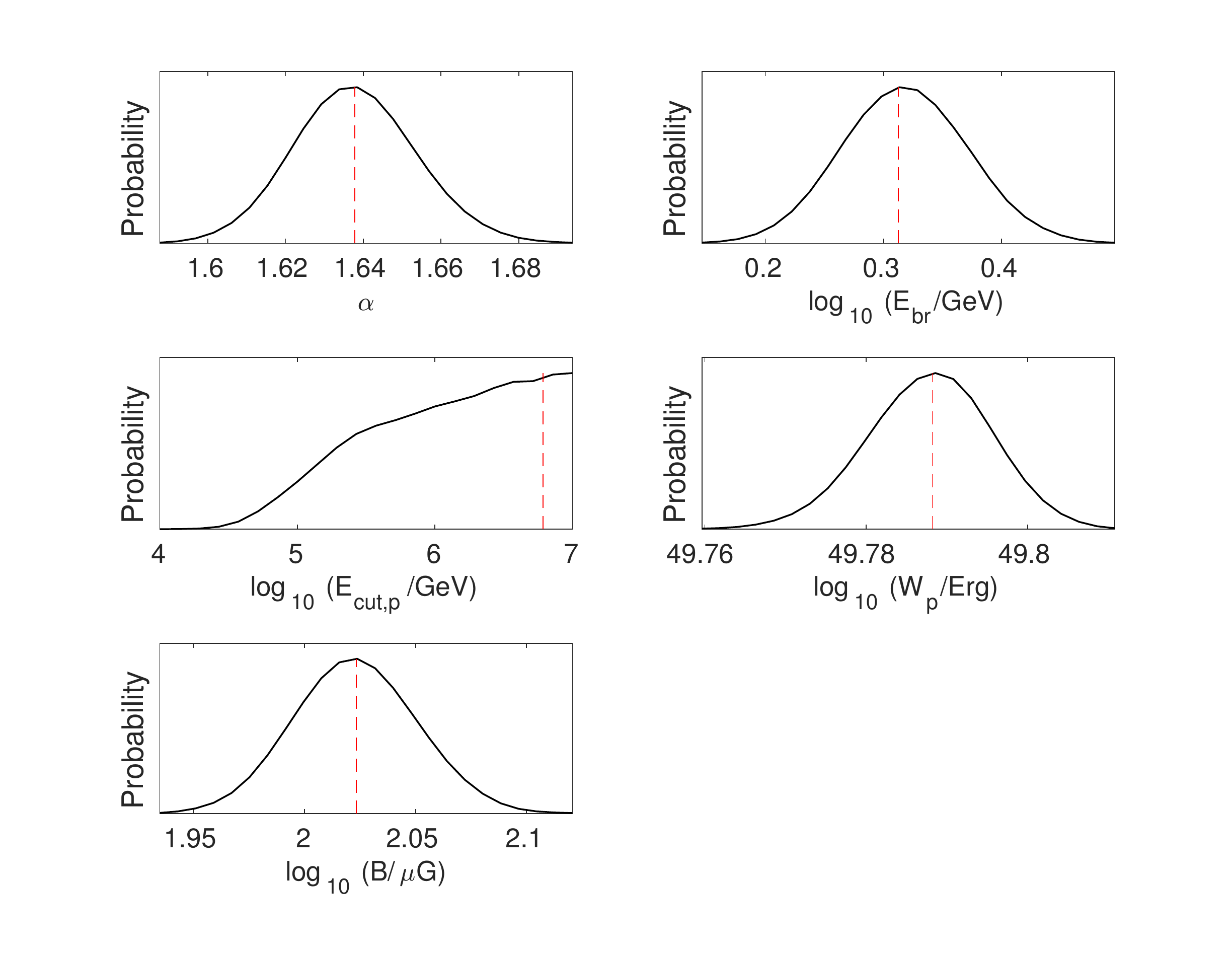}
\figsetgrpnote{ 1D probability distribution of the parameters for W51C }
\figsetgrpend

\figsetgrpstart
\figsetgrpnum{2.17}
\figsetgrptitle{G73.9+0.9 (a)
}
\figsetplot{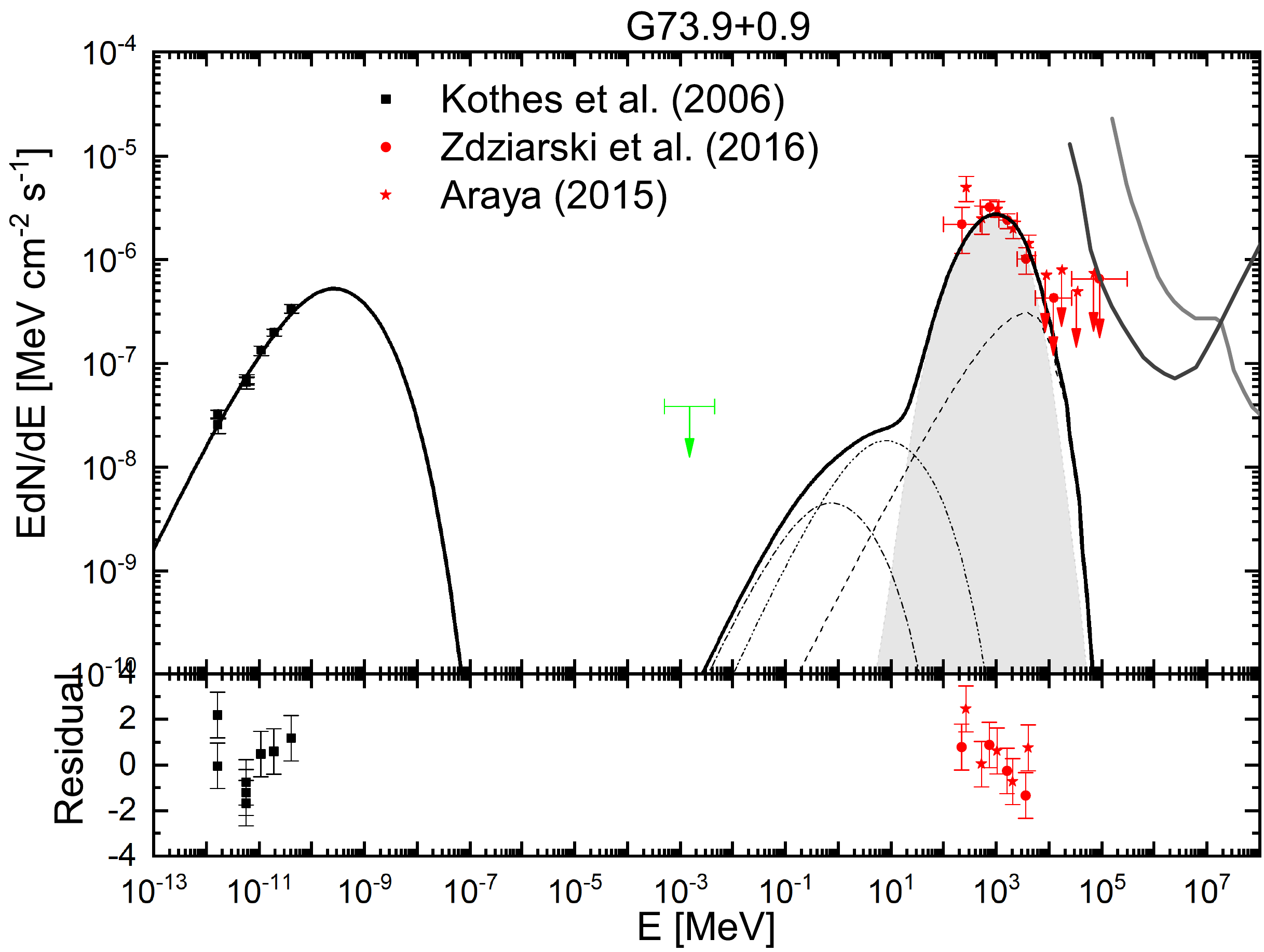}
\figsetgrpnote{The best fit to the spectral energy distribution (SED) for G73.9+0.9 }
\figsetgrpend

\figsetgrpstart
\figsetgrpnum{2.18}
\figsetgrptitle{G73.9+0.9 (b)
}
\figsetplot{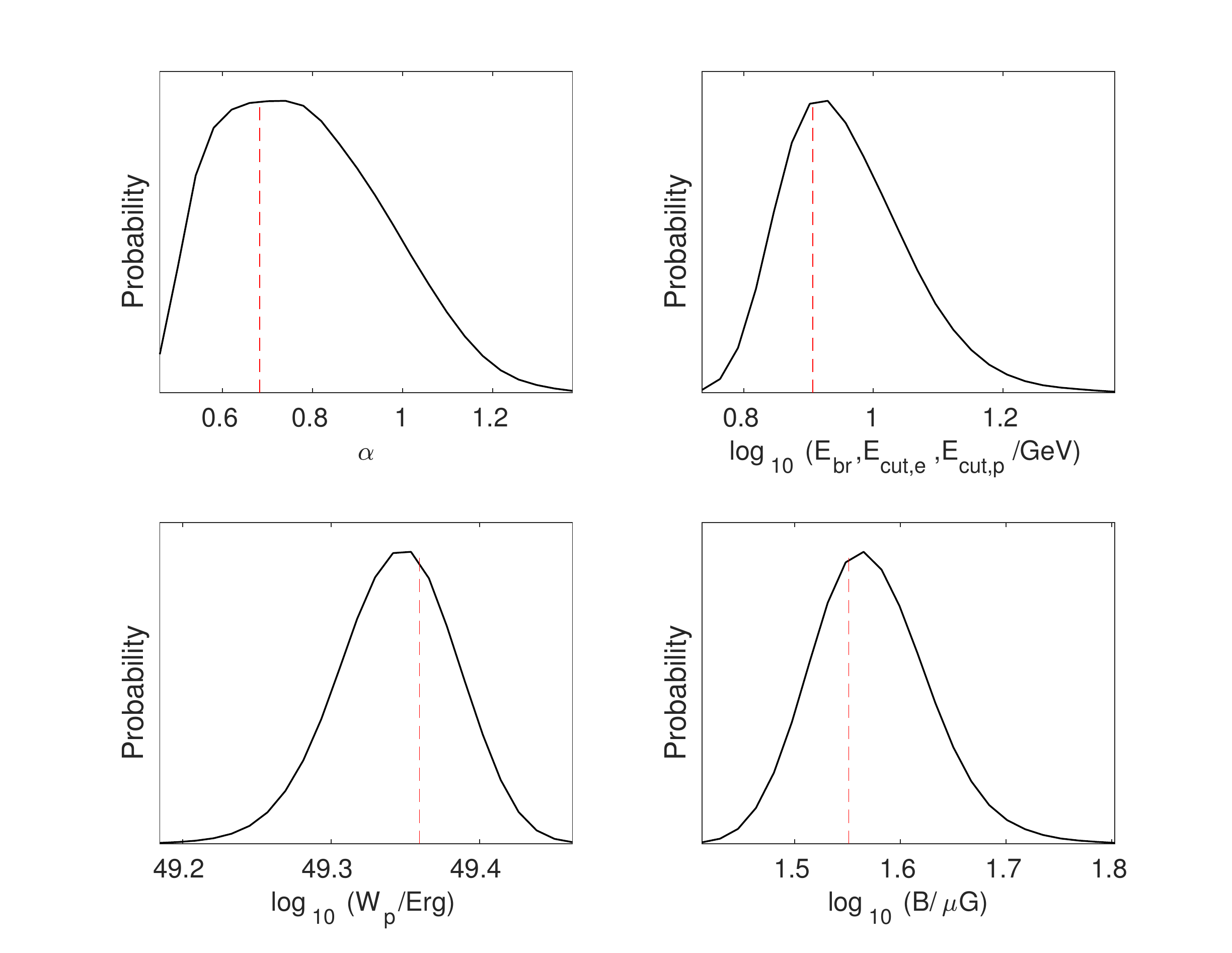}
\figsetgrpnote{1D probability distribution of the parameters for G73.0+0.9}
\figsetgrpend

\figsetgrpstart
\figsetgrpnum{2.19}
\figsetgrptitle{Cygnus Loop (a)
}
\figsetplot{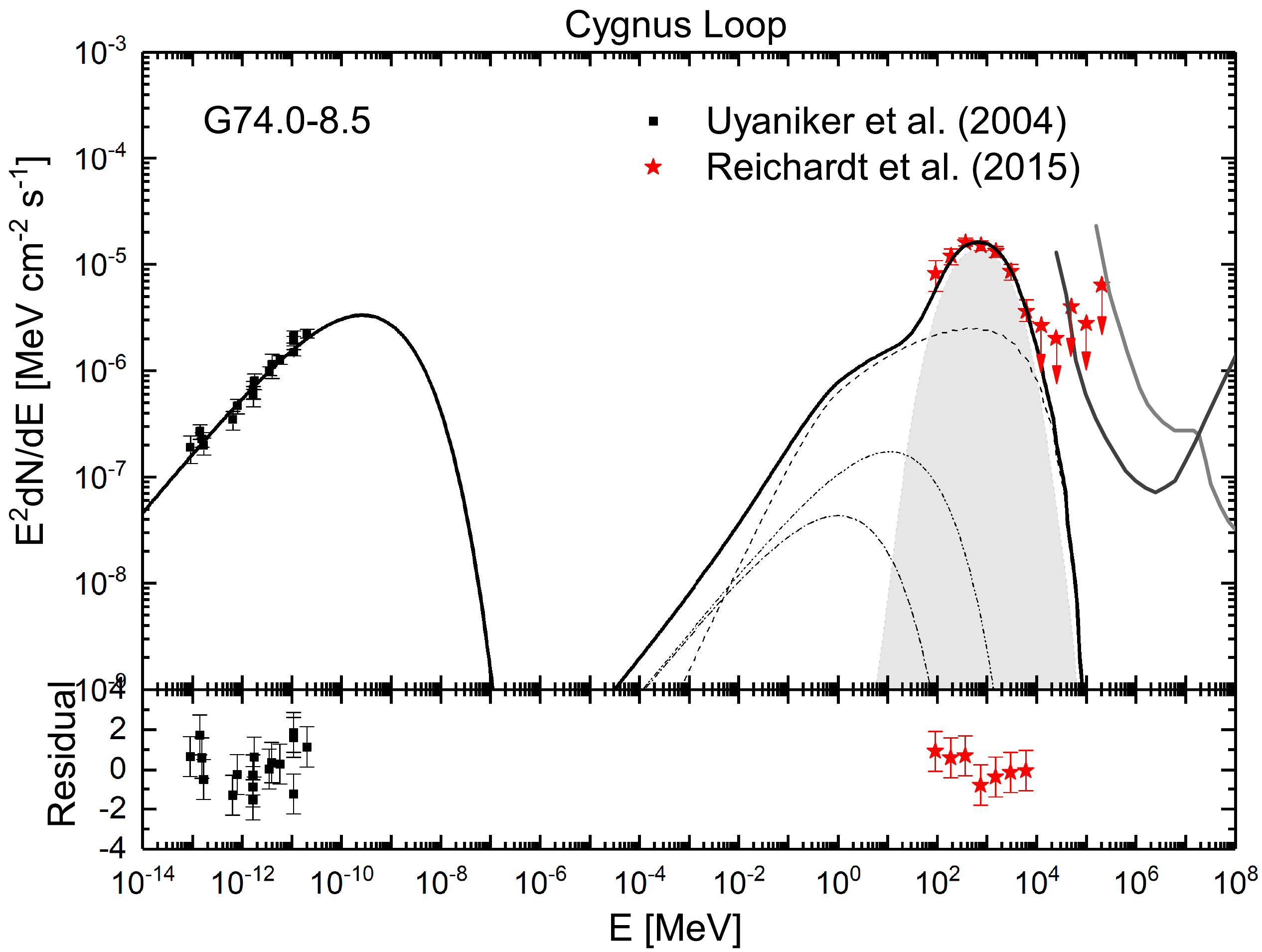}
\figsetgrpnote{The best fit to the spectral energy distribution (SED) for Cygnus Loop}
\figsetgrpend

\figsetgrpstart
\figsetgrpnum{2.20}
\figsetgrptitle{Cygnus Loop (b)
}
\figsetplot{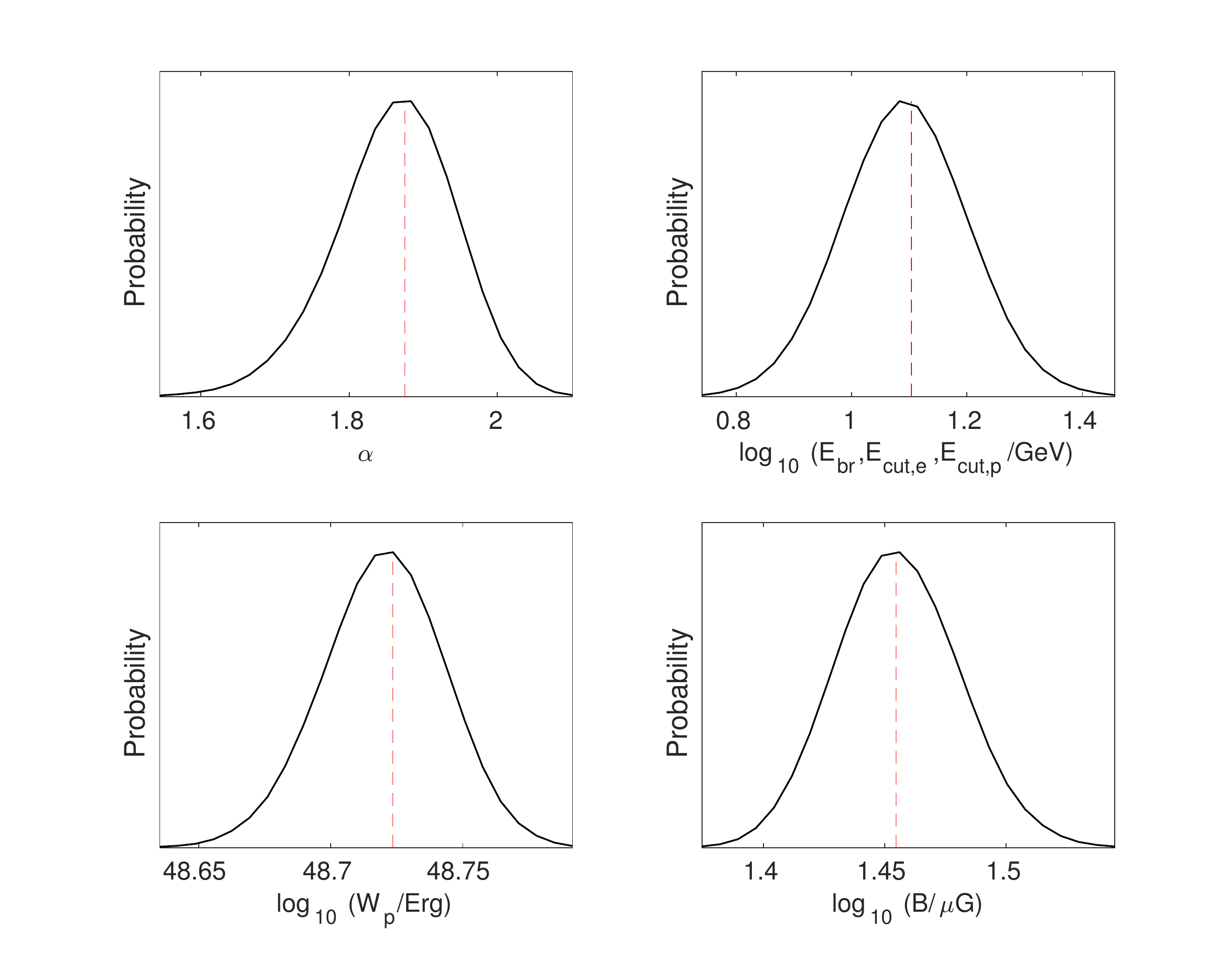}
\figsetgrpnote{1D probability distribution of the parameters for Cygnus Loop}
\figsetgrpend

\figsetgrpstart
\figsetgrpnum{2.21}
\figsetgrptitle{$\gamma$ Cygni (a)
}
\figsetplot{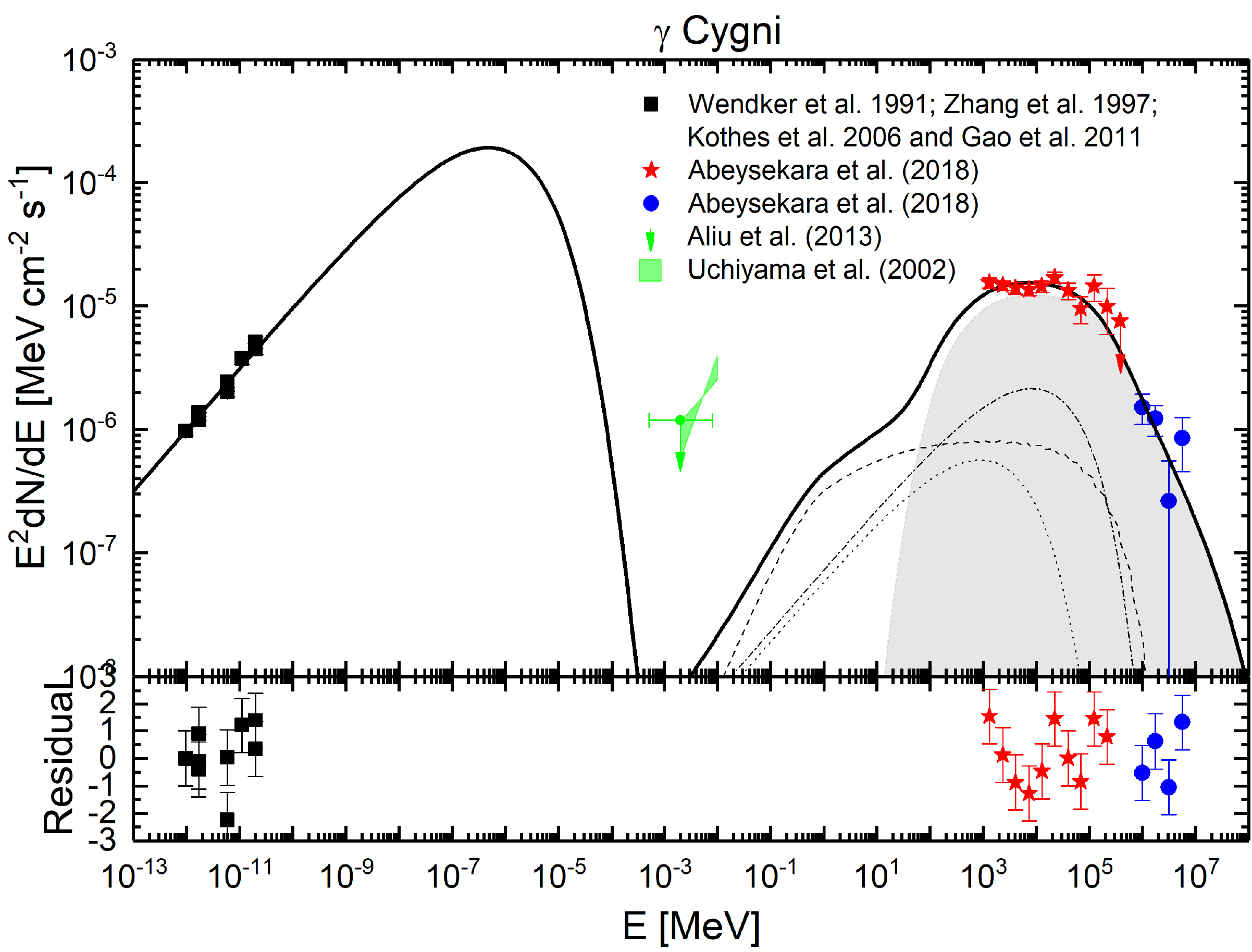}
\figsetgrpnote{The best fit to the spectral energy distribution (SED)  for G78.2+2.1 }
\figsetgrpend

\figsetgrpstart
\figsetgrpnum{2.22}
\figsetgrptitle{$\gamma$ Cygni (b)
}
\figsetplot{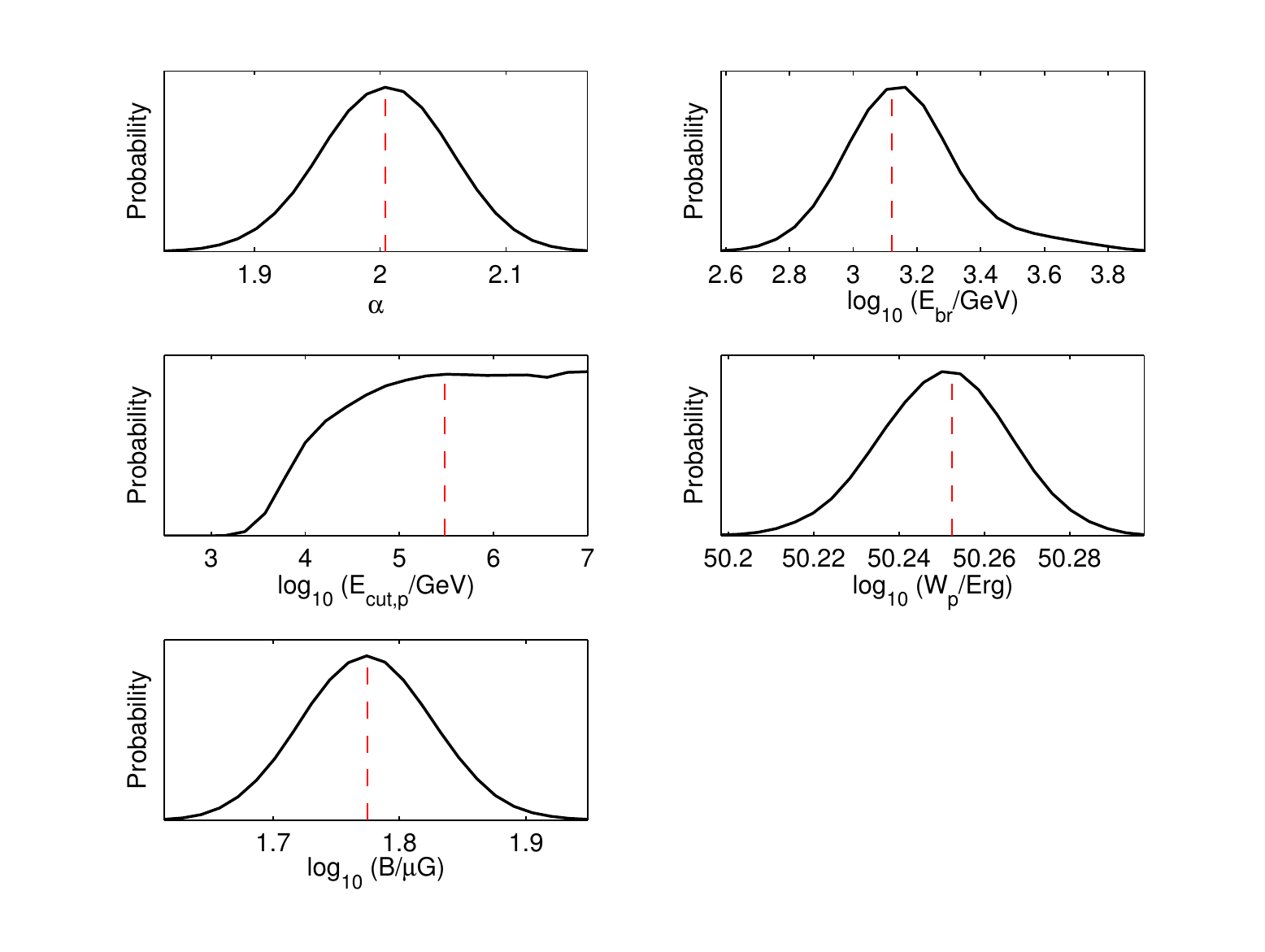}
\figsetgrpnote{1D probability distribution of the parameters for G78.2+2.1 }
\figsetgrpend

\figsetgrpstart
\figsetgrpnum{2.23}
\figsetgrptitle{HB 21 (a)
}
\figsetplot{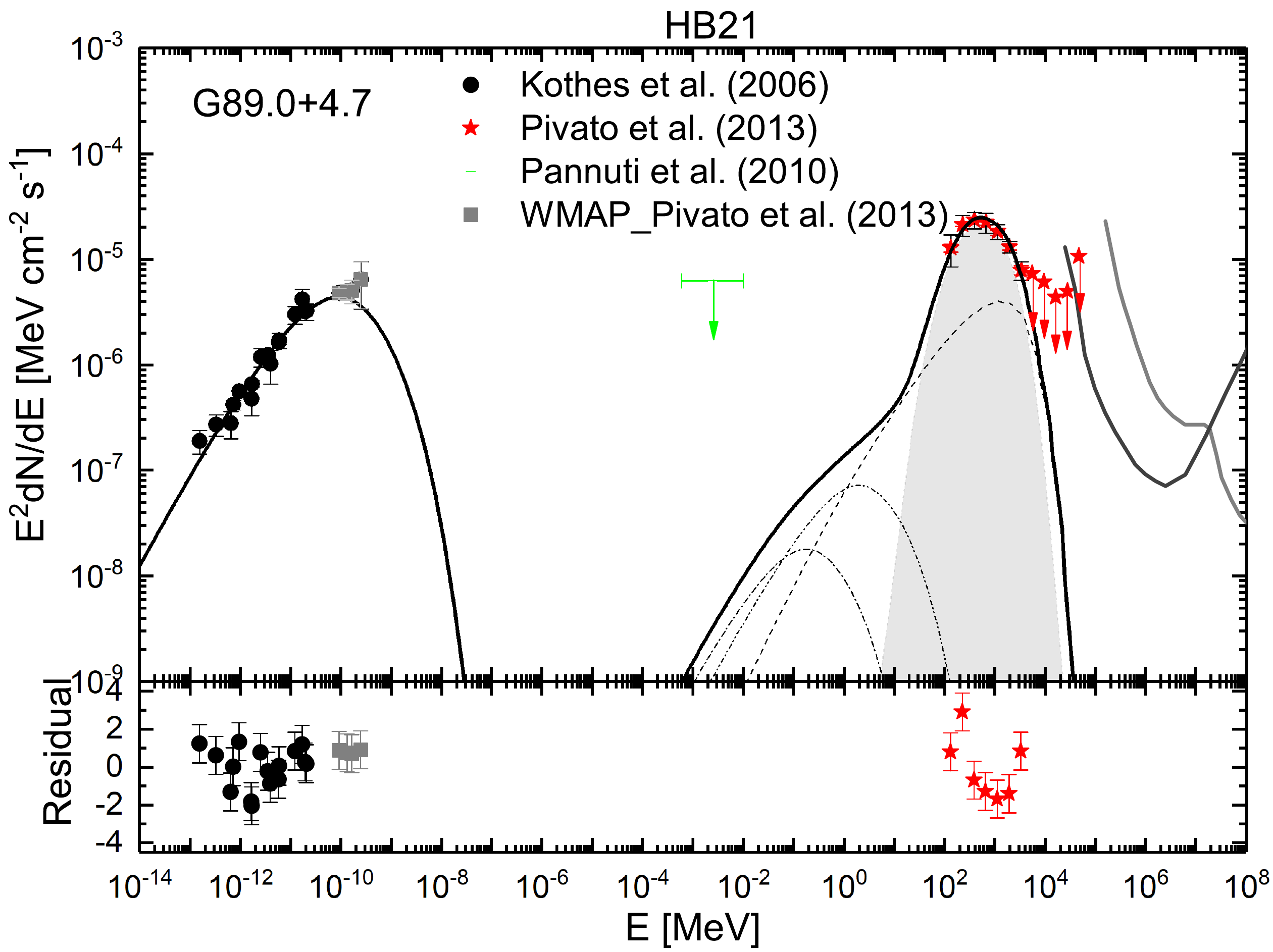}
\figsetgrpnote{The best fit to the spectral energy distribution (SED) for HB 21 }
\figsetgrpend

\figsetgrpstart
\figsetgrpnum{2.24}
\figsetgrptitle{HB 21 (b)
}
\figsetplot{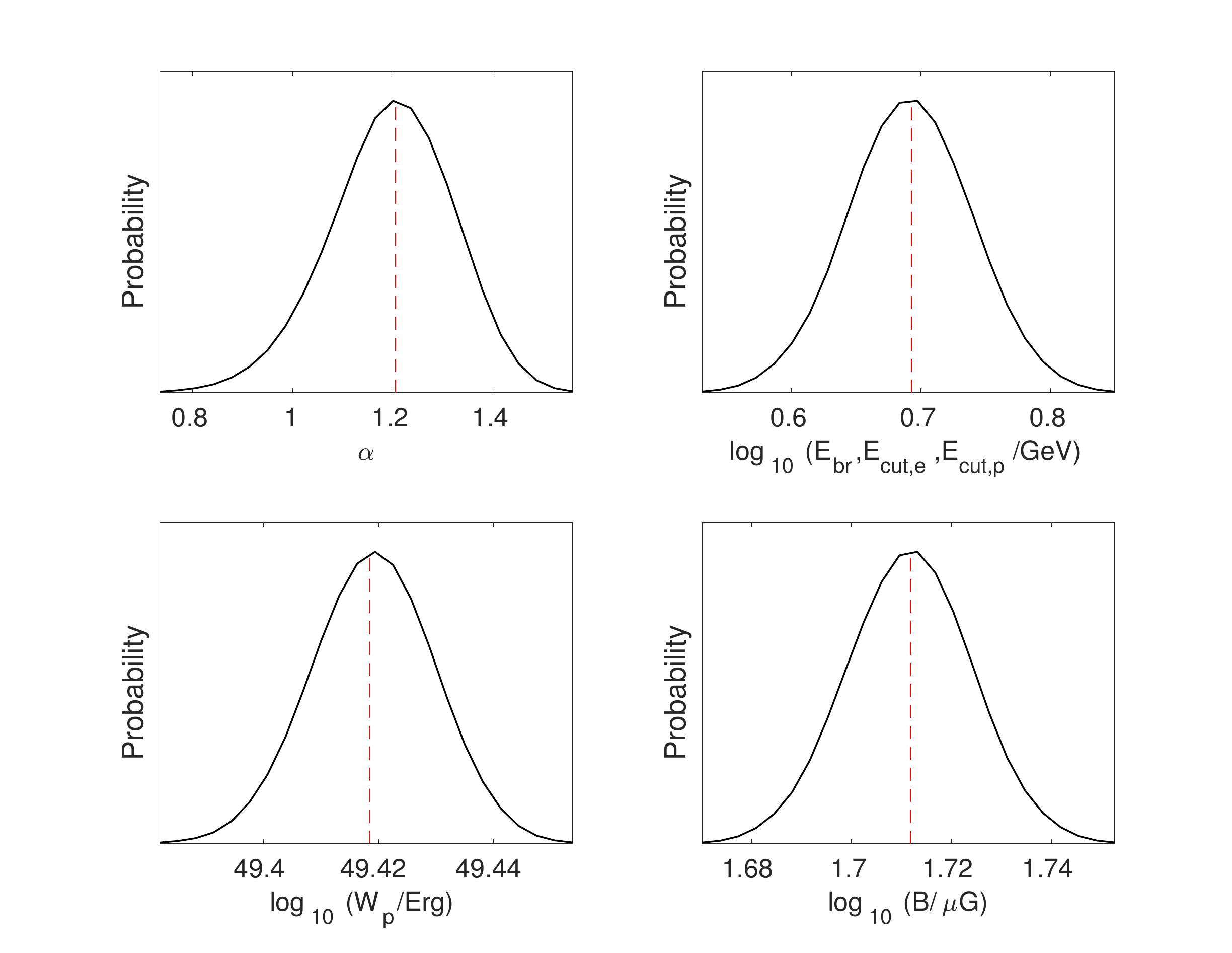}
\figsetgrpnote{1D probability distribution of the parameters for HB 21}
\figsetgrpend

\figsetgrpstart
\figsetgrpnum{2.25}
\figsetgrptitle{CTB109 (a)
}
\figsetplot{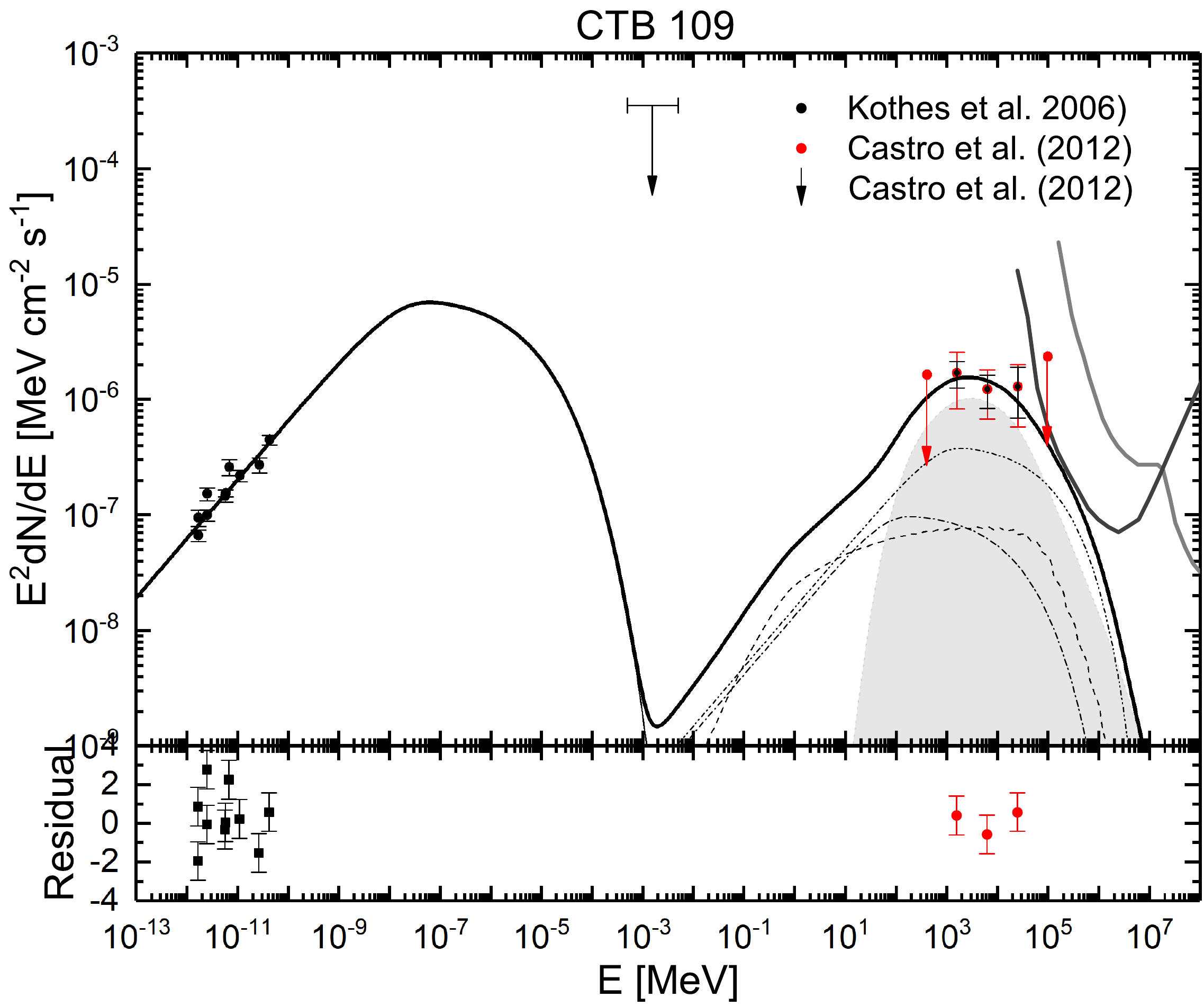}
\figsetgrpnote{The best fit to the spectral energy distribution (SED) for CTB 109 }
\figsetgrpend

\figsetgrpstart
\figsetgrpnum{2.26}
\figsetgrptitle{CTB109 (b)
}
\figsetplot{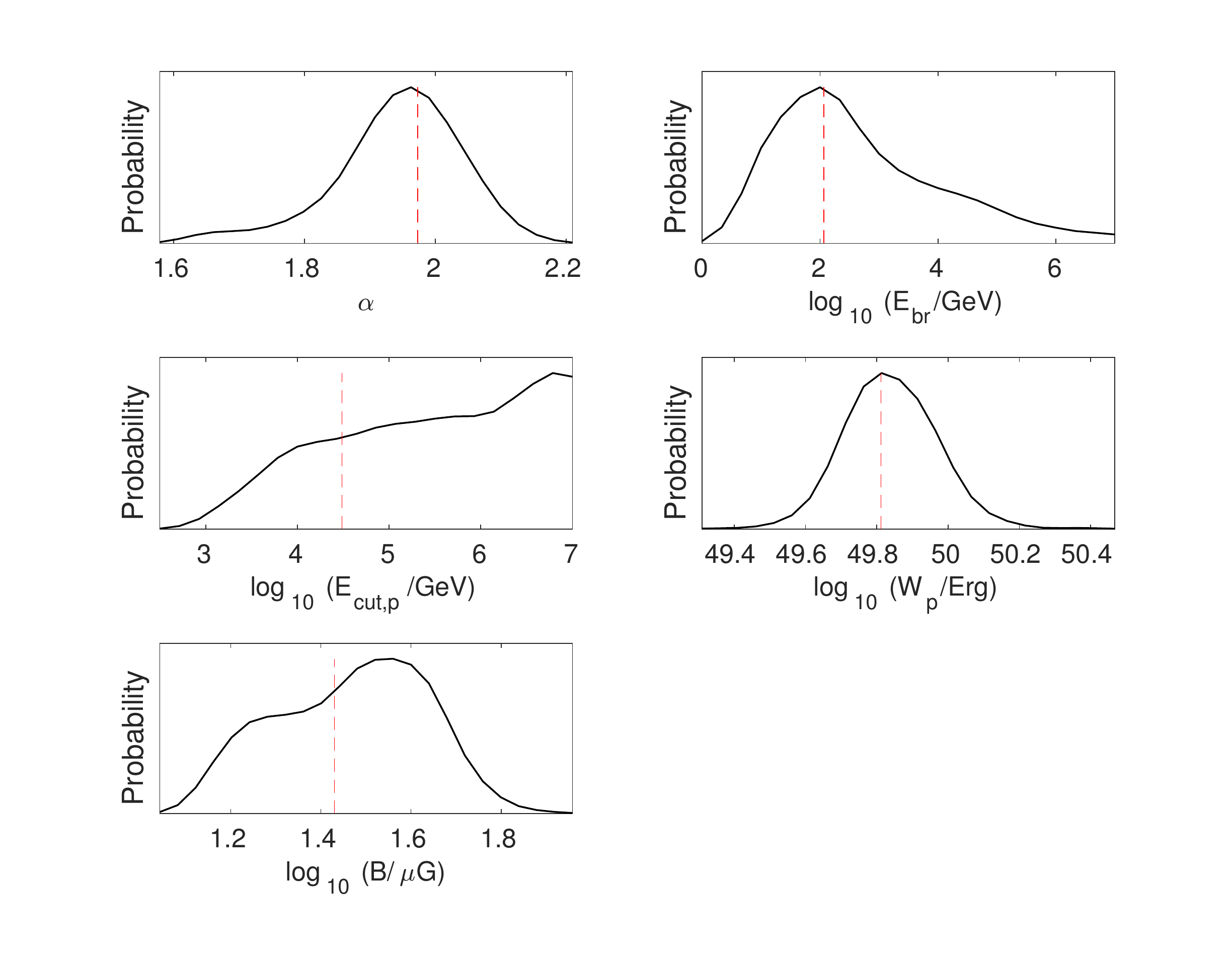}
\figsetgrpnote{1D probability distribution of the parameters for CTB 109}
\figsetgrpend

\figsetgrpstart
\figsetgrpnum{2.27}
\figsetgrptitle{Tycho (a)
}
\figsetplot{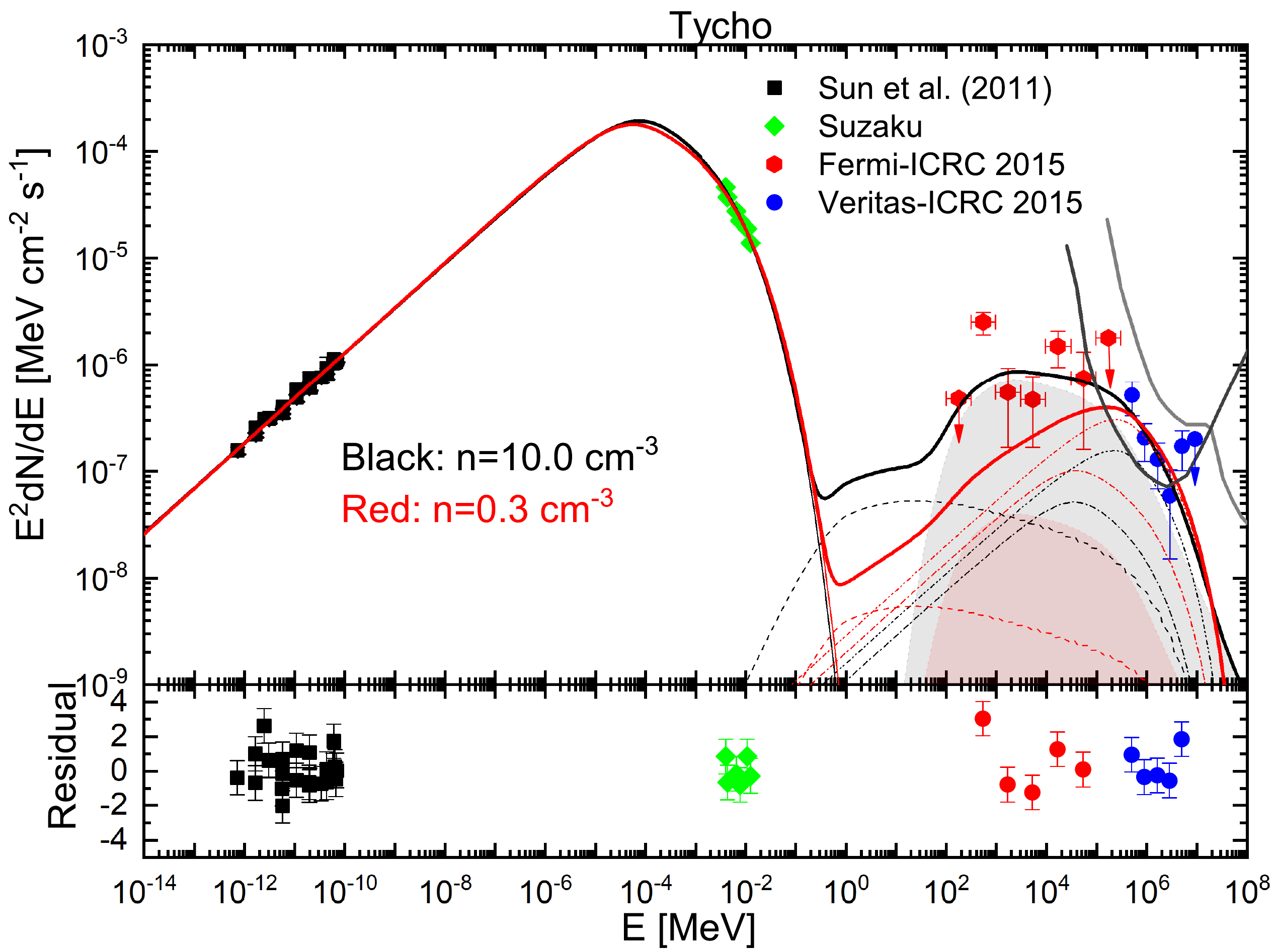}
\figsetgrpnote{The best fit to the spectral energy distribution (SED) for Tycho }
\figsetgrpend

\figsetgrpstart
\figsetgrpnum{2.28}
\figsetgrptitle{Tycho (b)
}
\figsetplot{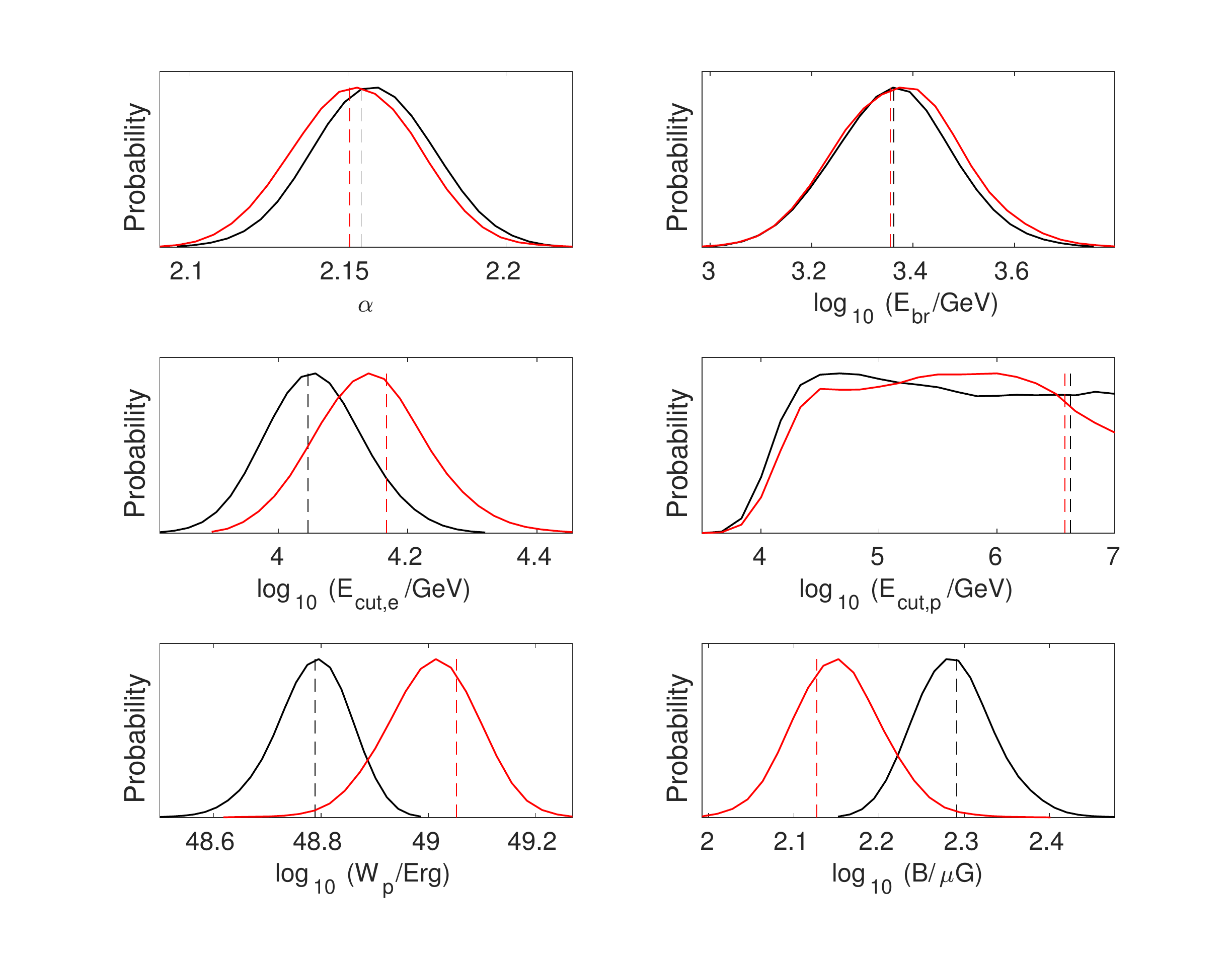}
\figsetgrpnote{1D probability distribution of the parameters for Tycho }
\figsetgrpend

\figsetgrpstart
\figsetgrpnum{2.29}
\figsetgrptitle{HB 3 (a)
}
\figsetplot{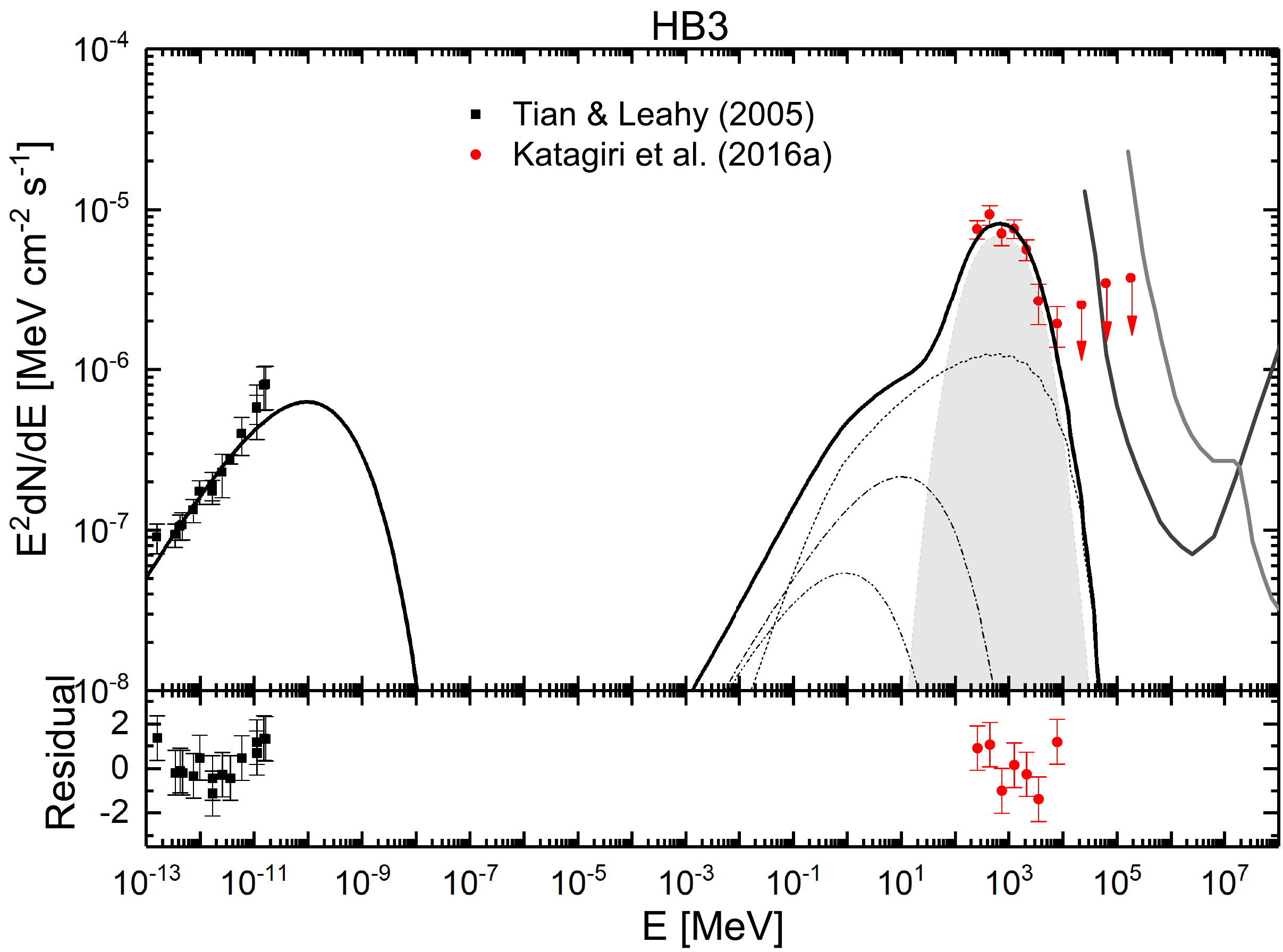}
\figsetgrpnote{The best fit to the spectral energy distribution (SED) for HB 3}
\figsetgrpend

\figsetgrpstart
\figsetgrpnum{2.30}
\figsetgrptitle{HB 3 (b)
}
\figsetplot{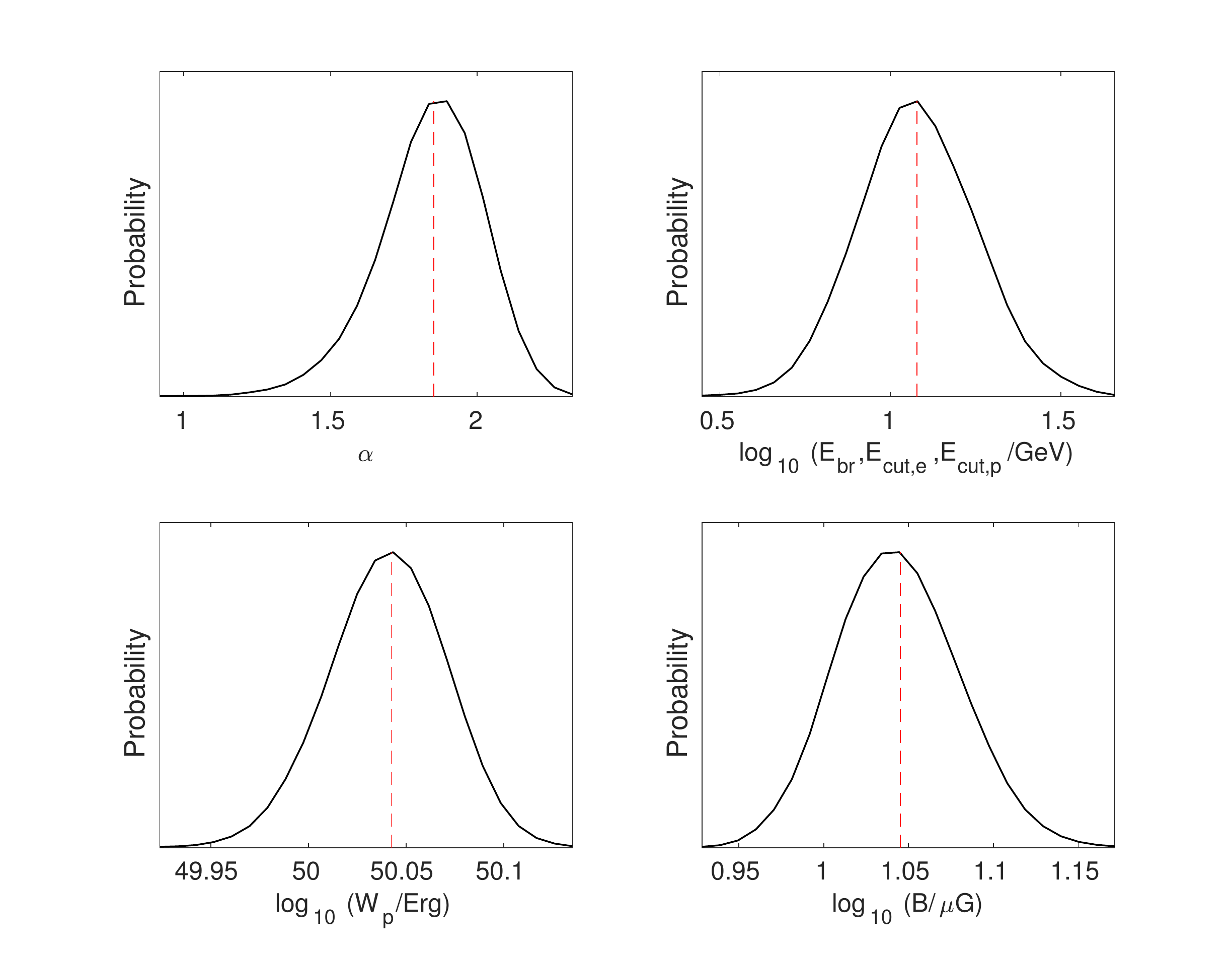}
\figsetgrpnote{ 1D probability distribution of the parameters for HB 3}
\figsetgrpend

\figsetgrpstart
\figsetgrpnum{2.31}
\figsetgrptitle{G150.3+4.5 (a)
}
\figsetplot{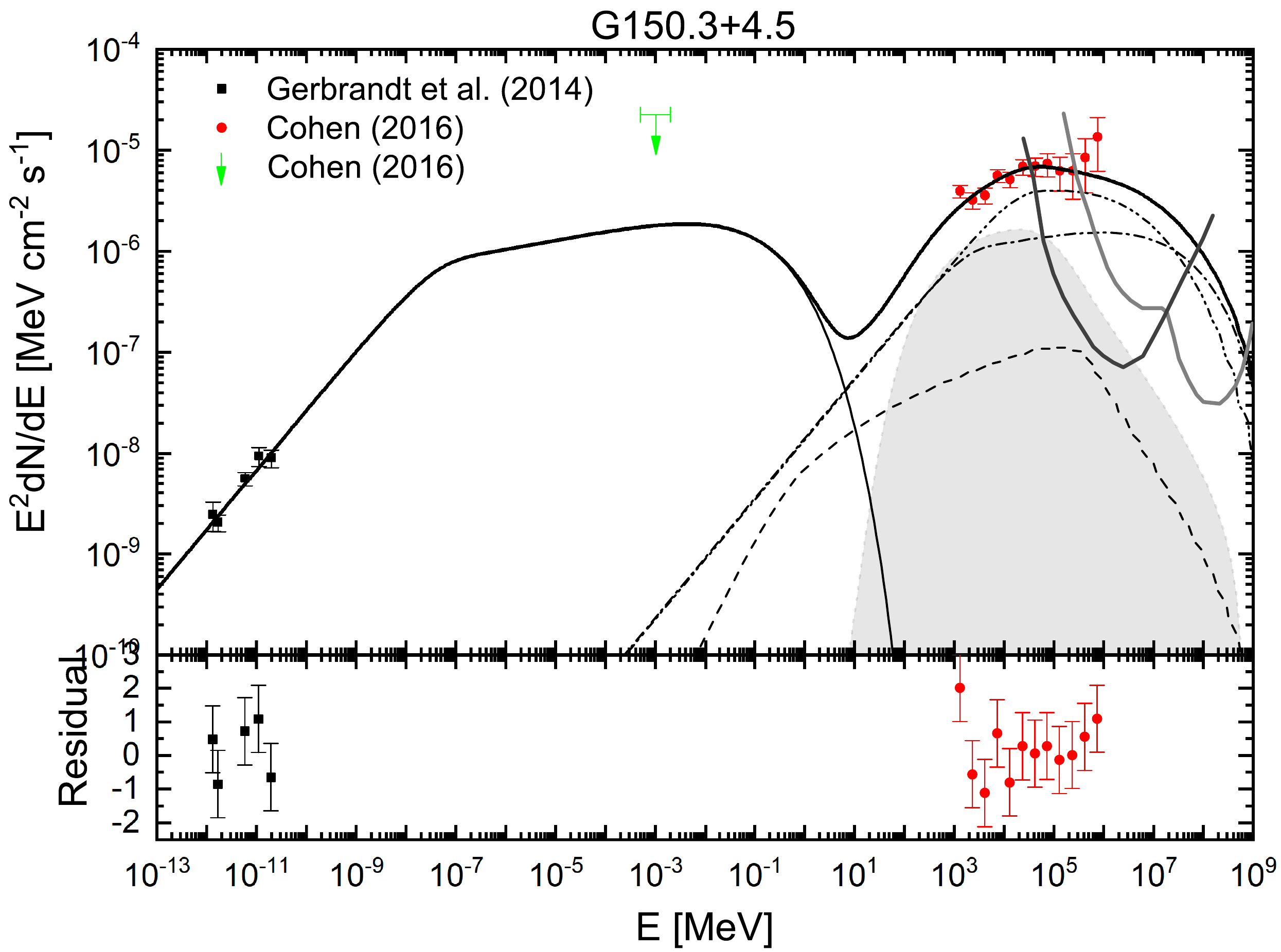}
\figsetgrpnote{The best fit to the spectral energy distribution (SED) for G150.3+4.5 }
\figsetgrpend

\figsetgrpstart
\figsetgrpnum{2.32}
\figsetgrptitle{G150.3+4.5 (b)
}
\figsetplot{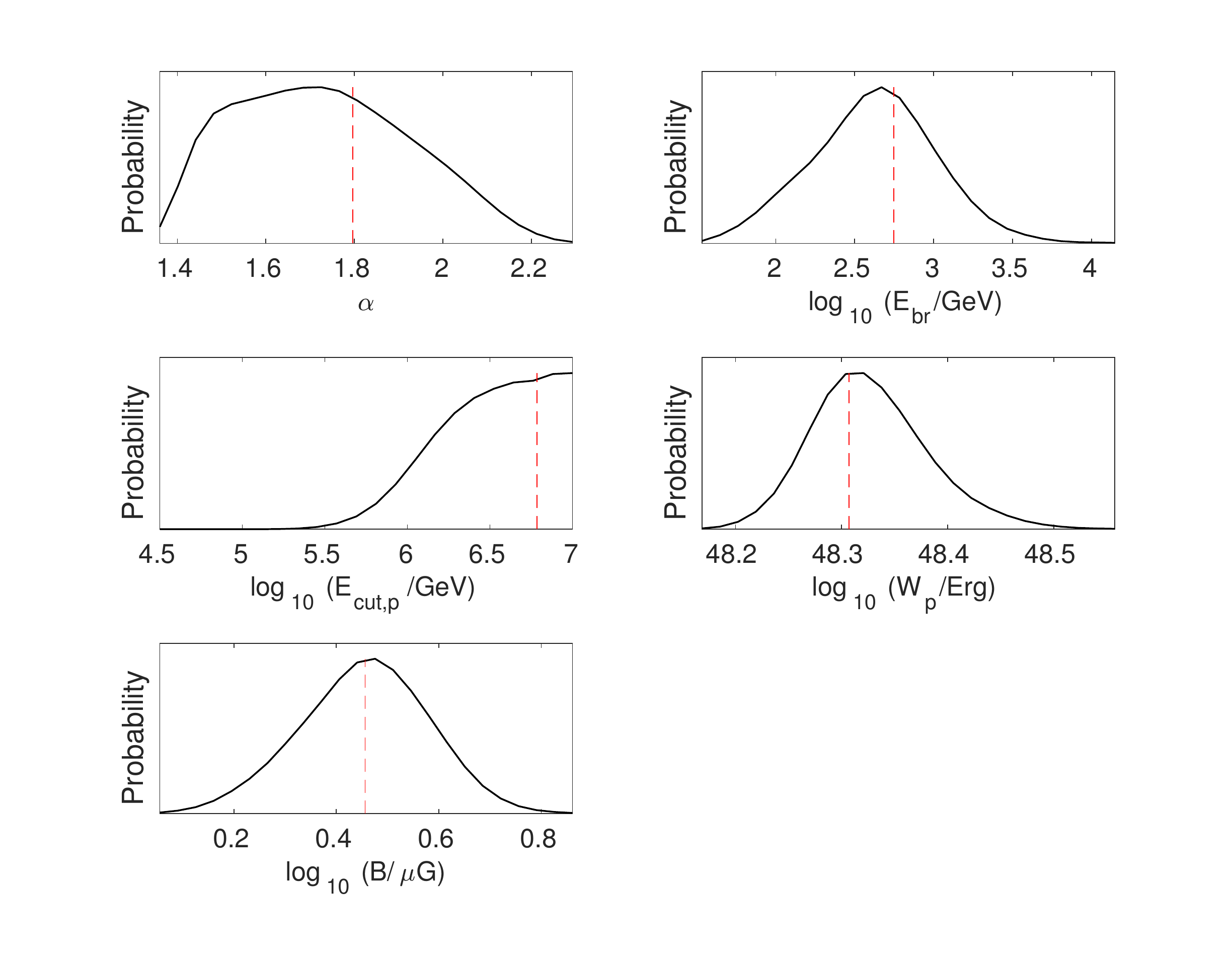}
\figsetgrpnote{1D probability distribution of the parameters for G150.3+4.5 }
\figsetgrpend

\figsetgrpstart
\figsetgrpnum{2.33}
\figsetgrptitle{HB 9 (a)
}
\figsetplot{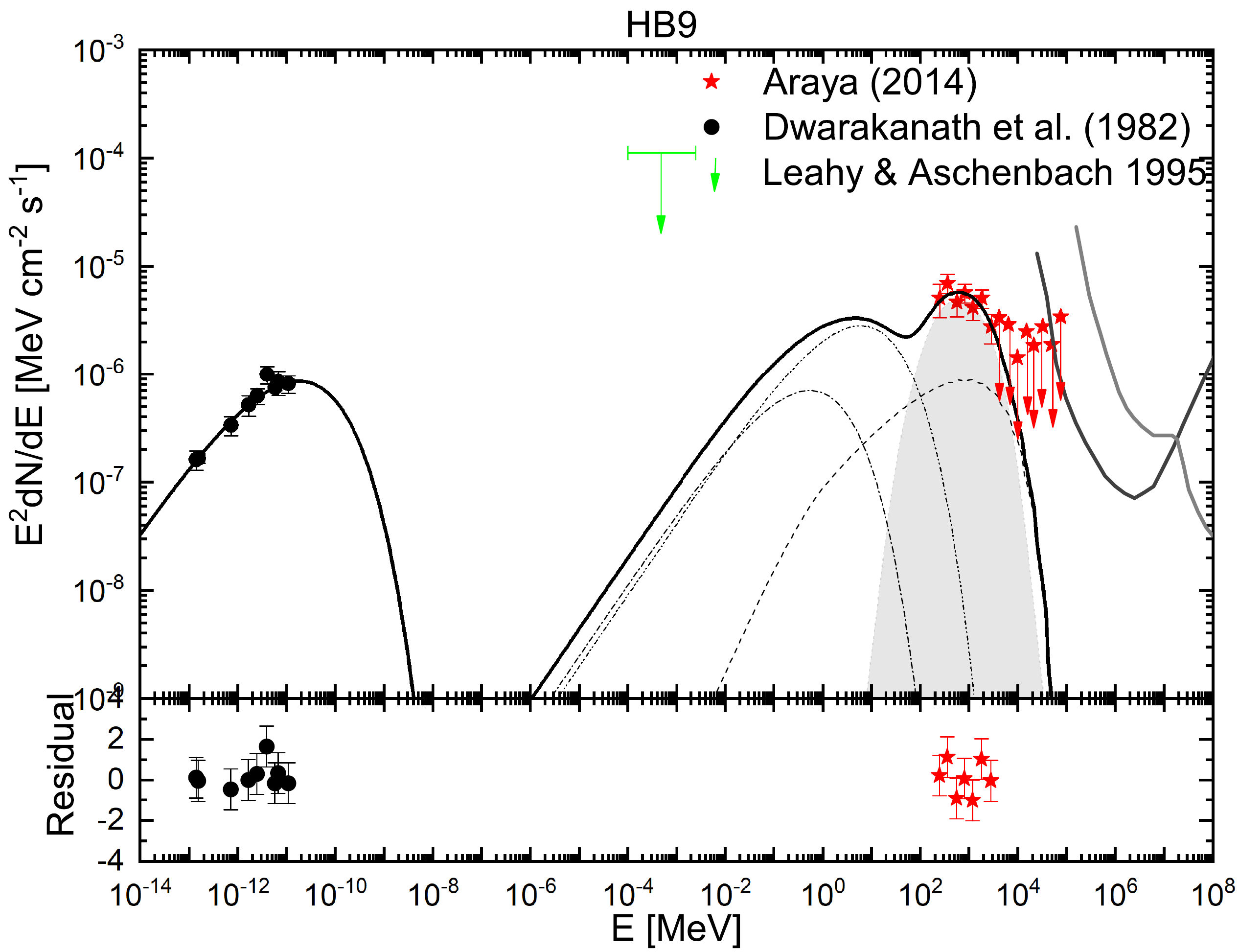}
\figsetgrpnote{The best fit to the spectral energy distribution (SED) for HB 9}
\figsetgrpend

\figsetgrpstart
\figsetgrpnum{2.34}
\figsetgrptitle{HB 9 (b)
}
\figsetplot{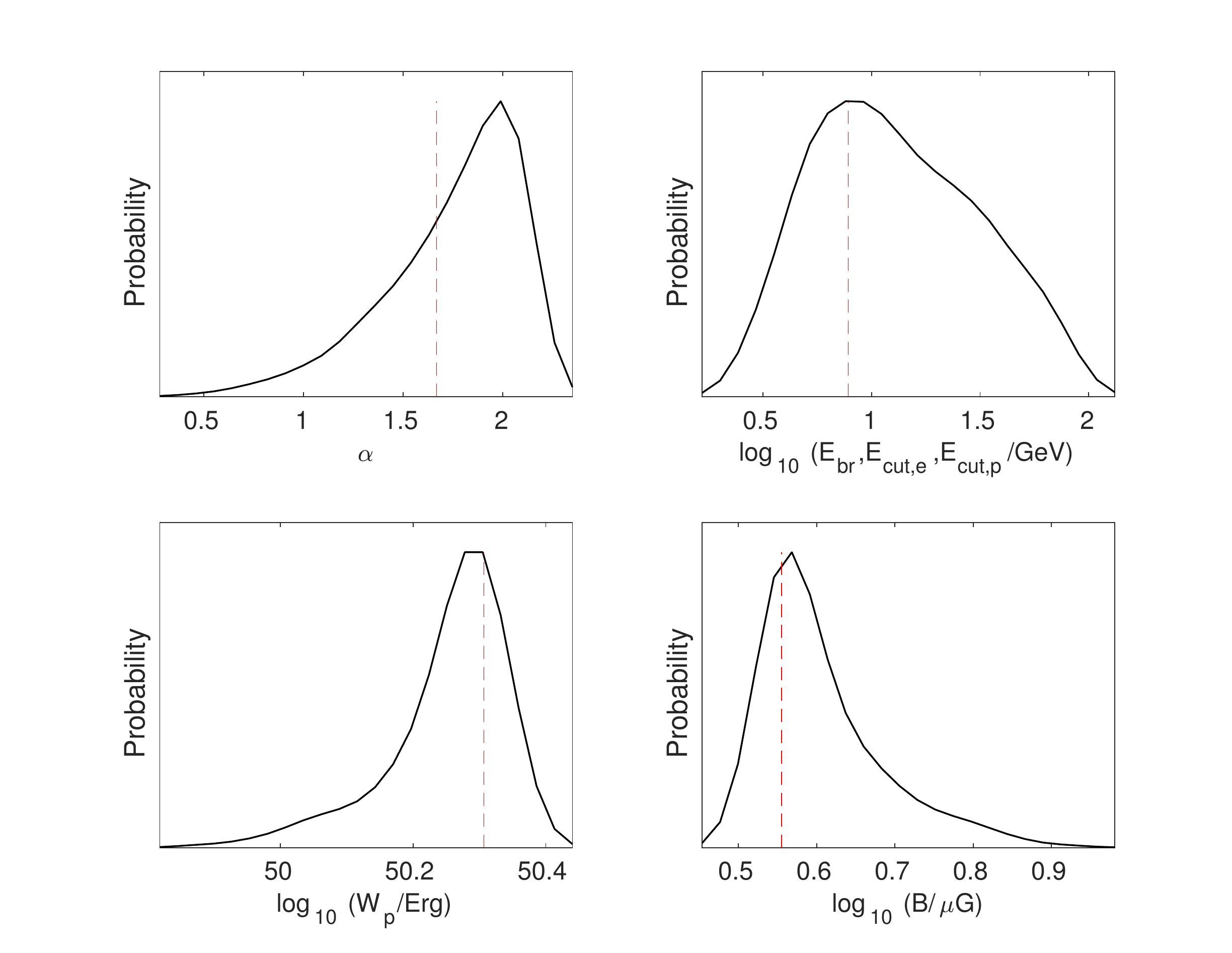}
\figsetgrpnote{ 1D probability distribution of the parameters for HB 9 }
\figsetgrpend

\figsetgrpstart
\figsetgrpnum{2.35}
\figsetgrptitle{G166.0+4.3 (a)
}
\figsetplot{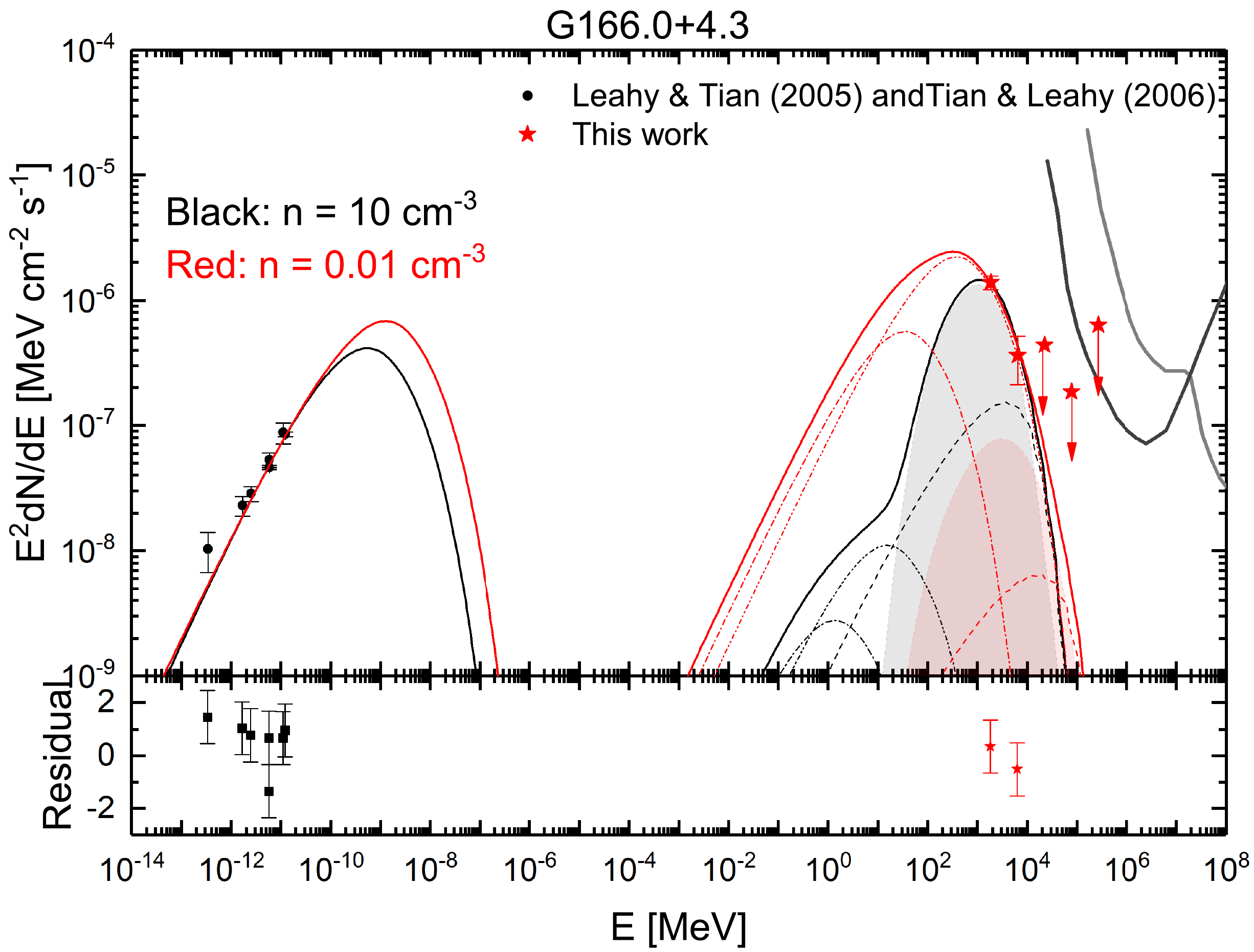}
\figsetgrpnote{The best fit to the spectral energy distribution (SED) for G166.0+4.3 }
\figsetgrpend

\figsetgrpstart
\figsetgrpnum{2.36}
\figsetgrptitle{G166.0+4.3 (b)
}
\figsetplot{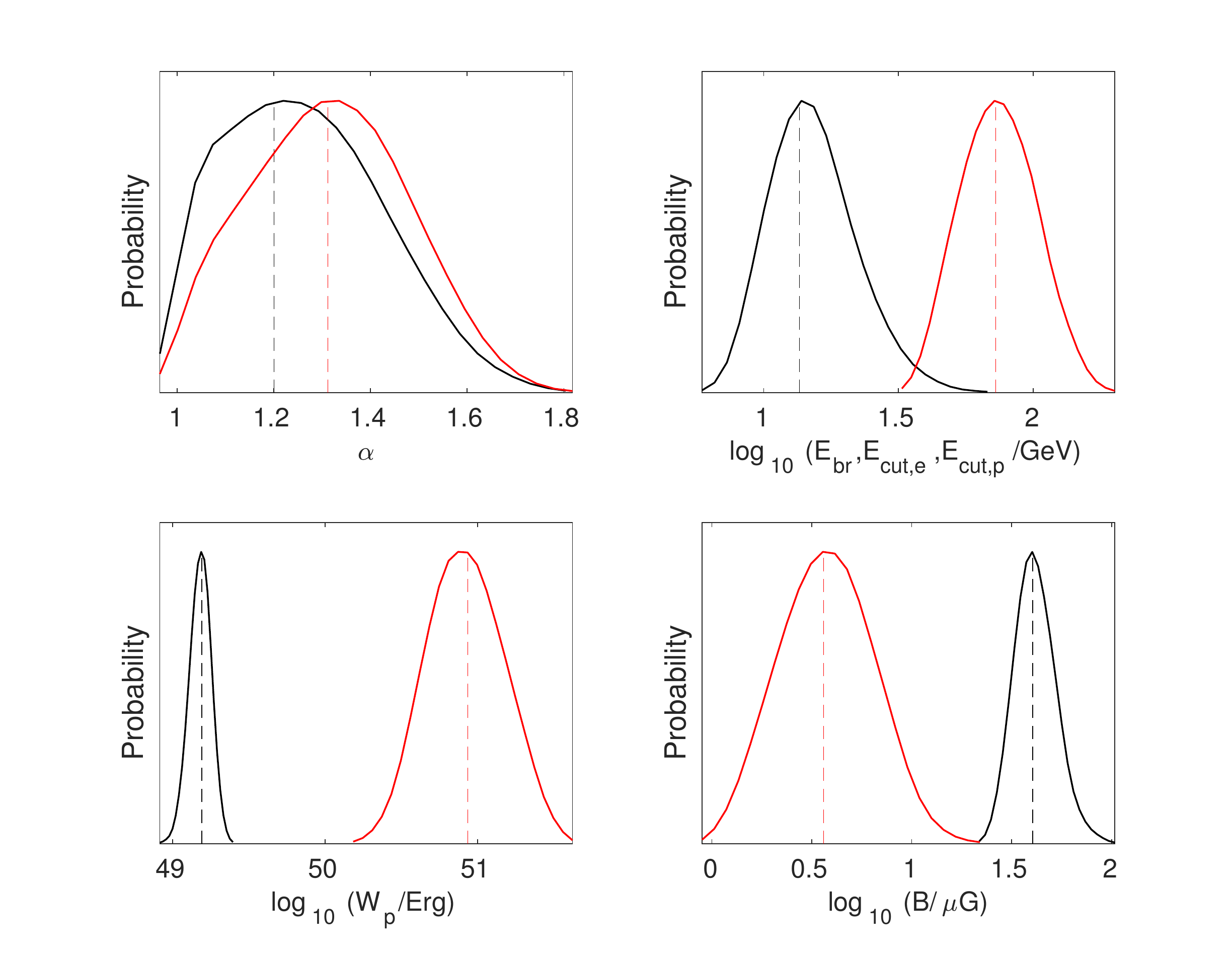}
\figsetgrpnote{1D probability distribution of the parameters for G166.0+4.3 }
\figsetgrpend

\figsetgrpstart
\figsetgrpnum{2.37}
\figsetgrptitle{S147 (a)
}
\figsetplot{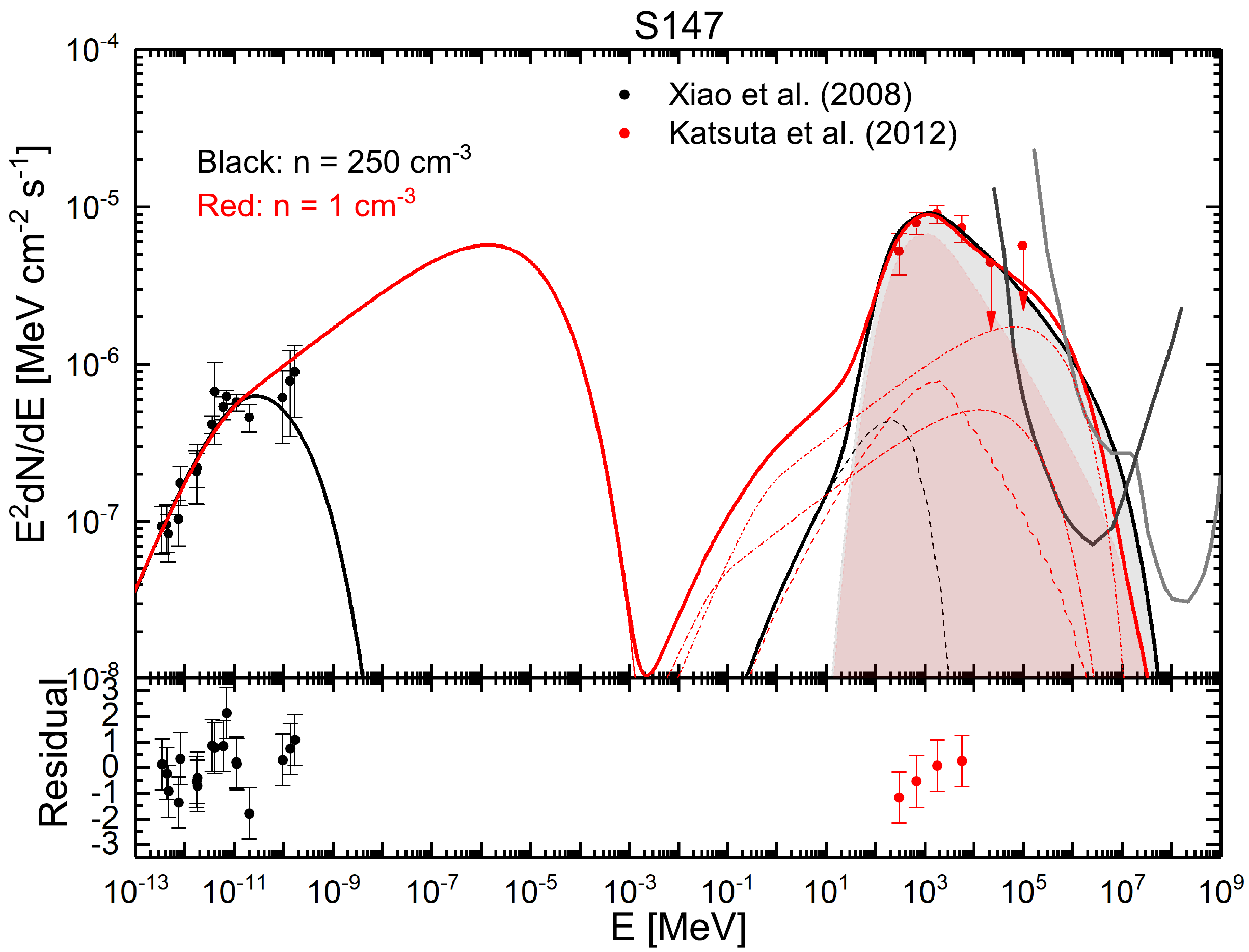}
\figsetgrpnote{The best fit to the spectral energy distribution (SED) for S147}
\figsetgrpend

\figsetgrpstart
\figsetgrpnum{2.38}
\figsetgrptitle{S147 (b)
}
\figsetplot{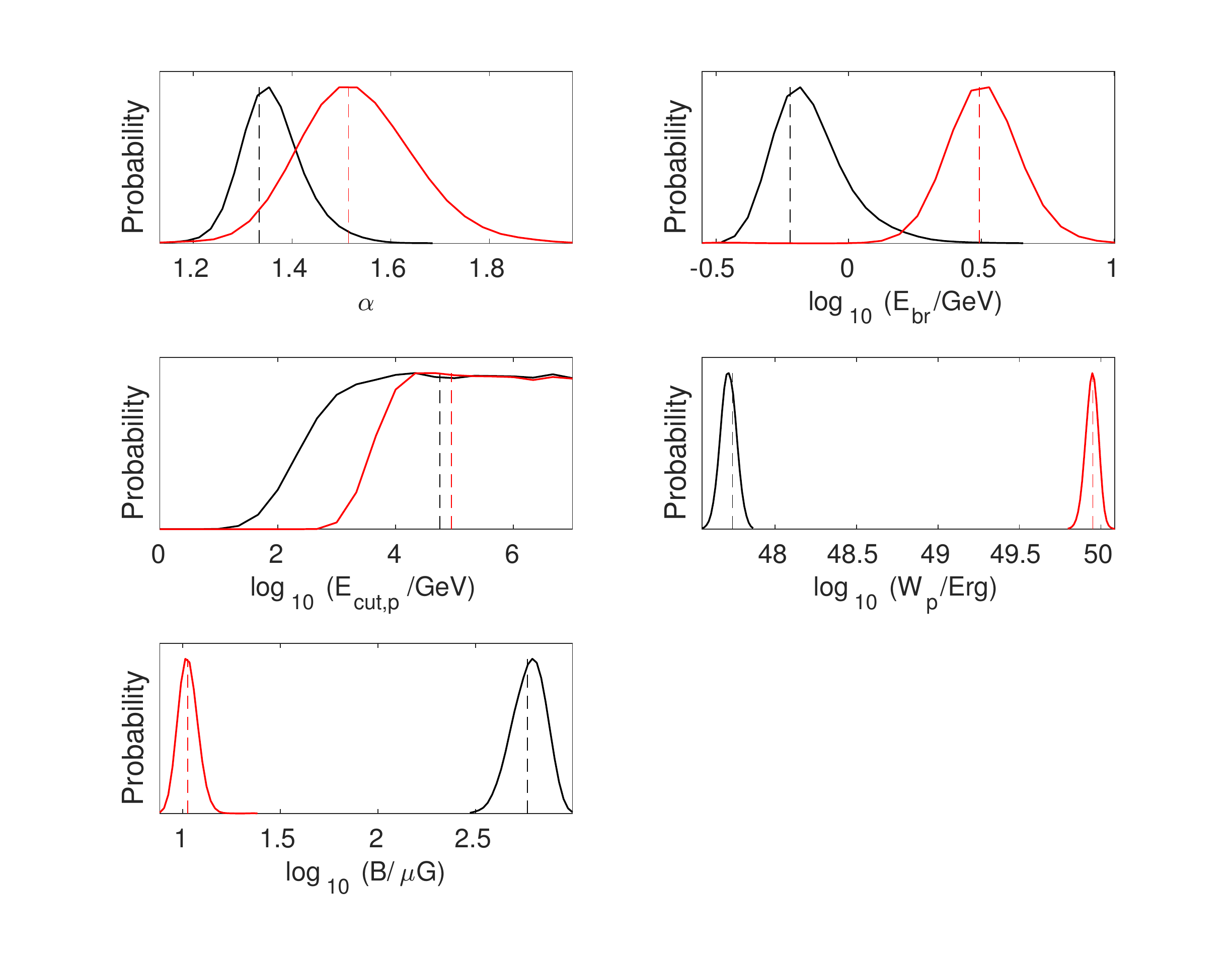}
\figsetgrpnote{1D probability distribution of the parameters for S147}
\figsetgrpend

\figsetgrpstart
\figsetgrpnum{2.39}
\figsetgrptitle{IC443 (a)
}
\figsetplot{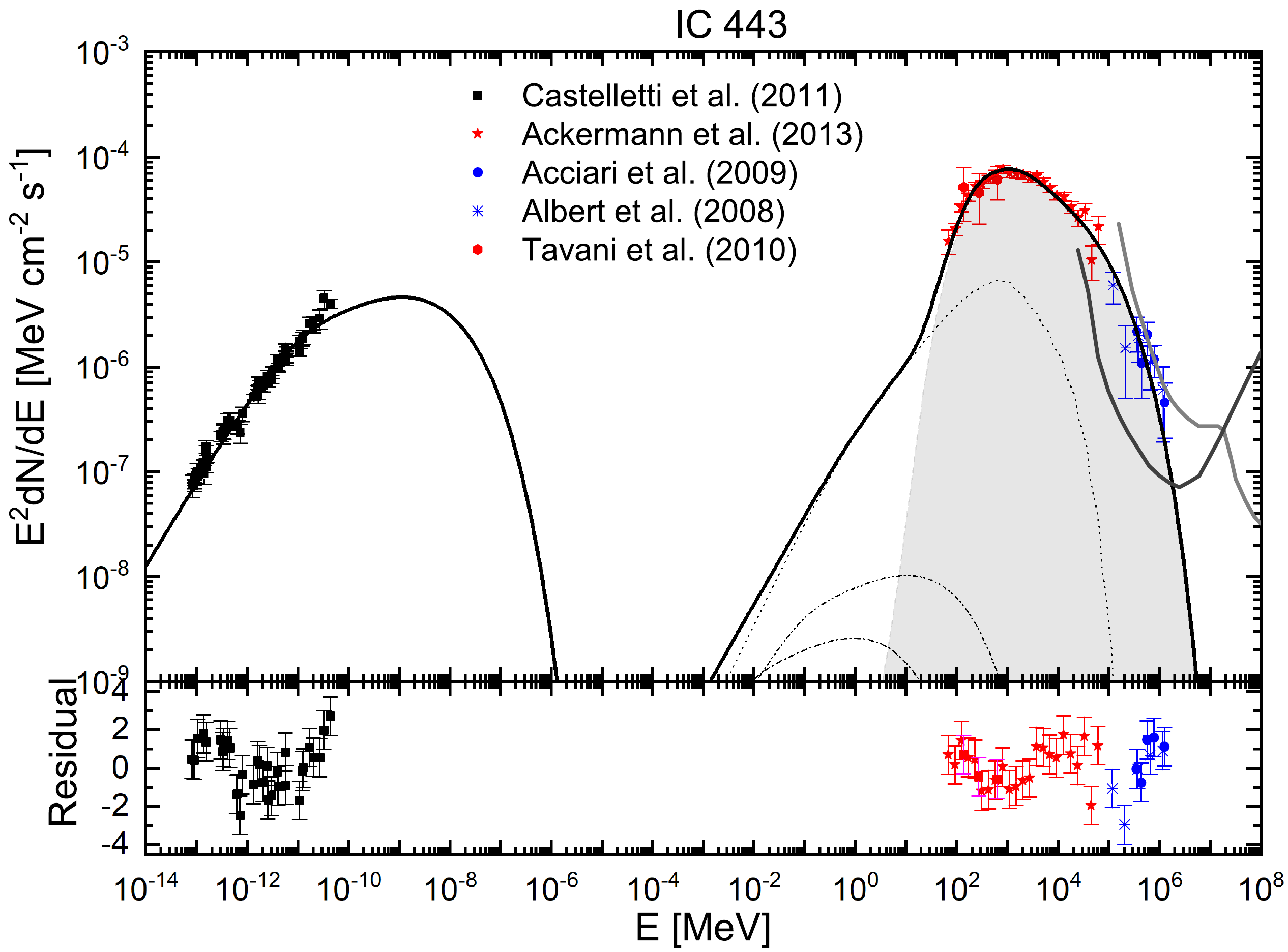}
\figsetgrpnote{The best fit to the spectral energy distribution (SED) for IC443 }
\figsetgrpend

\figsetgrpstart
\figsetgrpnum{2.40}
\figsetgrptitle{IC443 (b)
}
\figsetplot{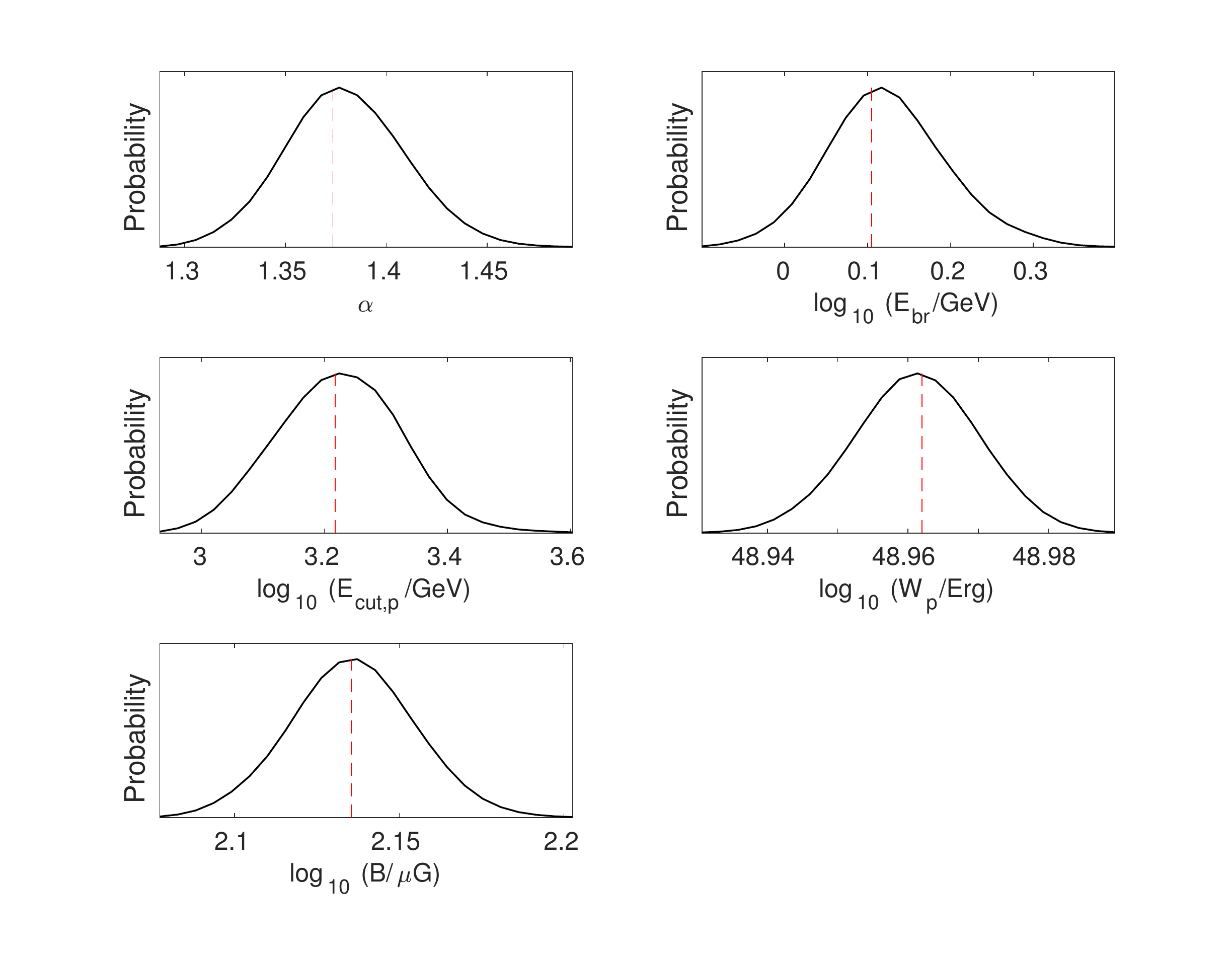}
\figsetgrpnote{1D probability distribution of the parameters for IC443}
\figsetgrpend

\figsetgrpstart
\figsetgrpnum{2.41}
\figsetgrptitle{G205.5+0.5 (a)
}
\figsetplot{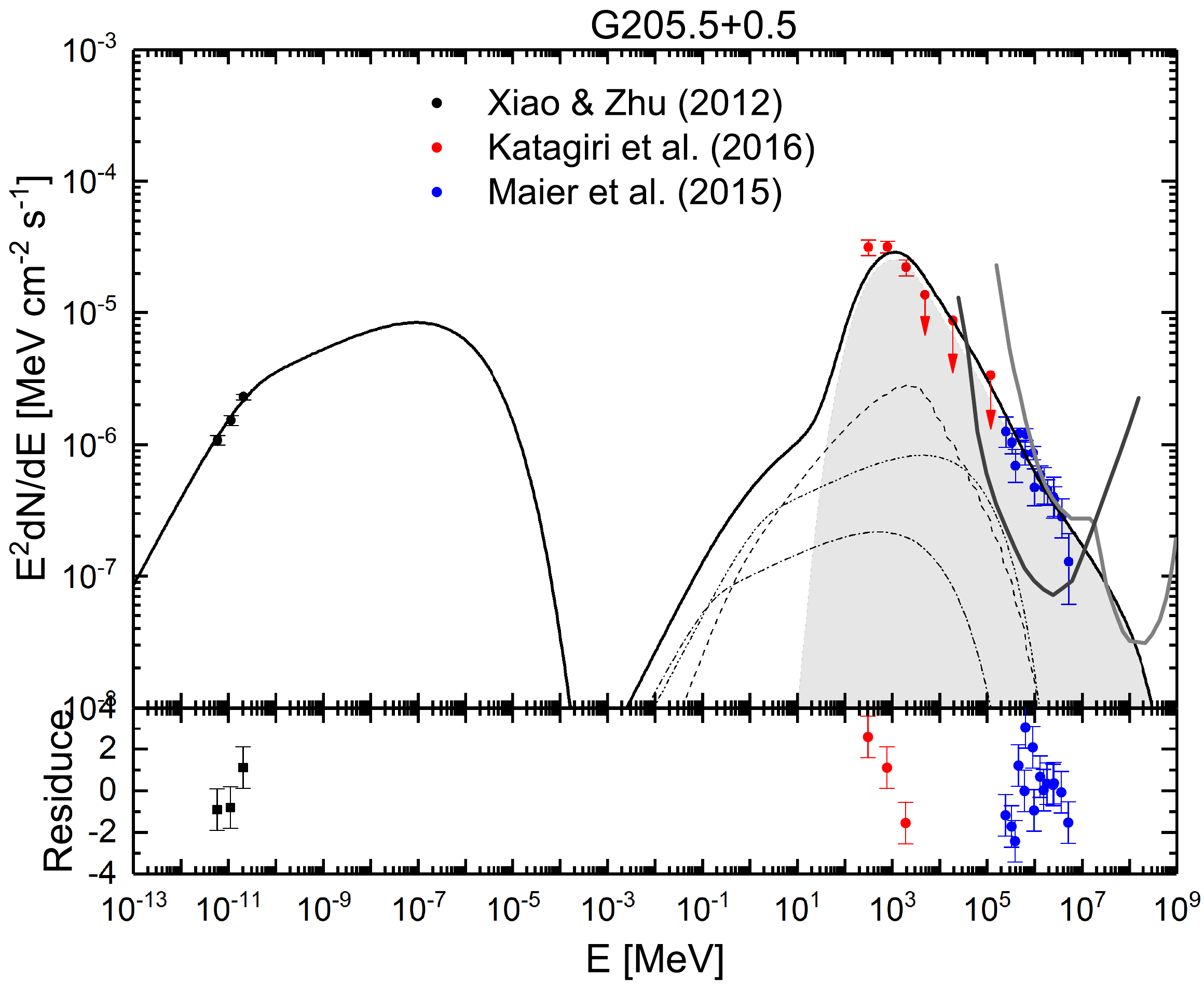}
\figsetgrpnote{The best fit to the spectral energy distribution (SED) for G205.5+0.5 }
\figsetgrpend

\figsetgrpstart
\figsetgrpnum{2.42}
\figsetgrptitle{G205.5+0.5 (b)
}
\figsetplot{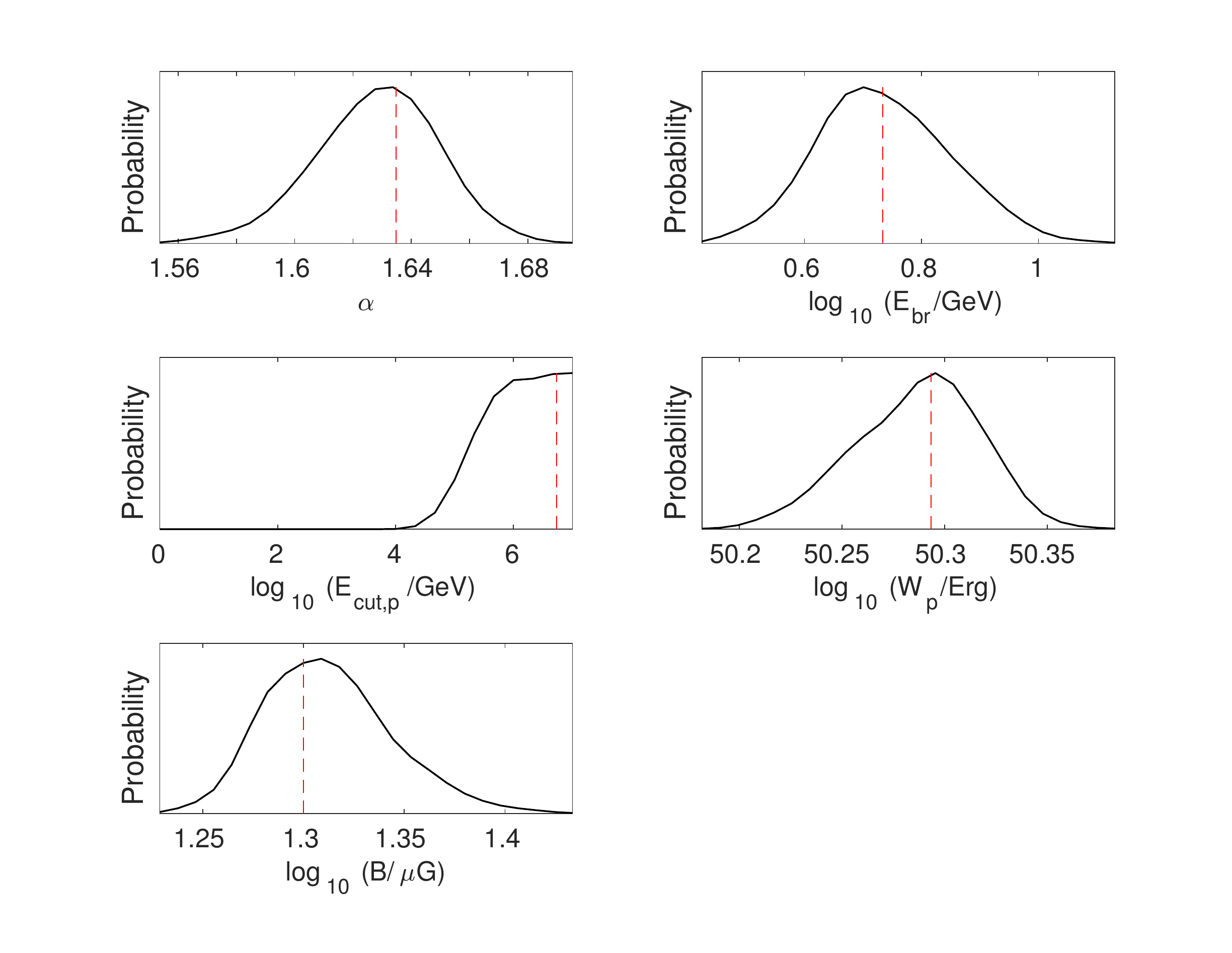}
\figsetgrpnote{ 1D probability distribution of the parameters for G205.5+0.5 }
\figsetgrpend

\figsetgrpstart
\figsetgrpnum{2.43}
\figsetgrptitle{Puppis A (a)
}
\figsetplot{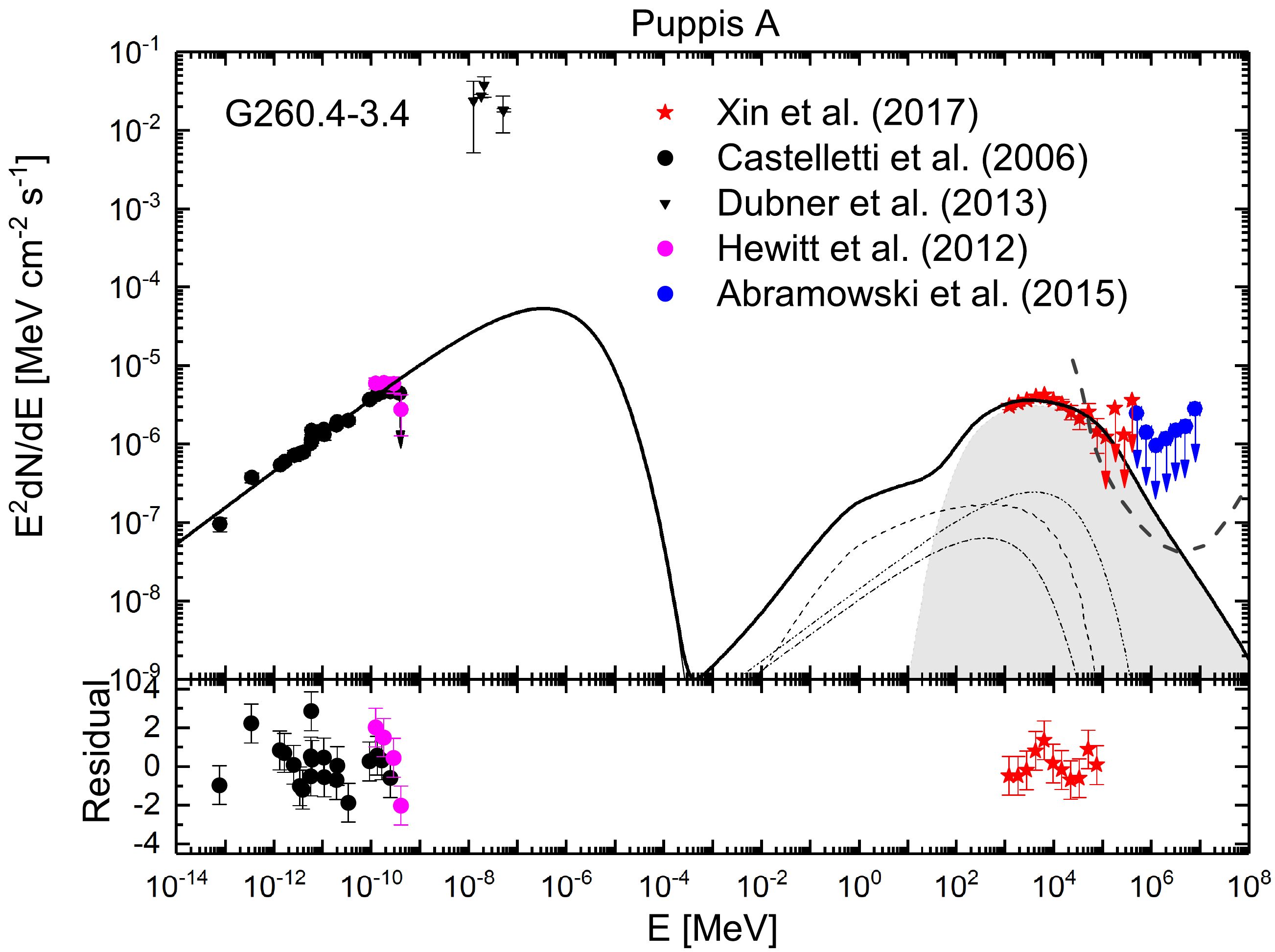}
\figsetgrpnote{The best fit to the spectral energy distribution (SED) for Puppis A}
\figsetgrpend

\figsetgrpstart
\figsetgrpnum{2.44}
\figsetgrptitle{Puppis A (b)
}
\figsetplot{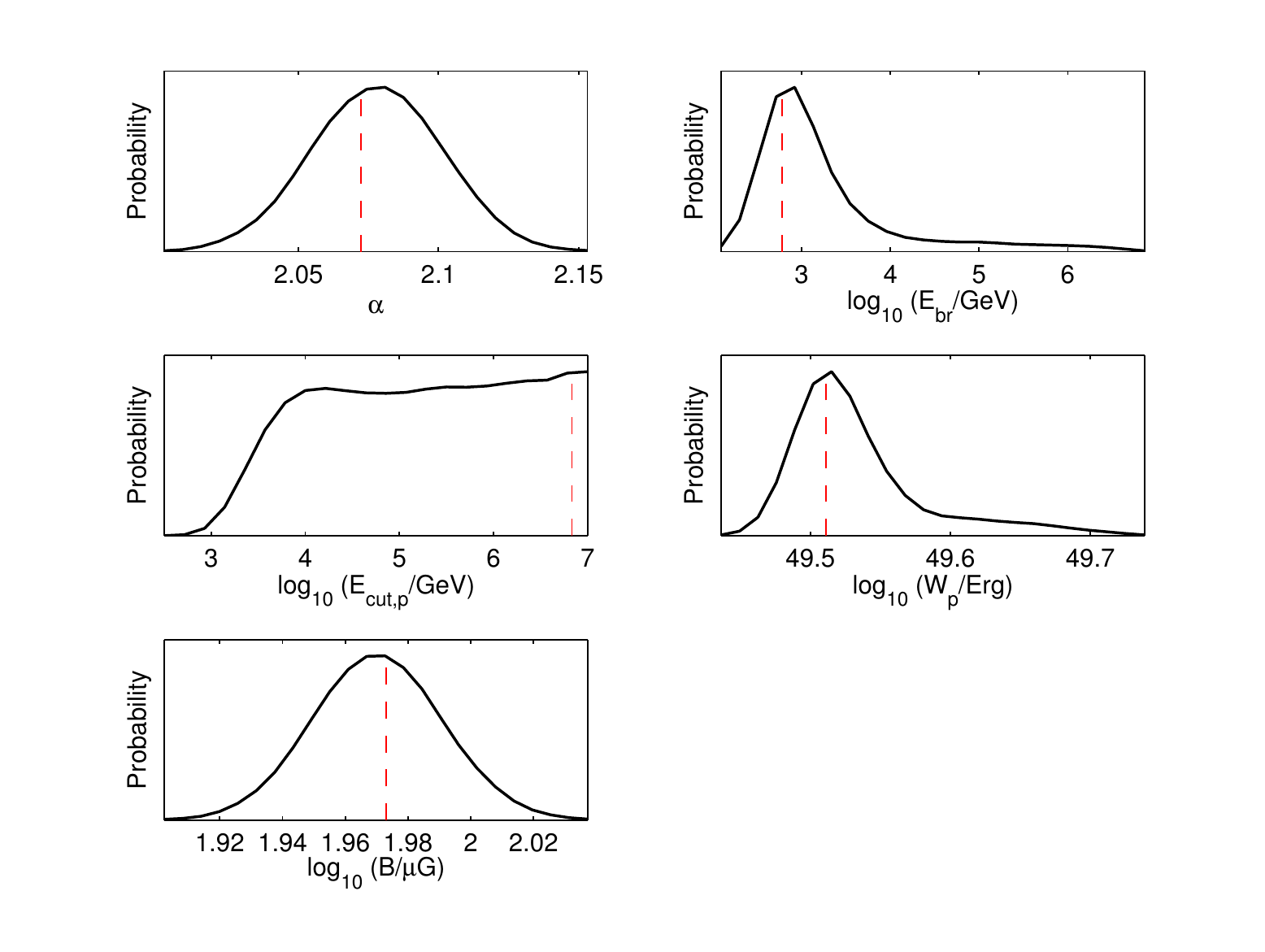}
\figsetgrpnote{1D probability distribution of the parameters for Puppis A }
\figsetgrpend

\figsetgrpstart
\figsetgrpnum{2.45}
\figsetgrptitle{G296.5+10.0 (a)
}
\figsetplot{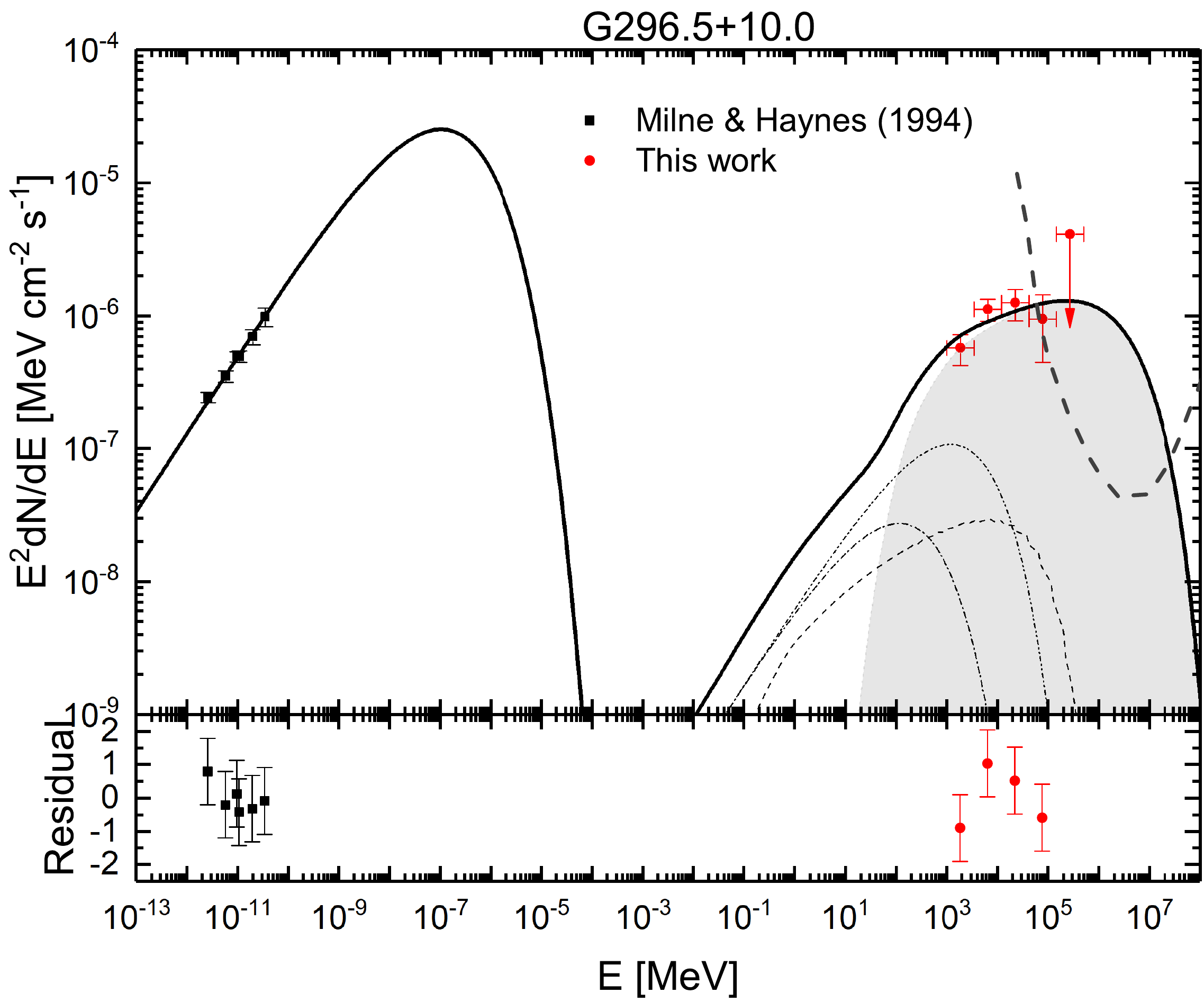}
\figsetgrpnote{The best fit to the spectral energy distribution (SED) for G296.5+10.0 }
\figsetgrpend

\figsetgrpstart
\figsetgrpnum{2.46}
\figsetgrptitle{G296.5+10.0 (b)
}
\figsetplot{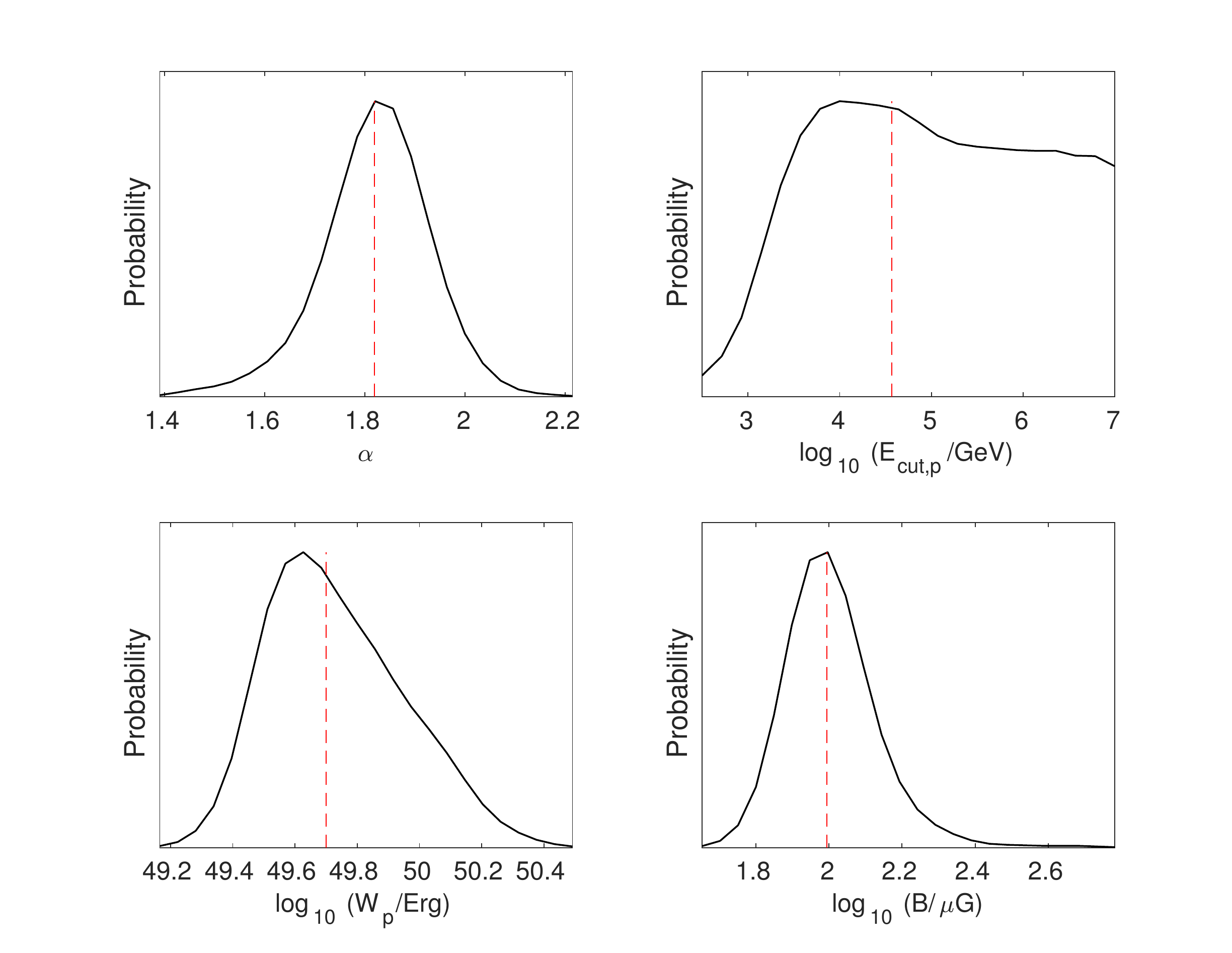}
\figsetgrpnote{1D probability distribution of the parameters for G296.5+10.0 }
\figsetgrpend

\figsetgrpstart
\figsetgrpnum{2.47}
\figsetgrptitle{RX J0852-4622 (a)
}
\figsetplot{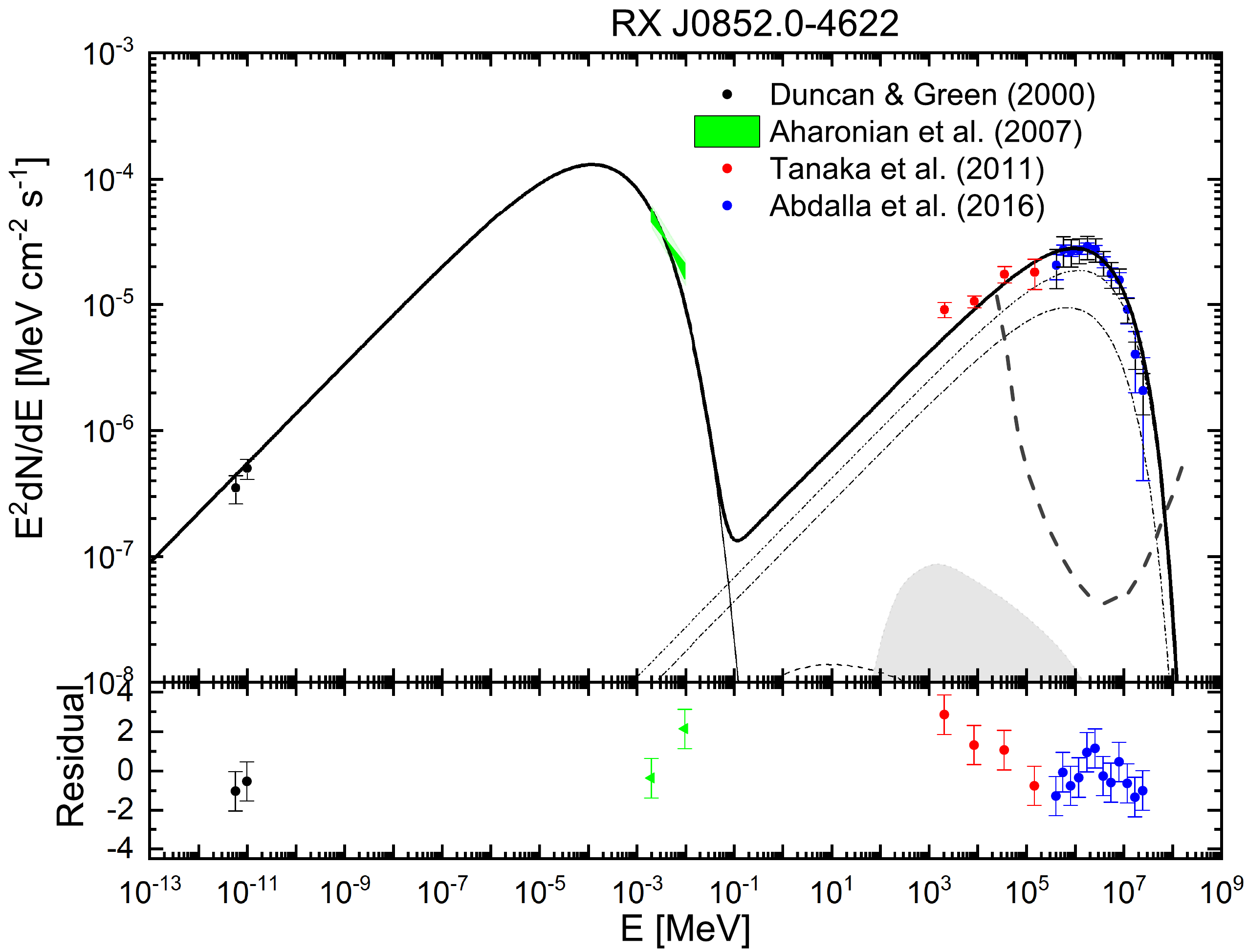}
\figsetgrpnote{The best fit to the spectral energy distribution (SED) for RX J0852-4622 }
\figsetgrpend

\figsetgrpstart
\figsetgrpnum{2.48}
\figsetgrptitle{RX J0852-4622 (b)
}
\figsetplot{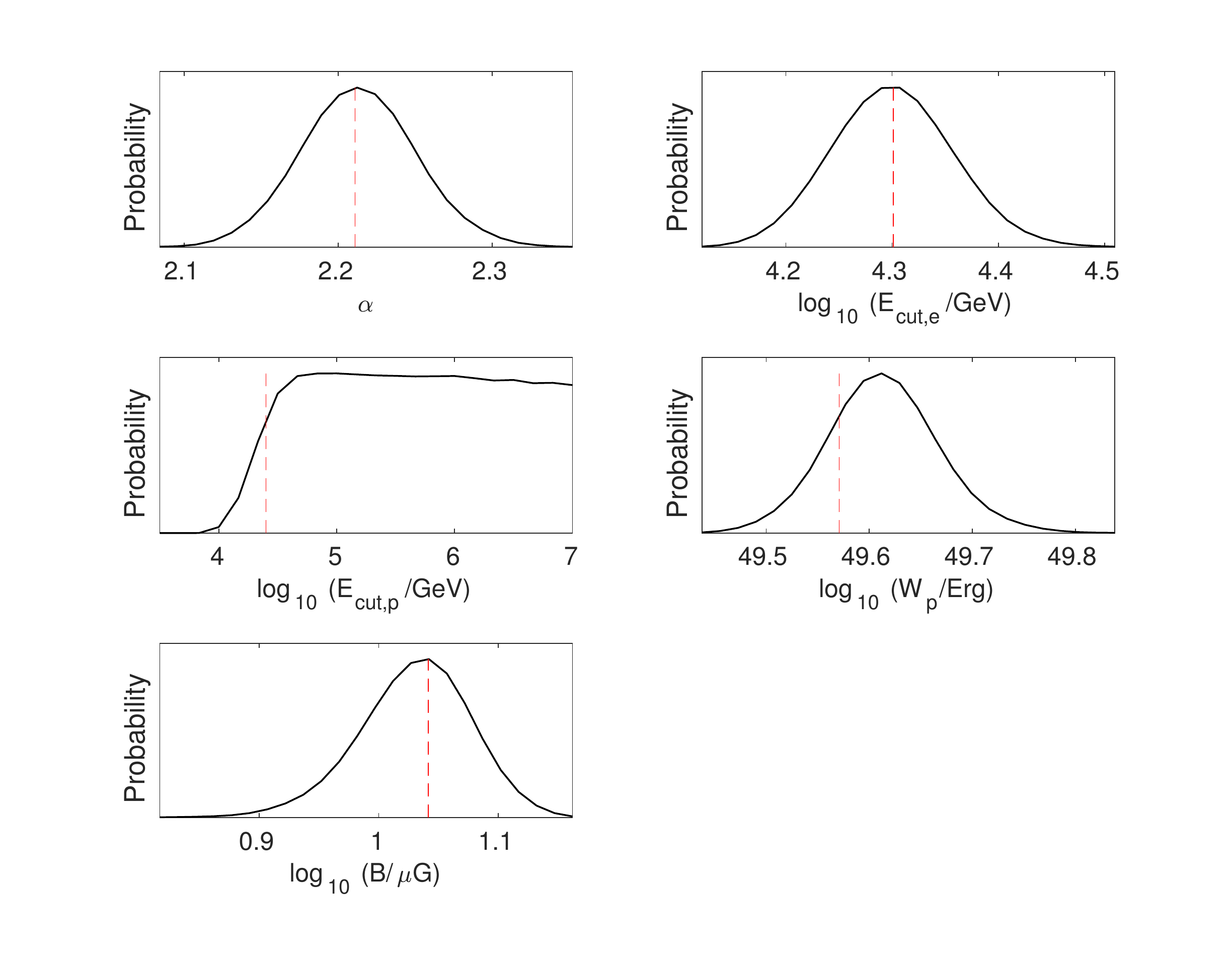}
\figsetgrpnote{1D probability distribution of the parameters for RX J0852-4622}
\figsetgrpend

\figsetgrpstart
\figsetgrpnum{2.49}
\figsetgrptitle{RX J0852-4622 ($\rm a^{'}$)
}
\figsetplot{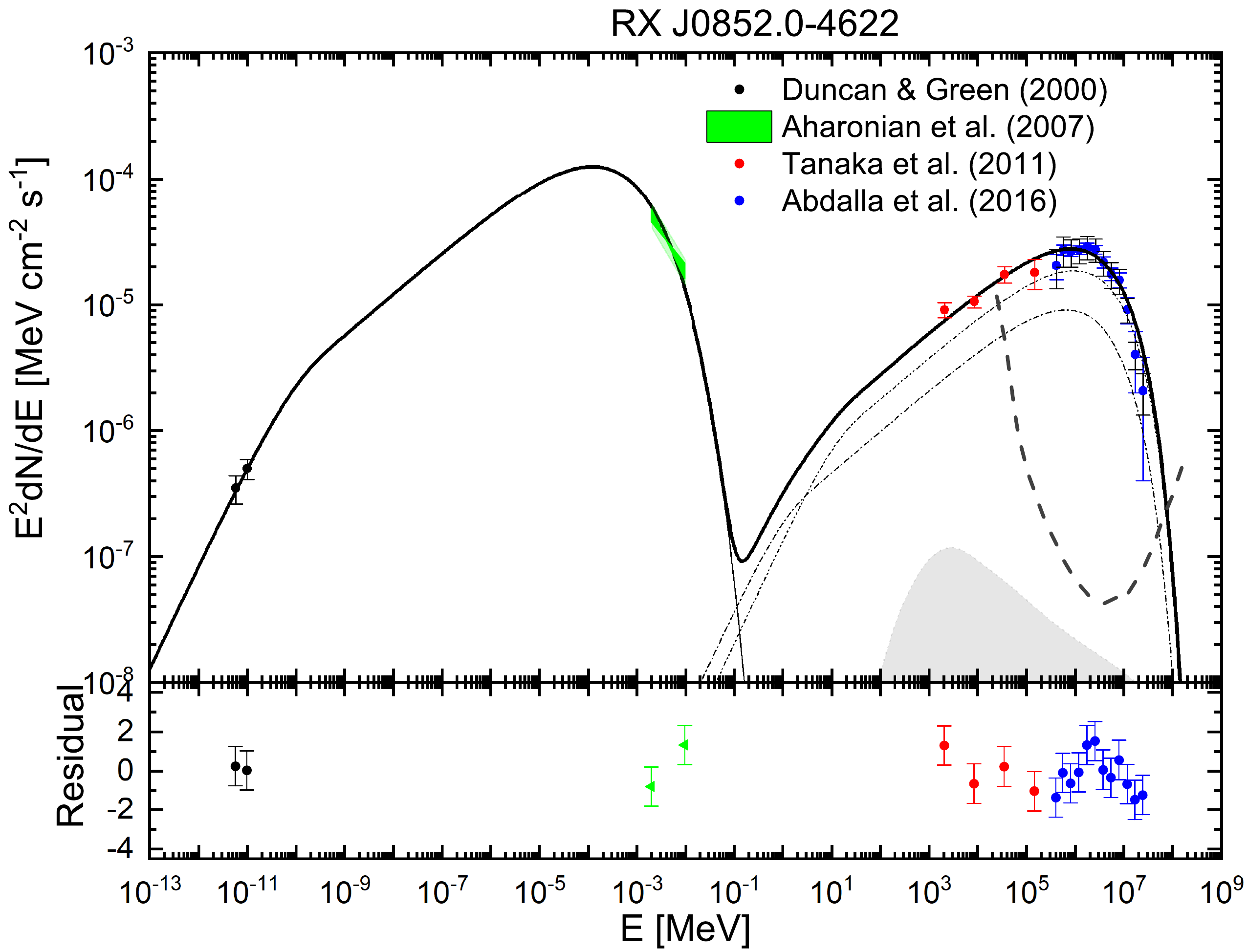}
\figsetgrpnote{The best fit to the spectral energy distribution (SED) for RX J0852-4622 }
\figsetgrpend

\figsetgrpstart
\figsetgrpnum{2.50}
\figsetgrptitle{RX J0852-4622 ($\rm b^{'}$)
}
\figsetplot{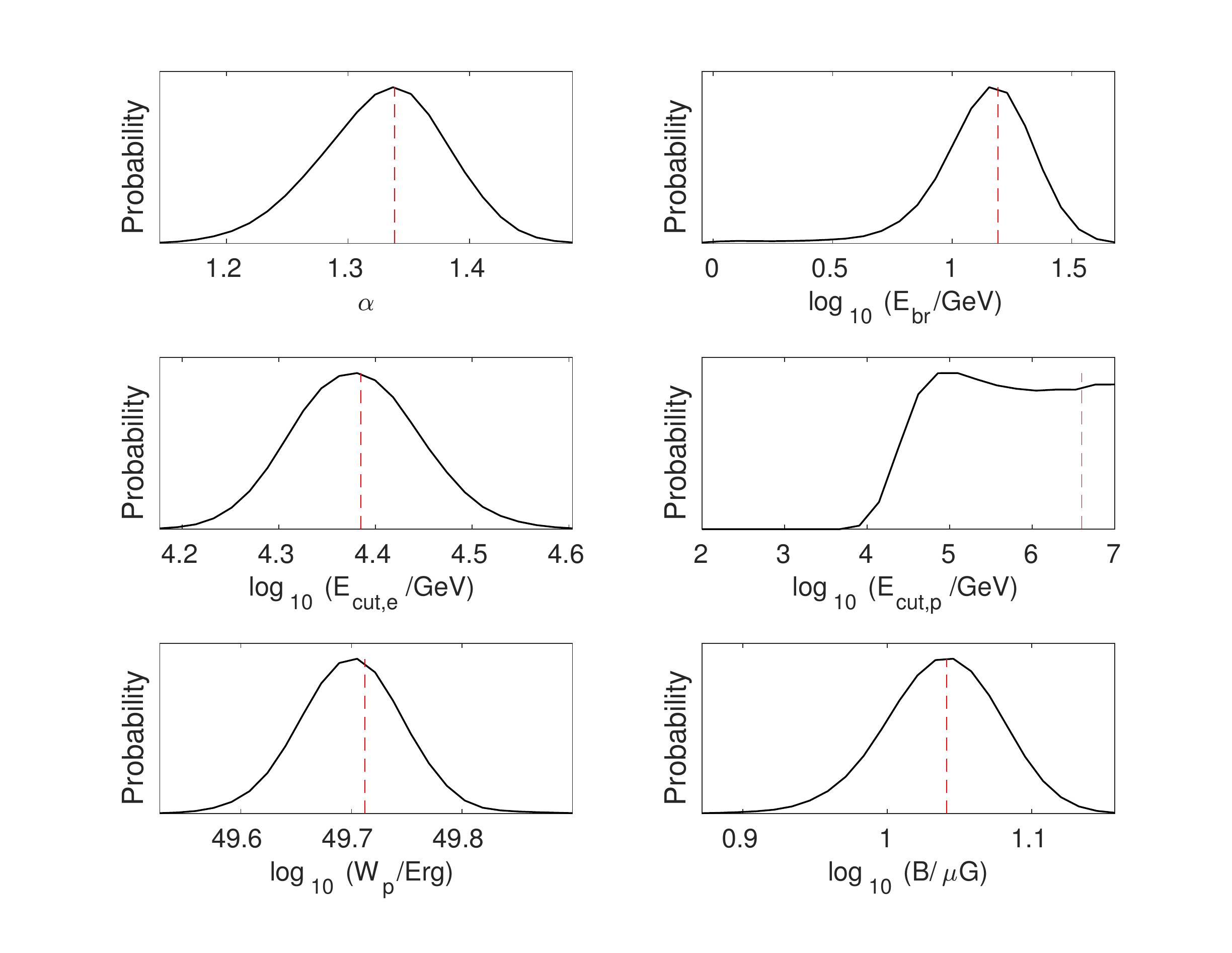}
\figsetgrpnote{1D probability distribution of the parameters for RX J0852-4622 }
\figsetgrpend

\figsetgrpstart
\figsetgrpnum{2.51}
\figsetgrptitle{Kes 17 (a)
}
\figsetplot{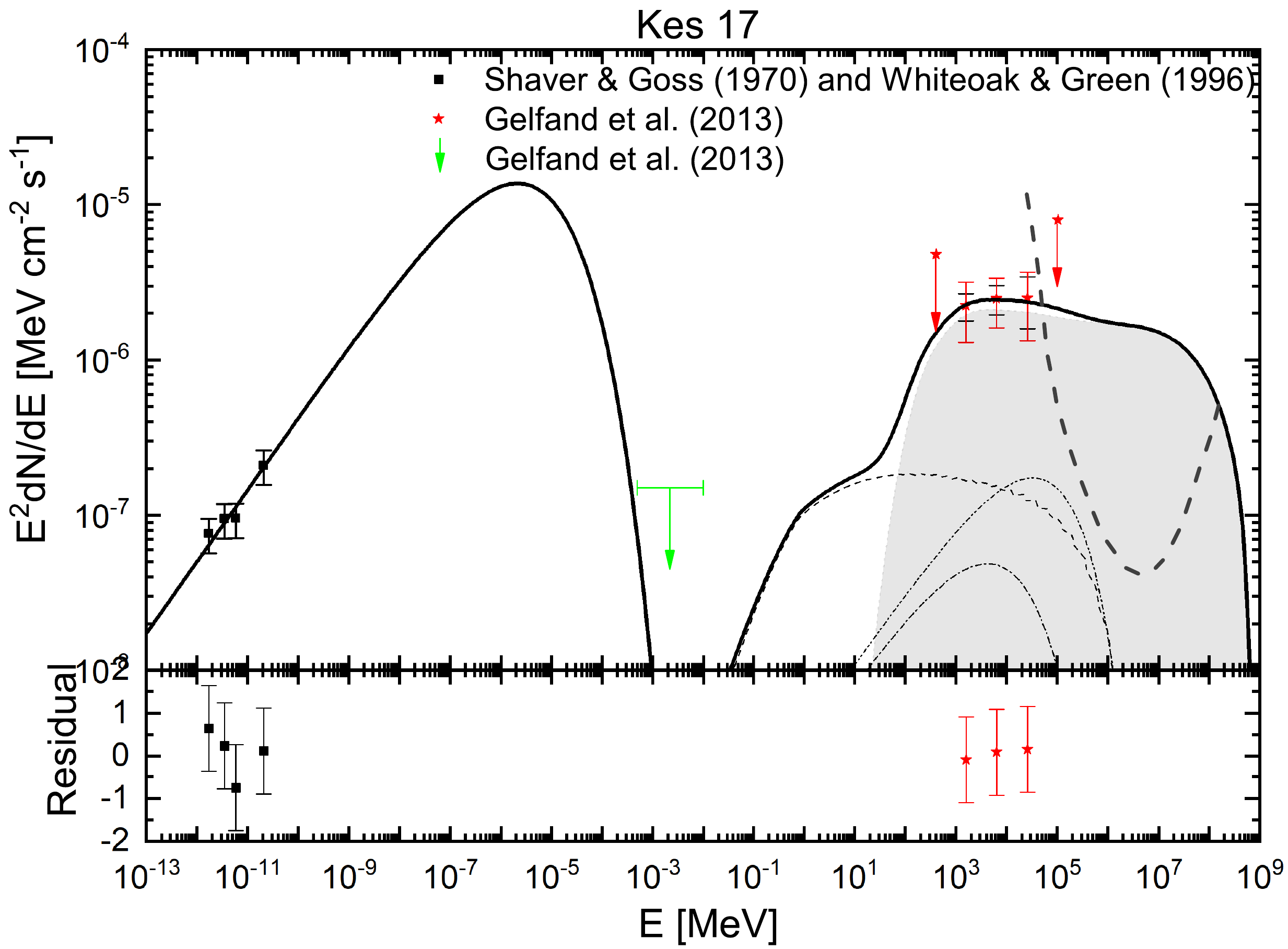}
\figsetgrpnote{The best fit to the spectral energy distribution (SED) for Kes 17 }
\figsetgrpend

\figsetgrpstart
\figsetgrpnum{2.52}
\figsetgrptitle{Kes 17 (b)
}
\figsetplot{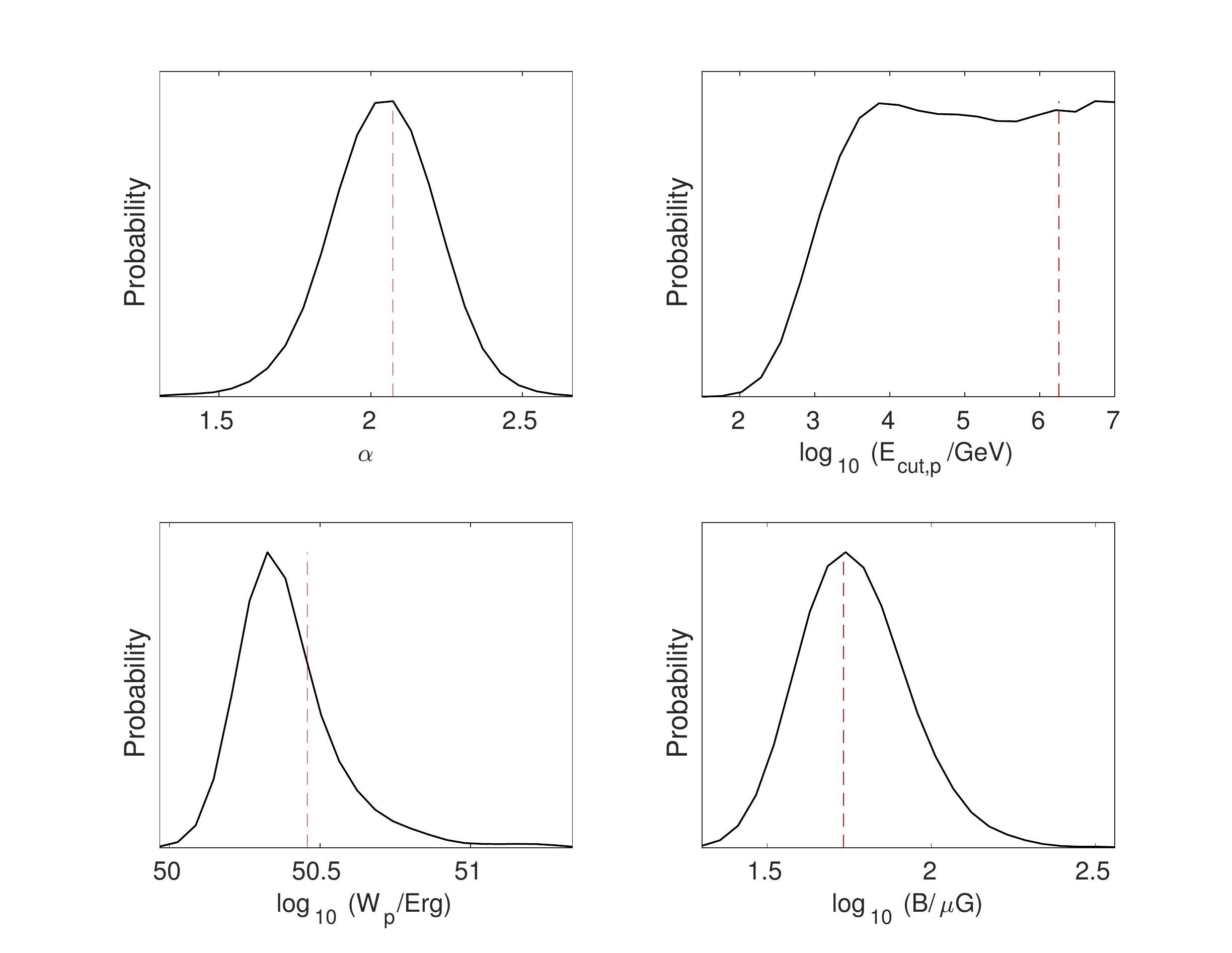}
\figsetgrpnote{1D probability distribution of the parameters for Kes 17}
\figsetgrpend

\figsetgrpstart
\figsetgrpnum{2.53}
\figsetgrptitle{RCW 86 (a)
}
\figsetplot{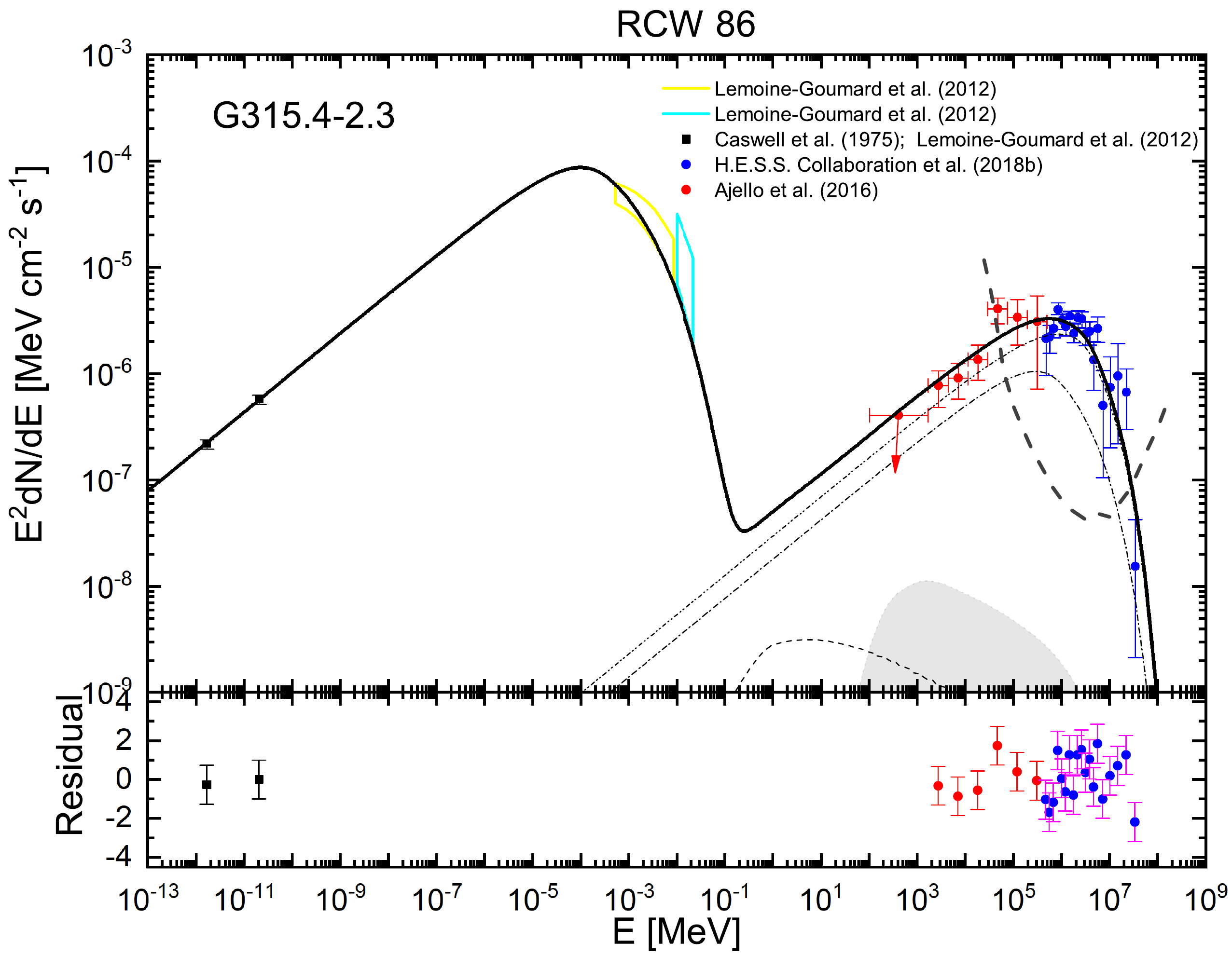}
\figsetgrpnote{The best fit to the spectral energy distribution (SED) for RCW 86}
\figsetgrpend

\figsetgrpstart
\figsetgrpnum{2.54}
\figsetgrptitle{RCW 86 (b)
}
\figsetplot{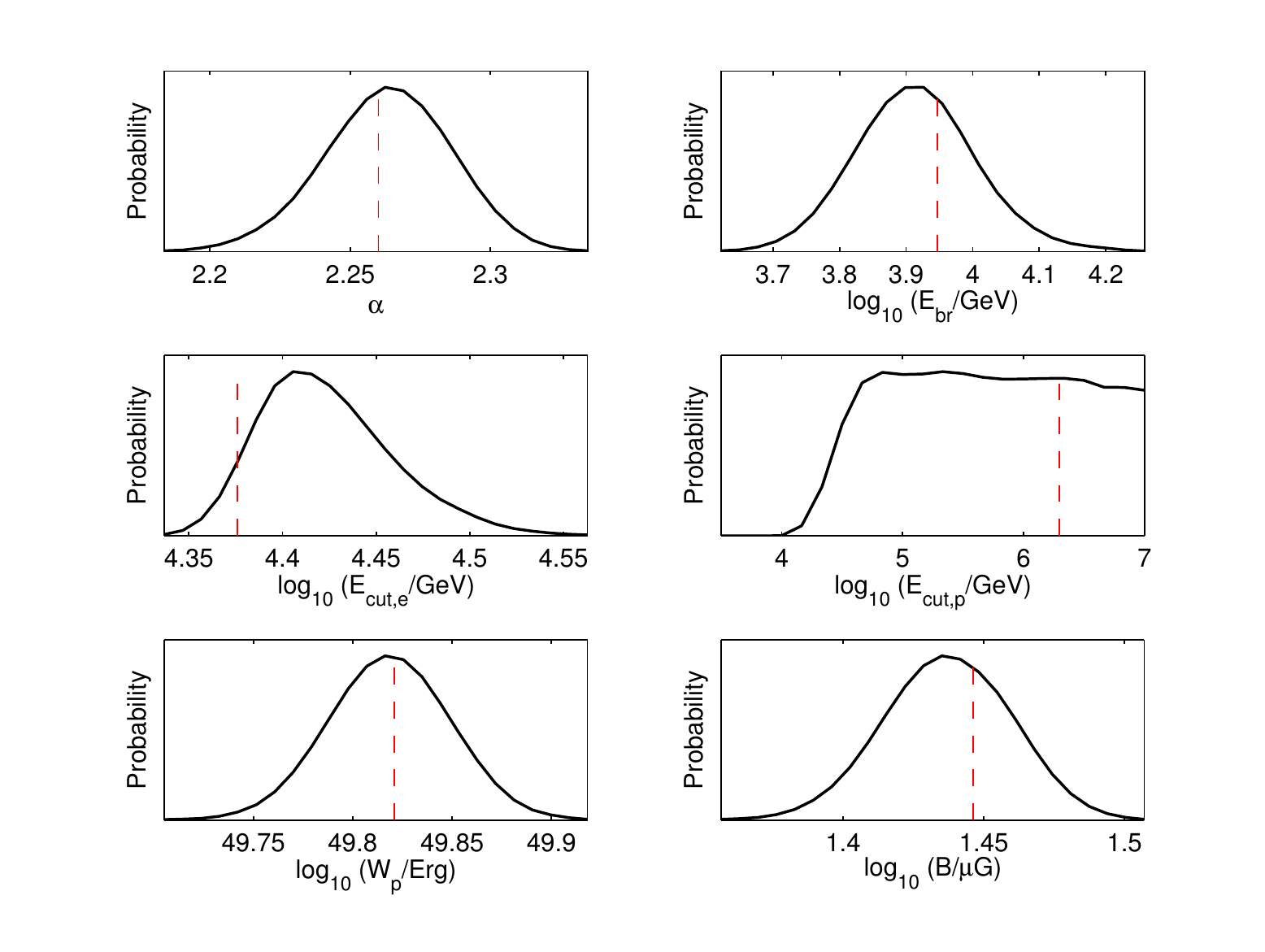}
\figsetgrpnote{1D probability distribution of the parameters for RCW 86}
\figsetgrpend

\figsetgrpstart
\figsetgrpnum{2.55}
\figsetgrptitle{MSH 15-56 (a)
}
\figsetplot{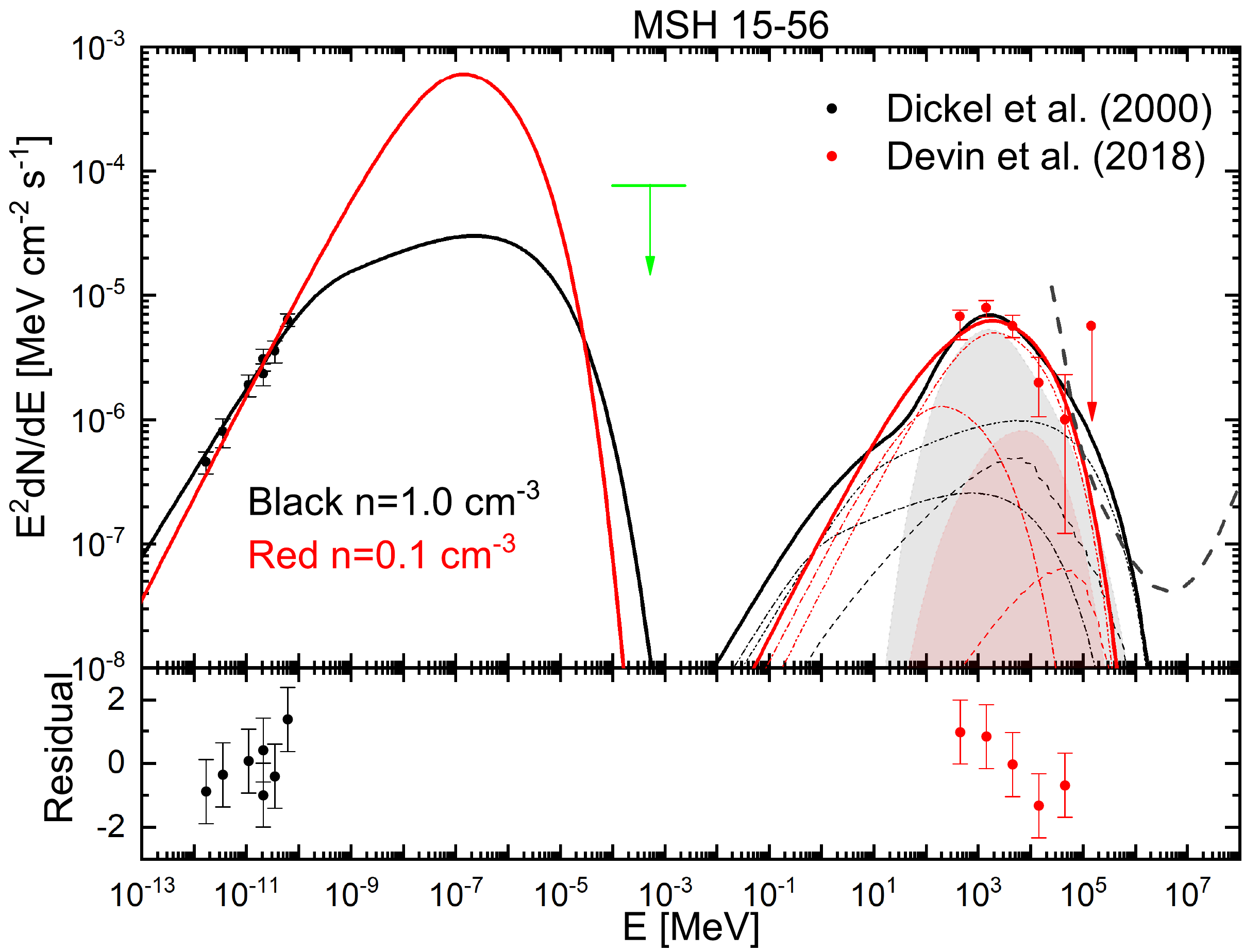}
\figsetgrpnote{The best fit to the spectral energy distribution (SED) for MSH 15-56 }
\figsetgrpend

\figsetgrpstart
\figsetgrpnum{2.56}
\figsetgrptitle{MSH 15-56 (b)
}
\figsetplot{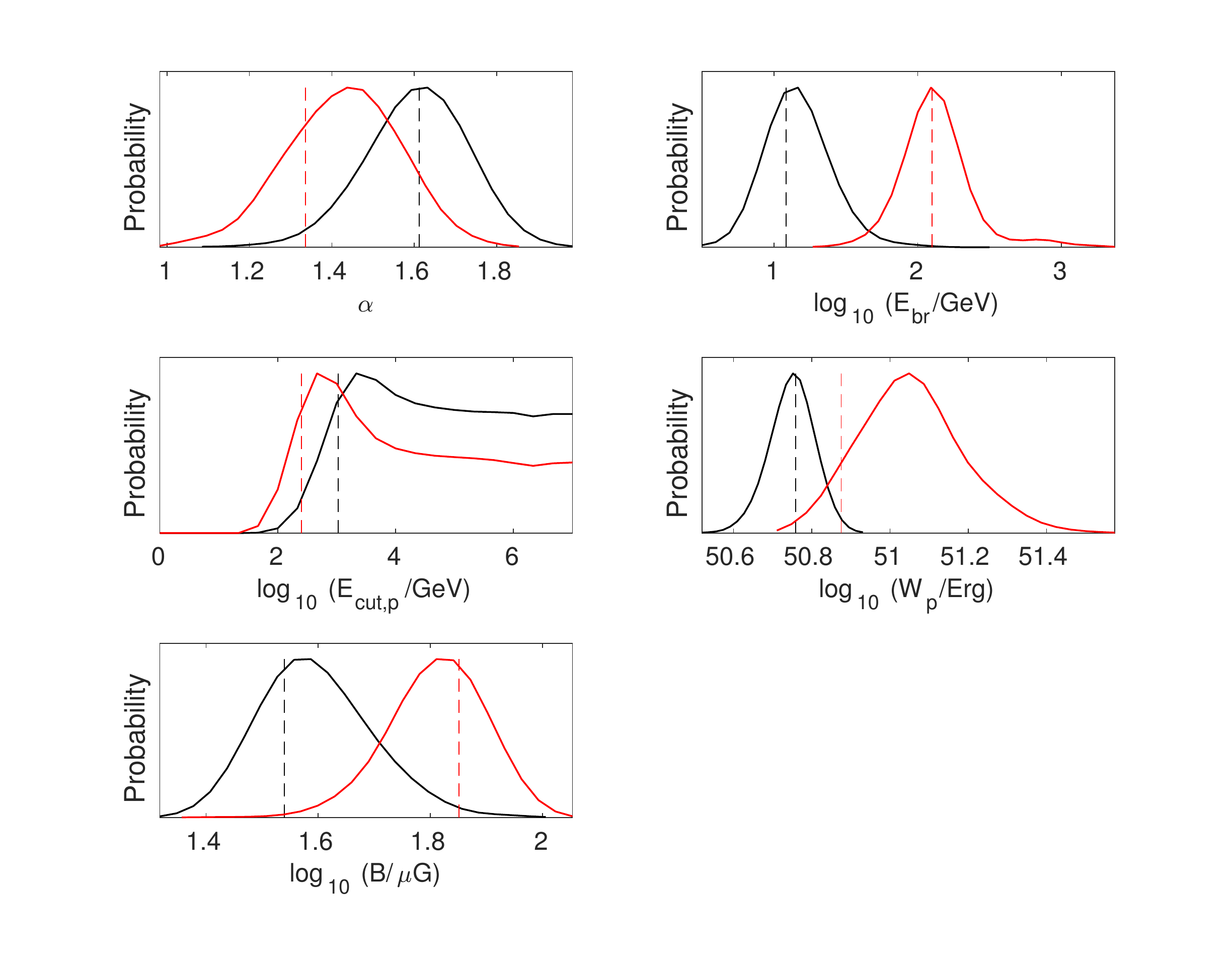}
\figsetgrpnote{1D probability distribution of the parameters for MSH 15-56}
\figsetgrpend

\figsetgrpstart
\figsetgrpnum{2.57}
\figsetgrptitle{SN1006 (a)
}
\figsetplot{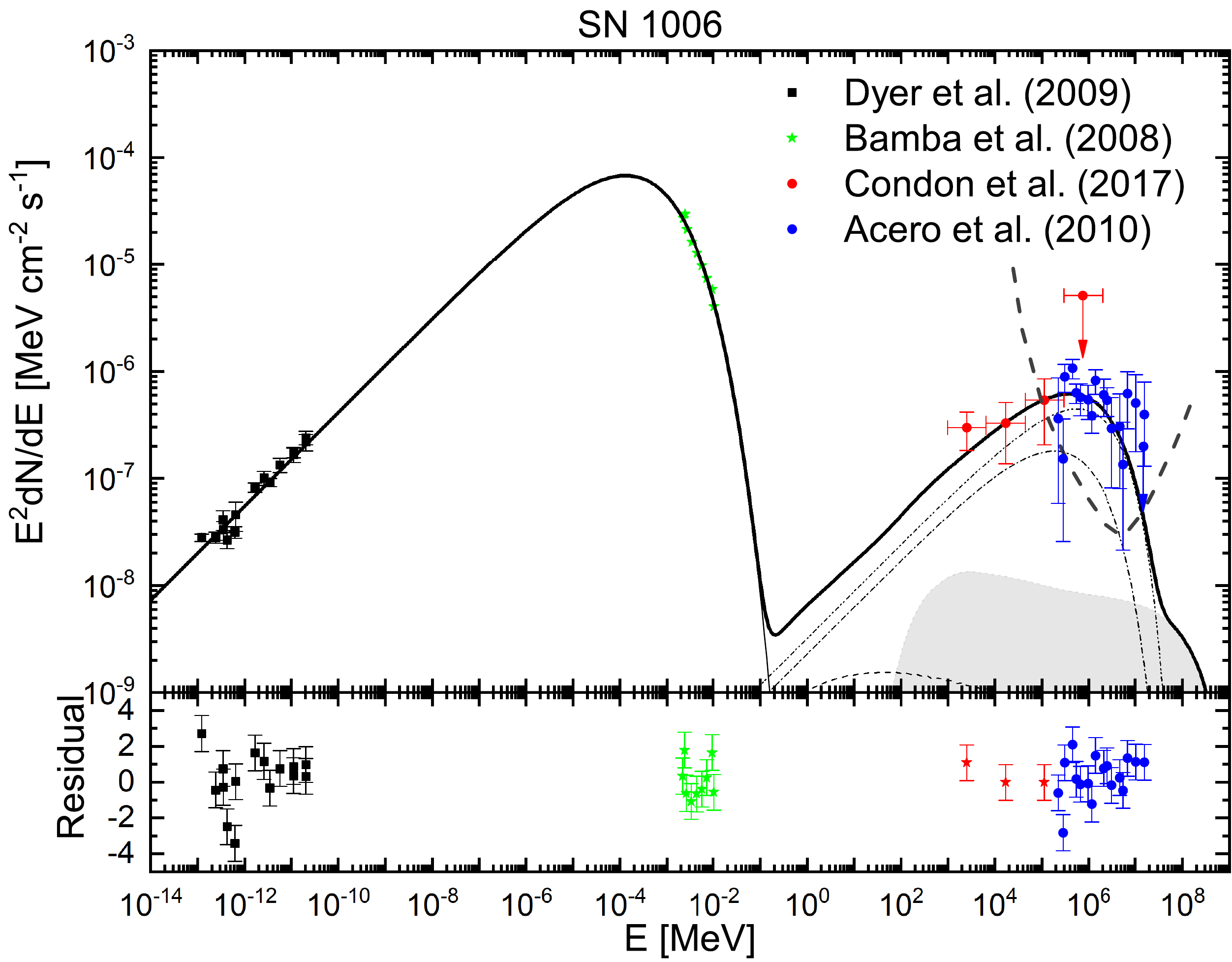}
\figsetgrpnote{The best fit to the spectral energy distribution (SED) for SN 1006 }
\figsetgrpend

\figsetgrpstart
\figsetgrpnum{2.58}
\figsetgrptitle{SN1006(b)
}
\figsetplot{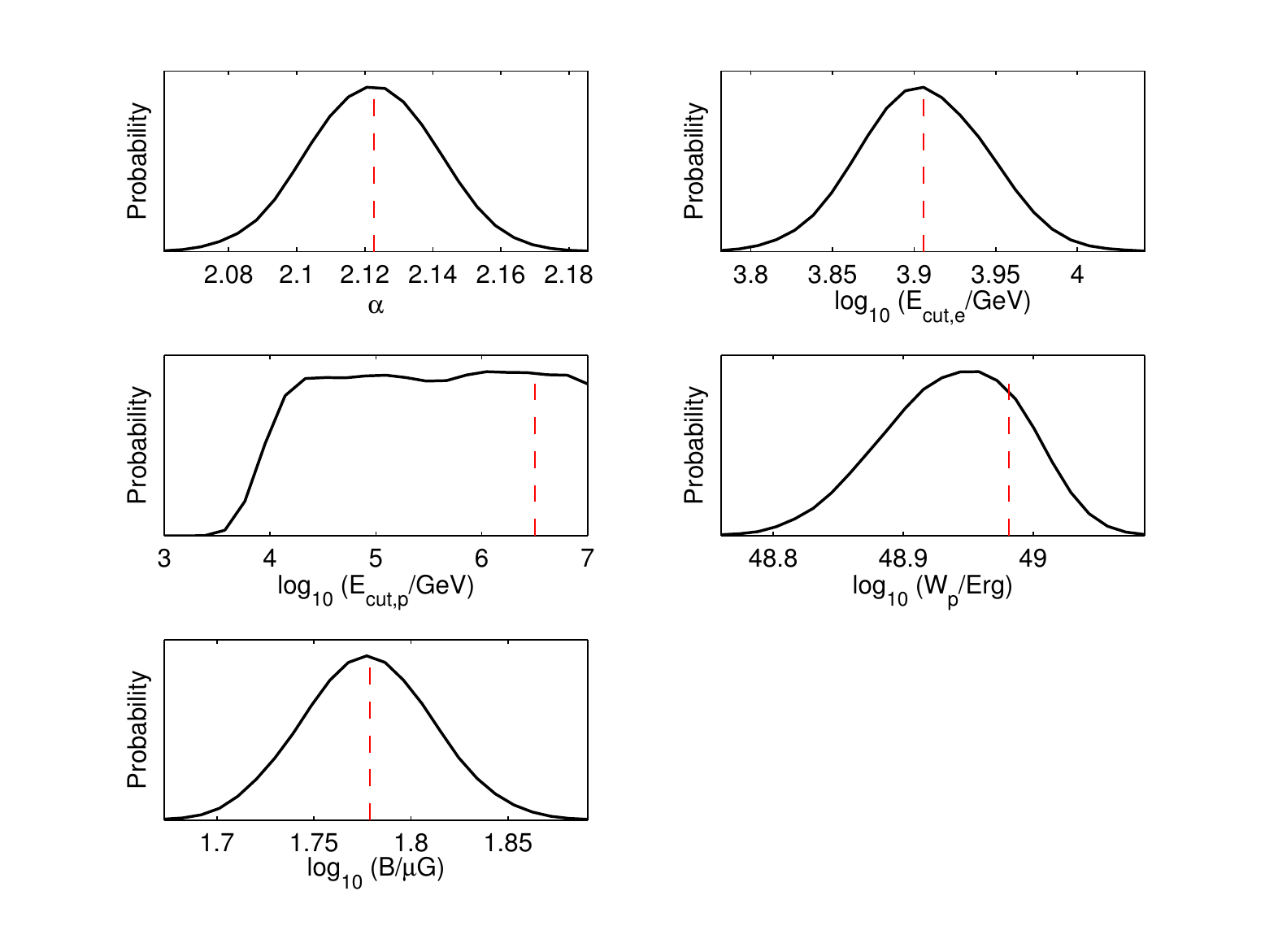}
\figsetgrpnote{1D probability distribution of the parameters for SN 1006 }
\figsetgrpend

\figsetgrpstart
\figsetgrpnum{2.59}
\figsetgrptitle{RCW 103 (a)
}
\figsetplot{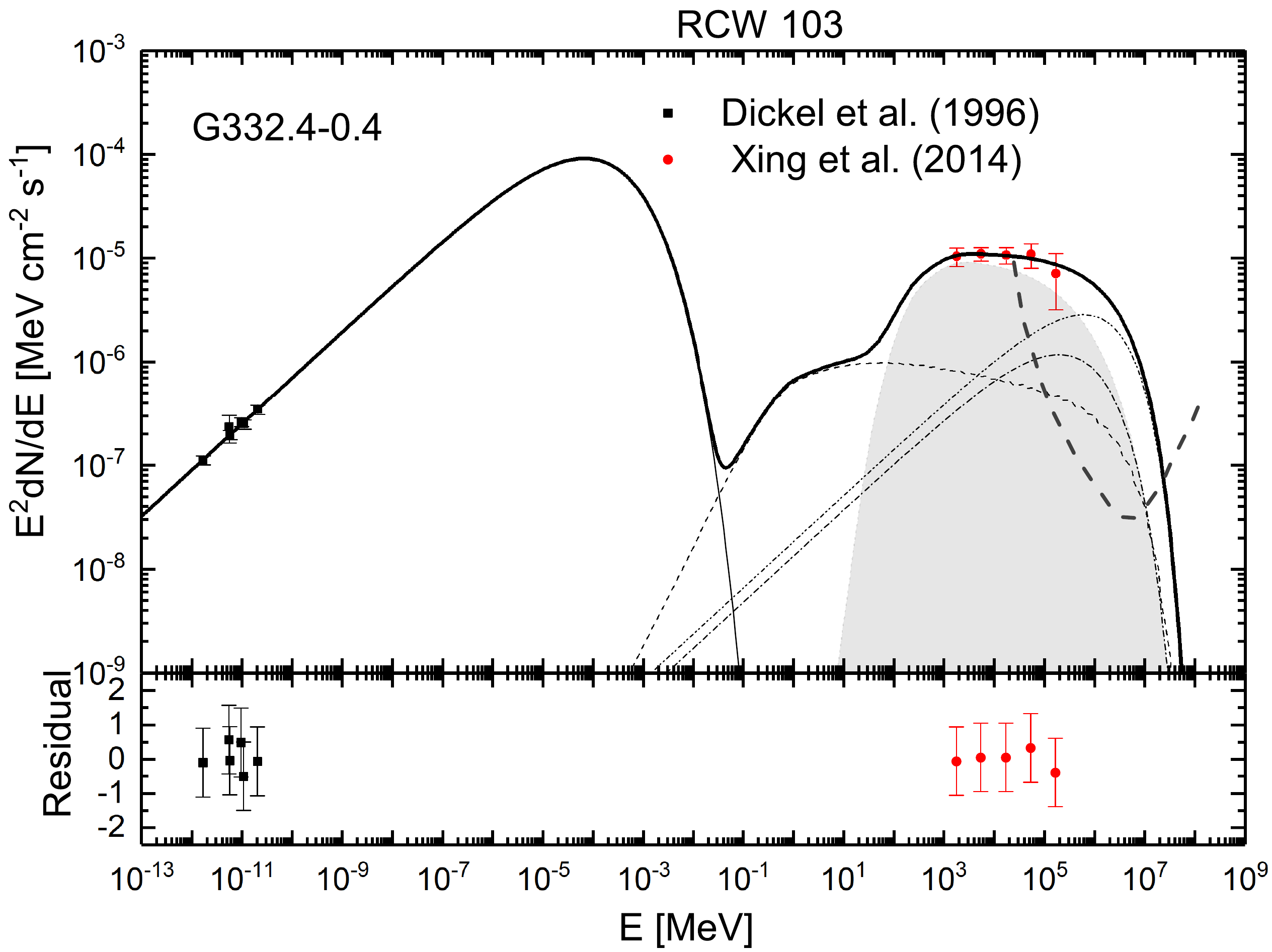}
\figsetgrpnote{The best fit to the spectral energy distribution (SED) for RCW 103}
\figsetgrpend

\figsetgrpstart
\figsetgrpnum{2.60}
\figsetgrptitle{RCW 103 (b)
}
\figsetplot{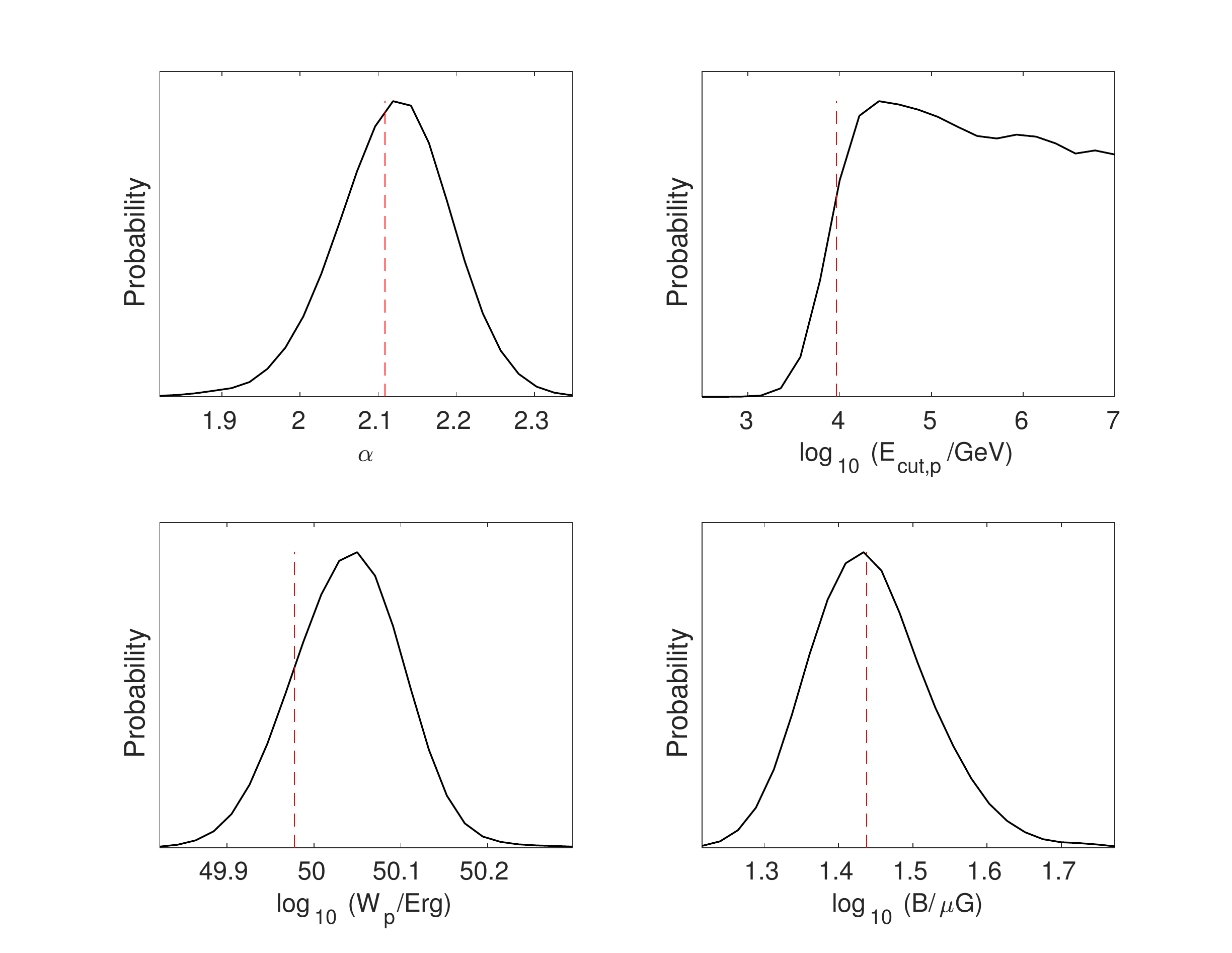}
\figsetgrpnote{1D probability distribution of the parameters for RCW 103}
\figsetgrpend

\figsetgrpstart
\figsetgrpnum{2.61}
\figsetgrptitle{CTB 33 (a)
}
\figsetplot{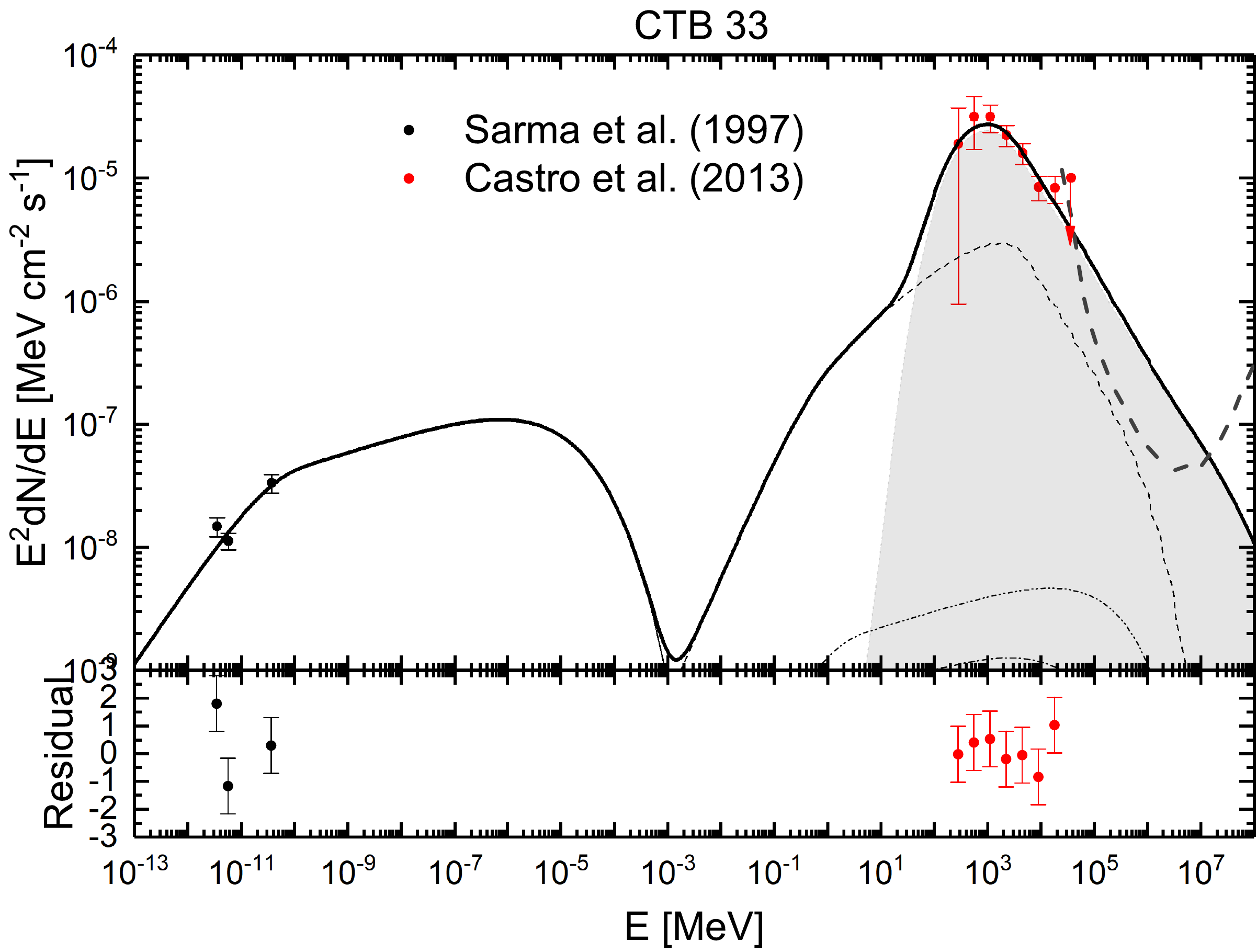}
\figsetgrpnote{The best fit to the spectral energy distribution (SED) for CTB 33}
\figsetgrpend

\figsetgrpstart
\figsetgrpnum{2.62}
\figsetgrptitle{CTB 33 (b)
}
\figsetplot{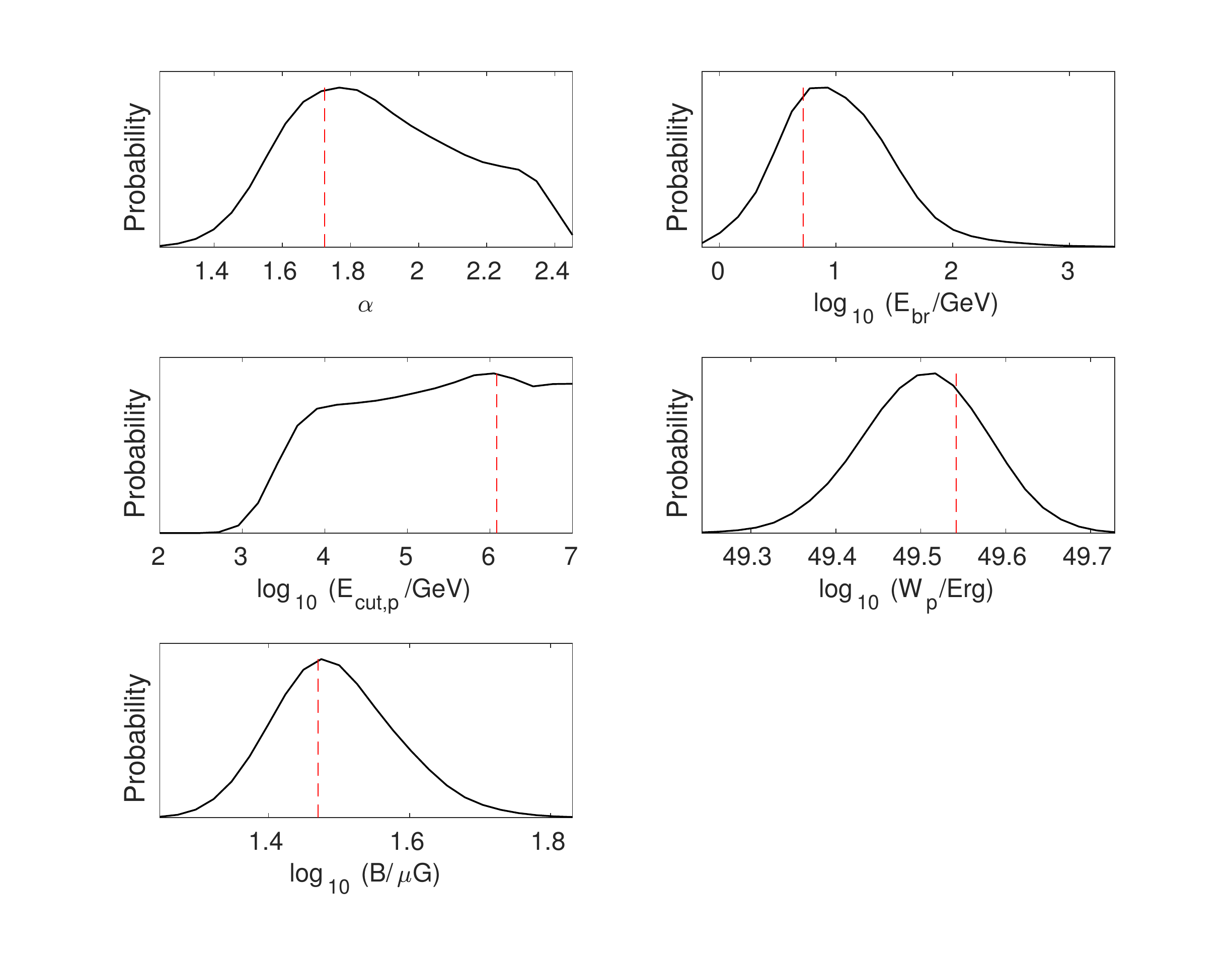}
\figsetgrpnote{1D probability distribution of the parameters for CTB 33}
\figsetgrpend

\figsetgrpstart
\figsetgrpnum{2.63}
\figsetgrptitle{RX J1713.7-3946 (a)
}
\figsetplot{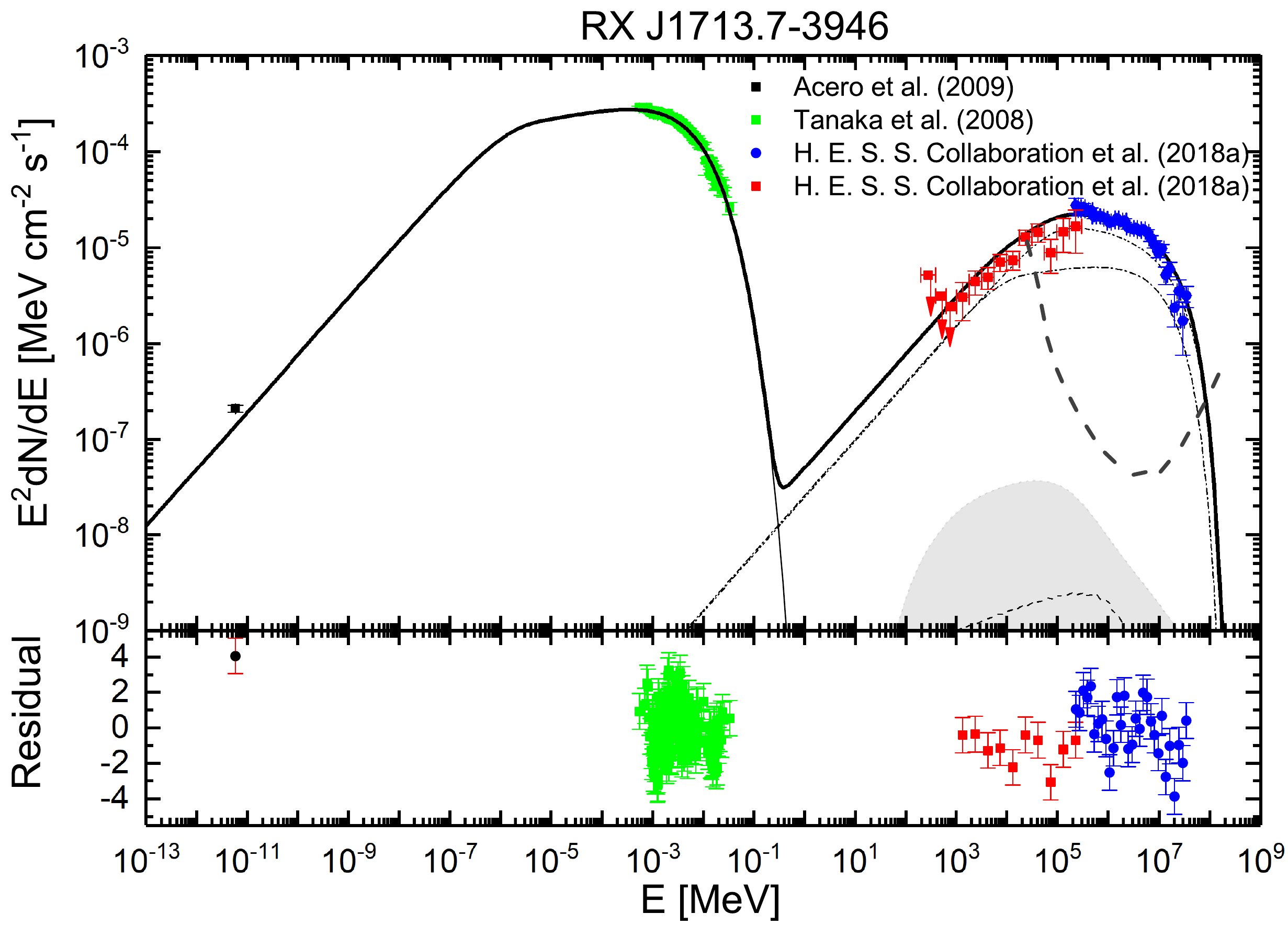}
\figsetgrpnote{The best fit to the spectral energy distribution (SED) for RX J1713.7-3946 }
\figsetgrpend

\figsetgrpstart
\figsetgrpnum{2.64}
\figsetgrptitle{RX J1713.7-3946 (b)
}
\figsetplot{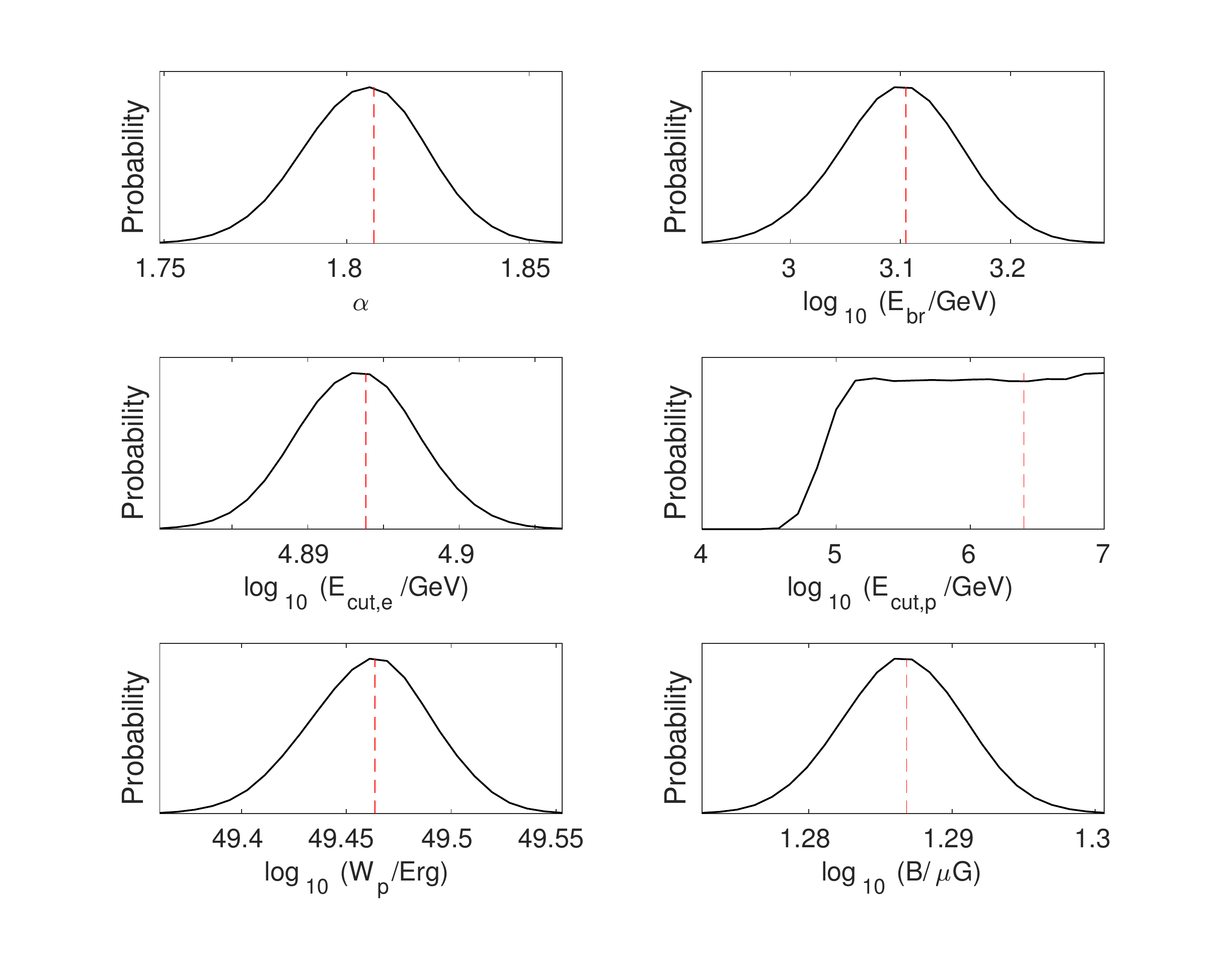}
\figsetgrpnote{1D probability distribution of the parameters for RX J1713.7-3946 }
\figsetgrpend

\figsetgrpstart
\figsetgrpnum{2.65}
\figsetgrptitle{CTB 37A (a)
}
\figsetplot{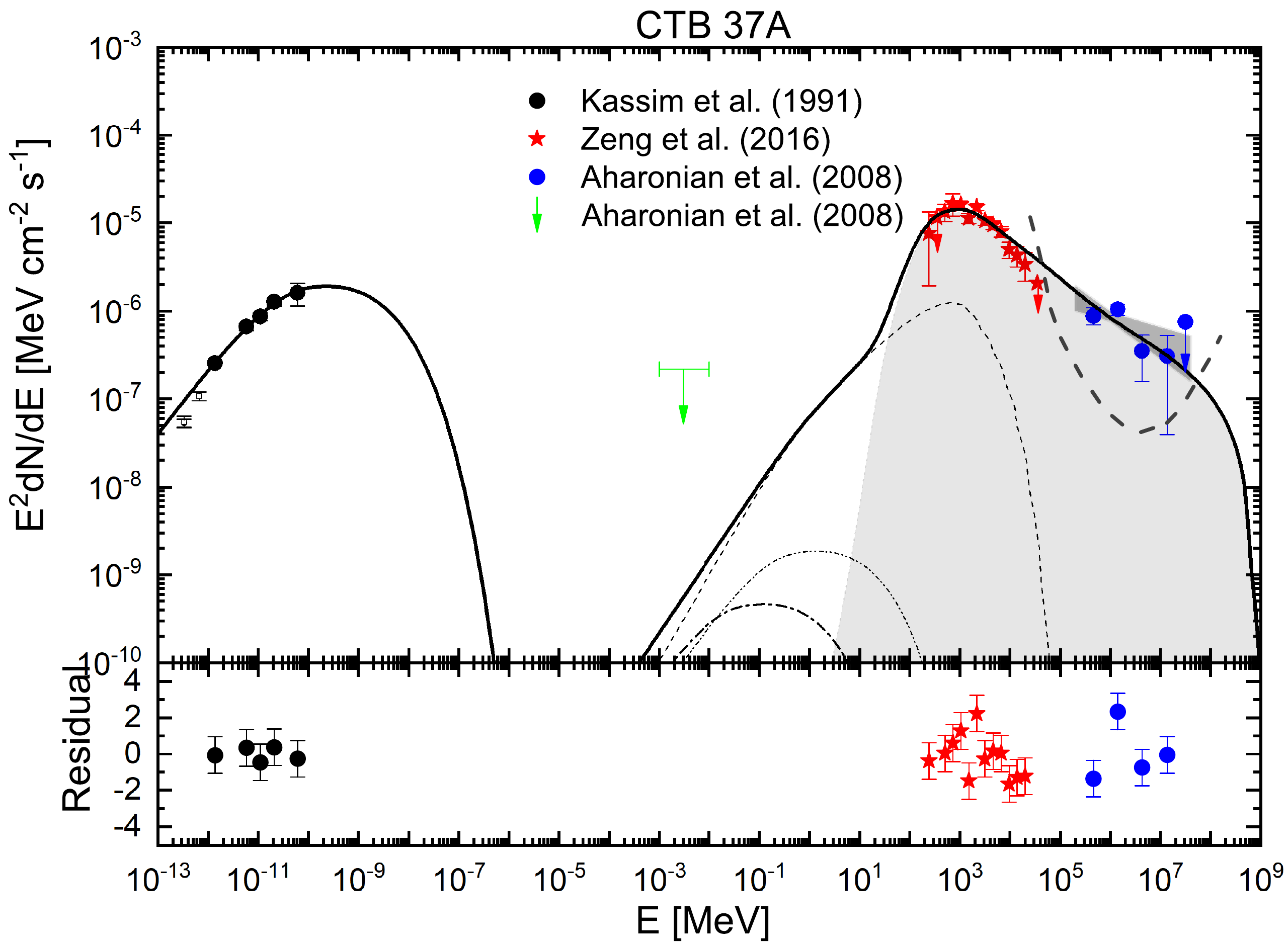}
\figsetgrpnote{The best fit to the spectral energy distribution (SED) for CTB 37A}
\figsetgrpend

\figsetgrpstart
\figsetgrpnum{2.66}
\figsetgrptitle{CTB 37A (b)
}
\figsetplot{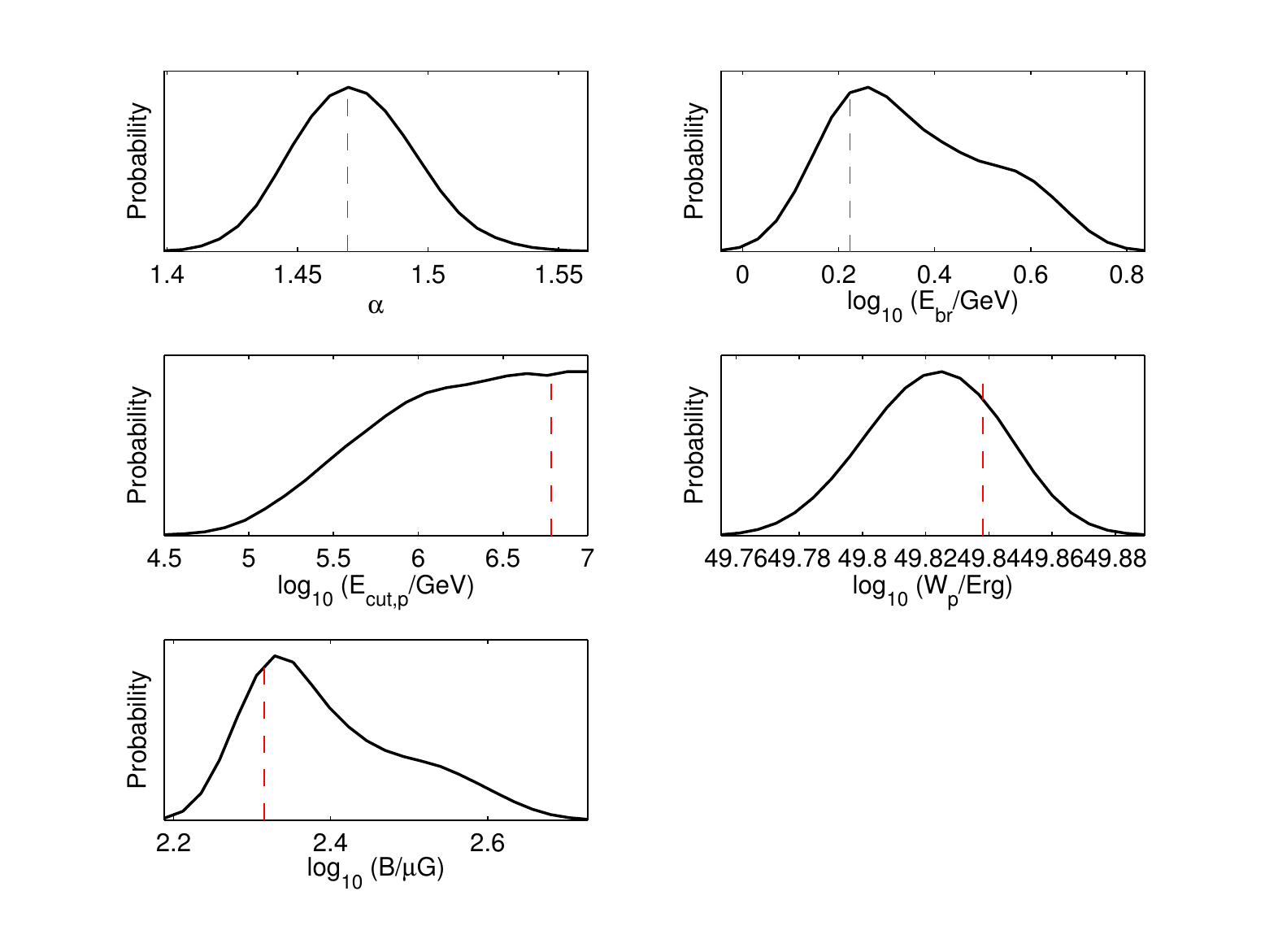}
\figsetgrpnote{1D probability distribution of the parameters for CTB 37A}
\figsetgrpend

\figsetgrpstart
\figsetgrpnum{2.67}
\figsetgrptitle{CTB 37B (a)
}
\figsetplot{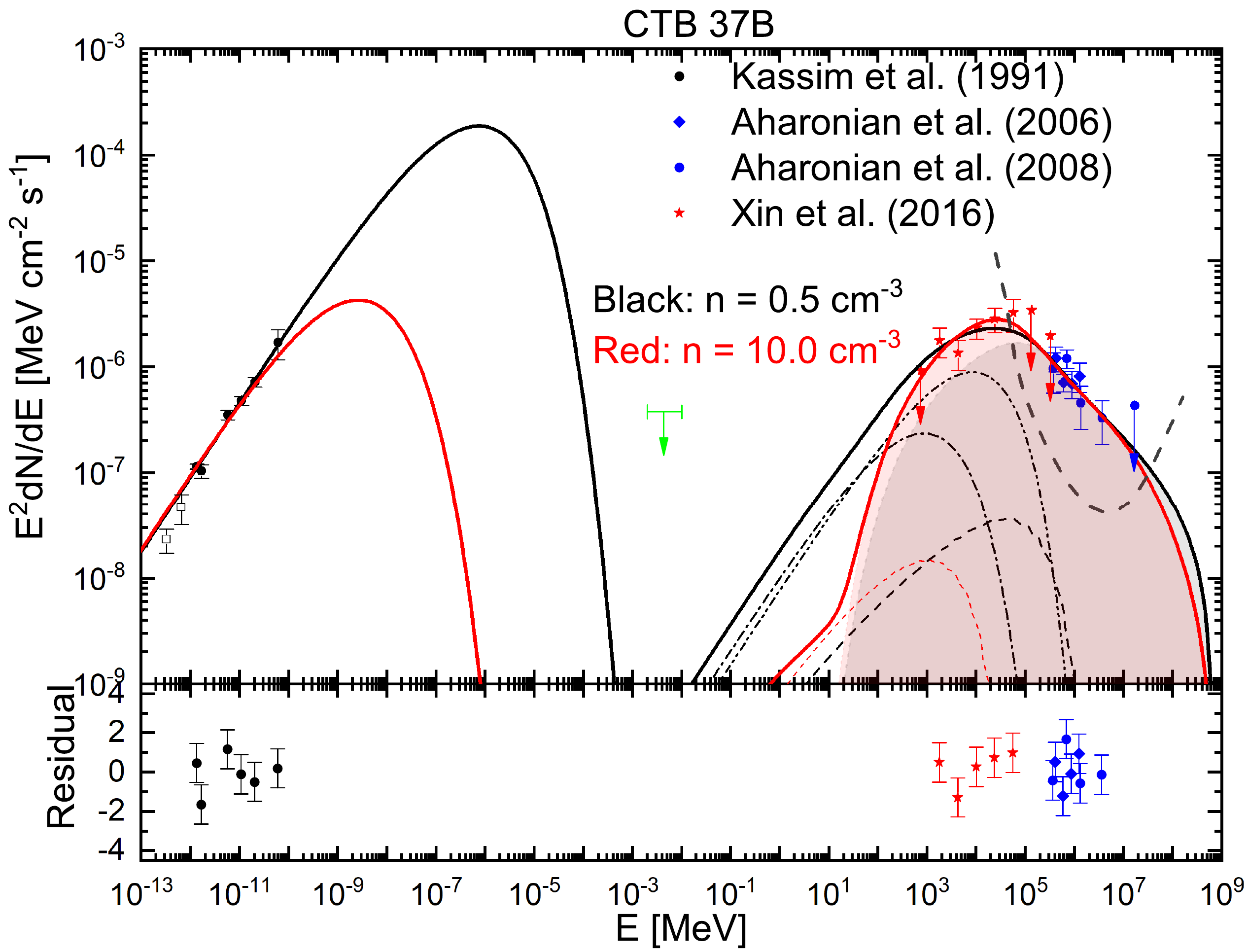}
\figsetgrpnote{The best fit to the spectral energy distribution (SED) for CTB 37B}
\figsetgrpend

\figsetgrpstart
\figsetgrpnum{2.68}
\figsetgrptitle{CTB 37B (b)
}
\figsetplot{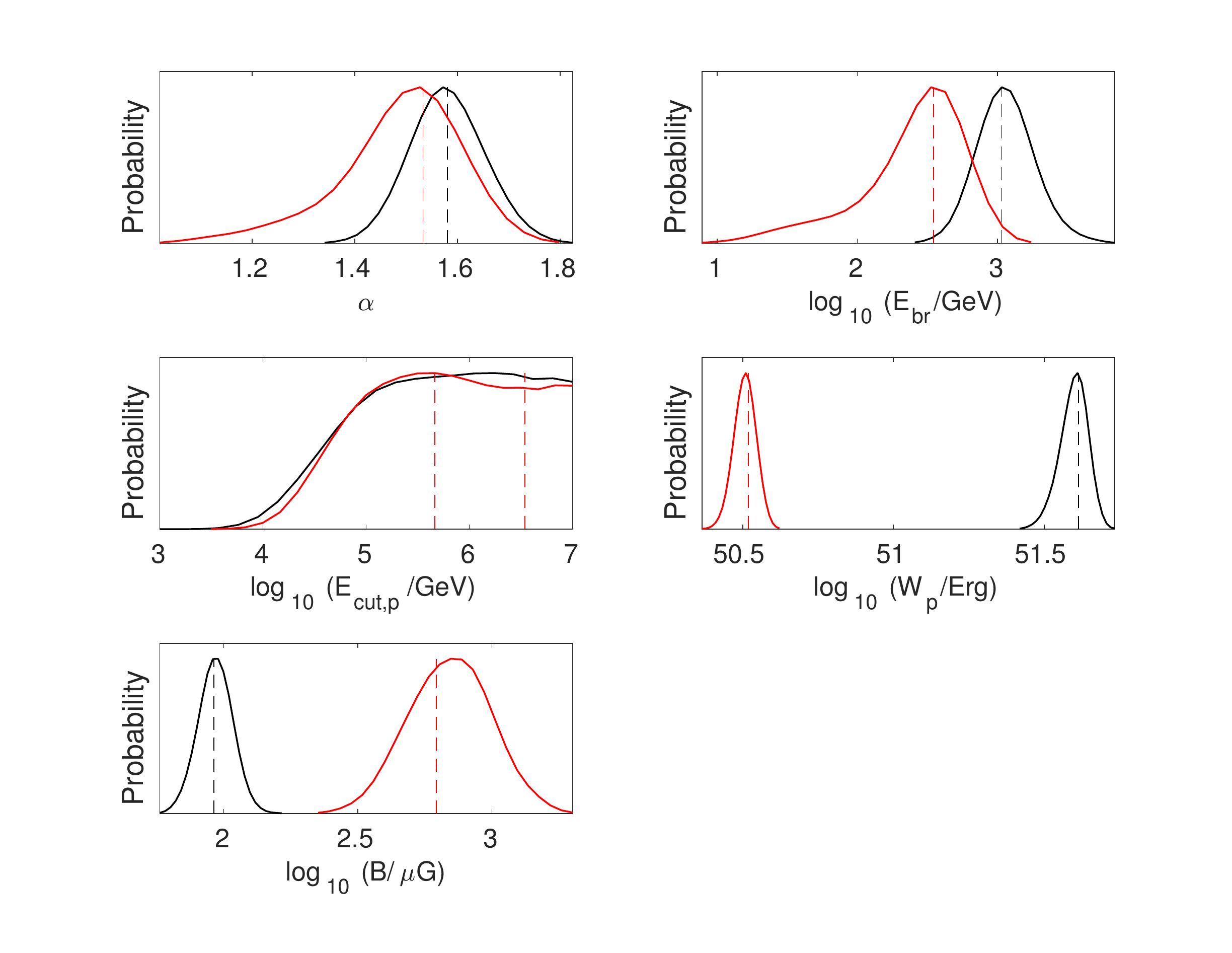}
\figsetgrpnote{1D probability distribution of the parameters for CTB 37B }
\figsetgrpend

\figsetgrpstart
\figsetgrpnum{2.69}
\figsetgrptitle{G349.7+0.2 (a)
}
\figsetplot{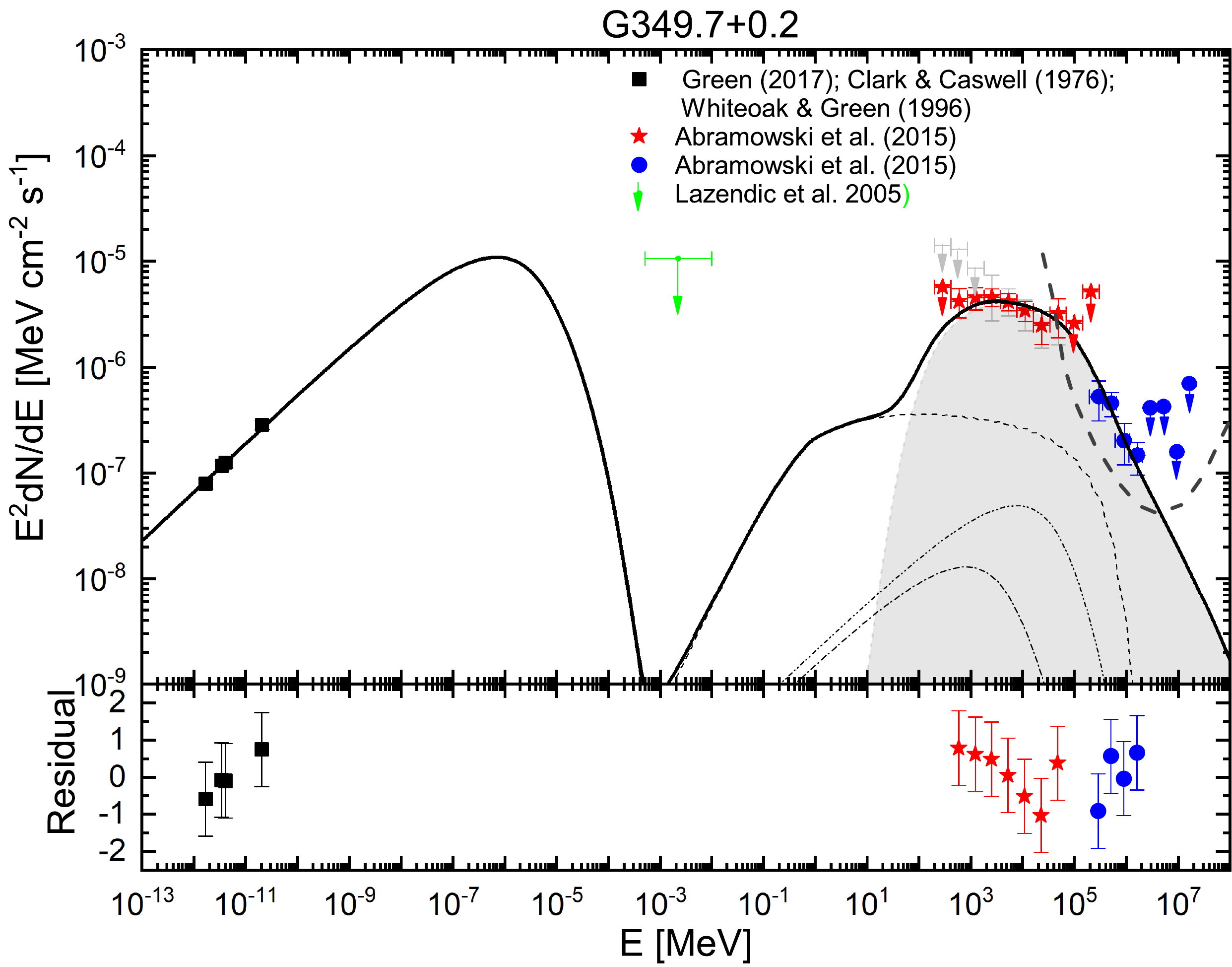}
\figsetgrpnote{The best fit to the spectral energy distribution (SED) for G349.7+0.2 }
\figsetgrpend

\figsetgrpstart
\figsetgrpnum{2.70}
\figsetgrptitle{G349.7+0.2 (b)
}
\figsetplot{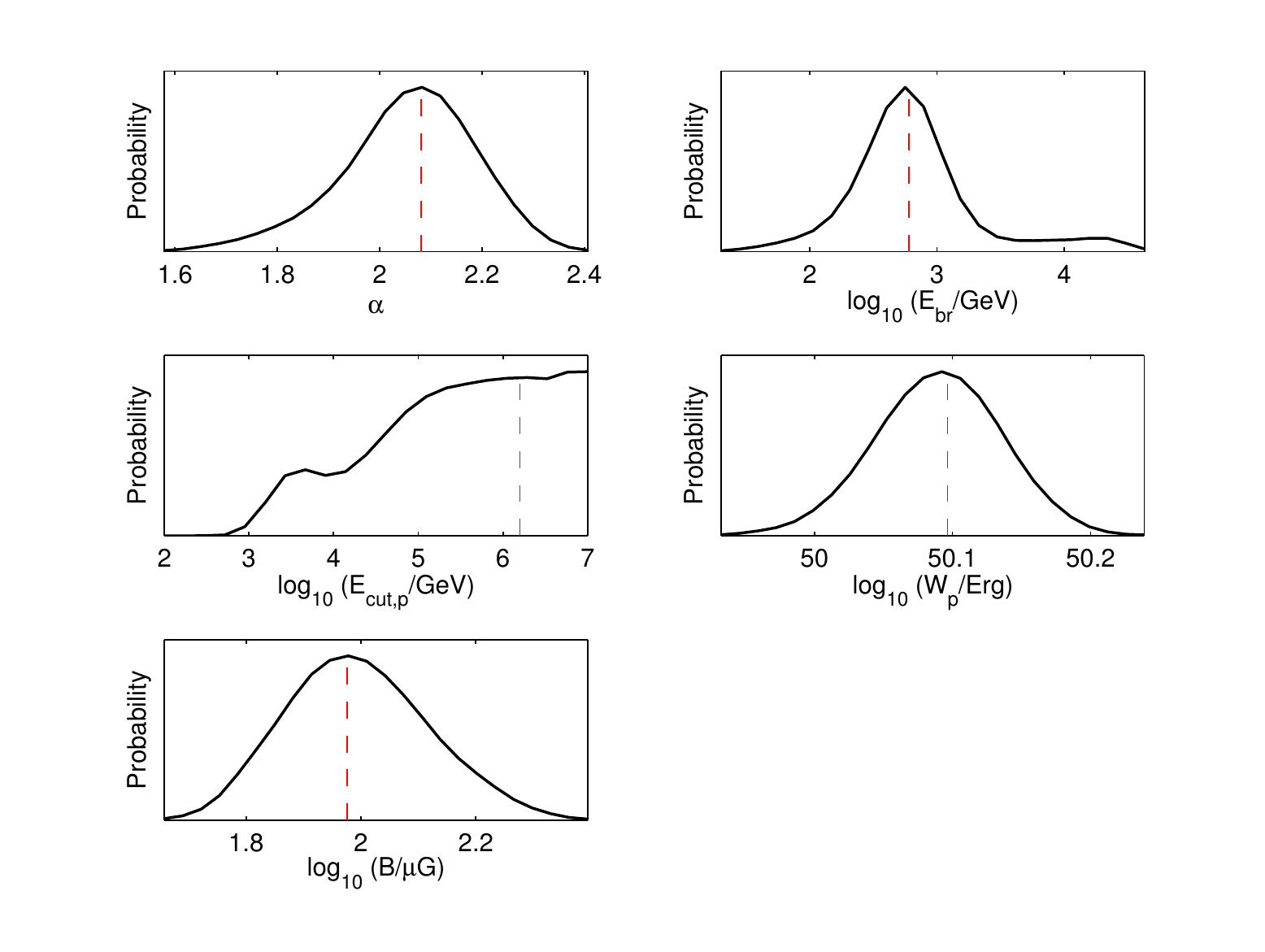}
\figsetgrpnote{1D probability distribution of the parameters for G349.7+0.2}
\figsetgrpend

\figsetgrpstart
\figsetgrpnum{2.71}
\figsetgrptitle{Hess J1731-347 (a)
}
\figsetplot{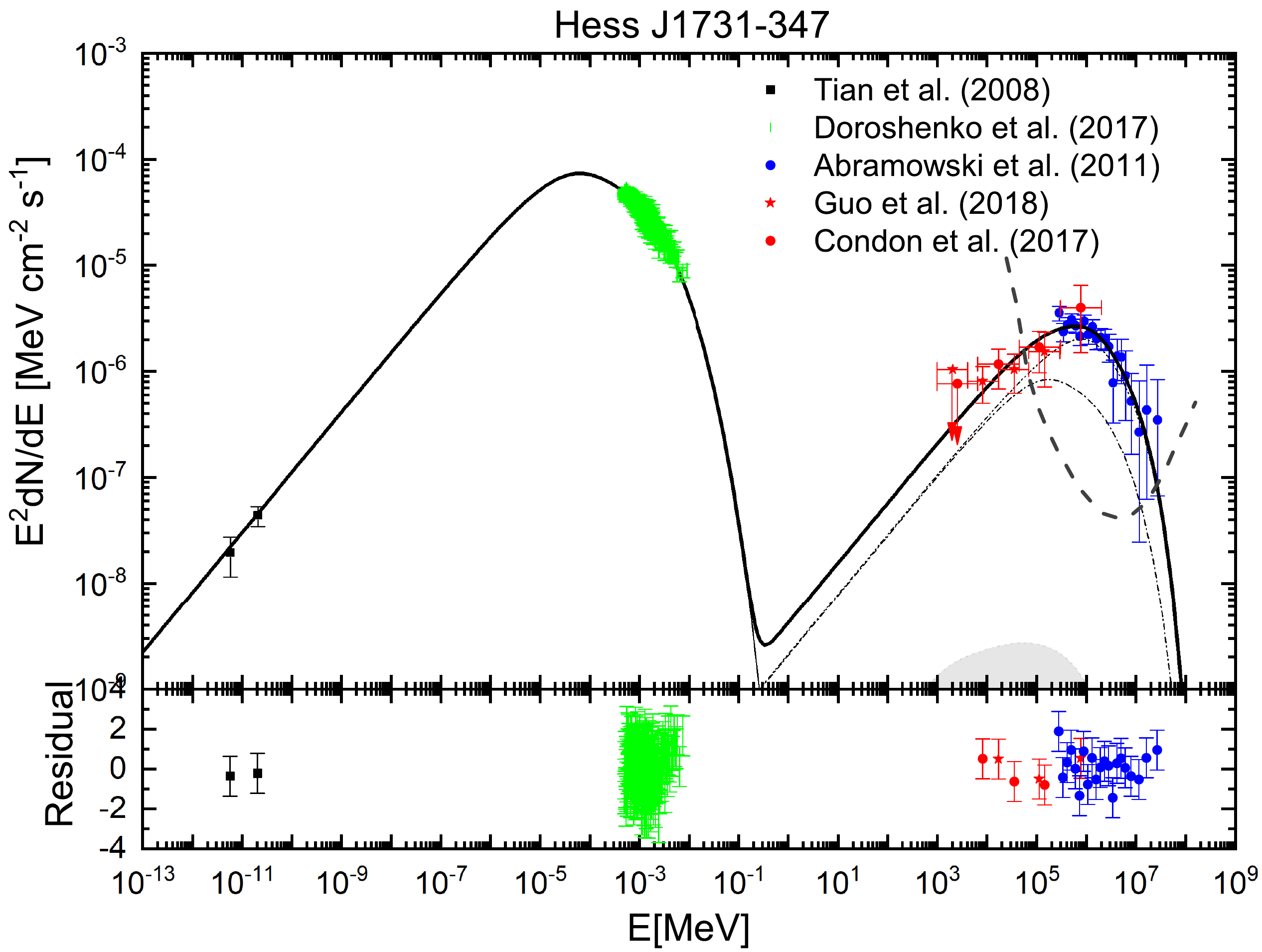}
\figsetgrpnote{The best fit to the spectral energy distribution (SED) for Hess J1731-347  }
\figsetgrpend

\figsetgrpstart
\figsetgrpnum{2.72}
\figsetgrptitle{Hess J1731-347 (b)
}
\figsetplot{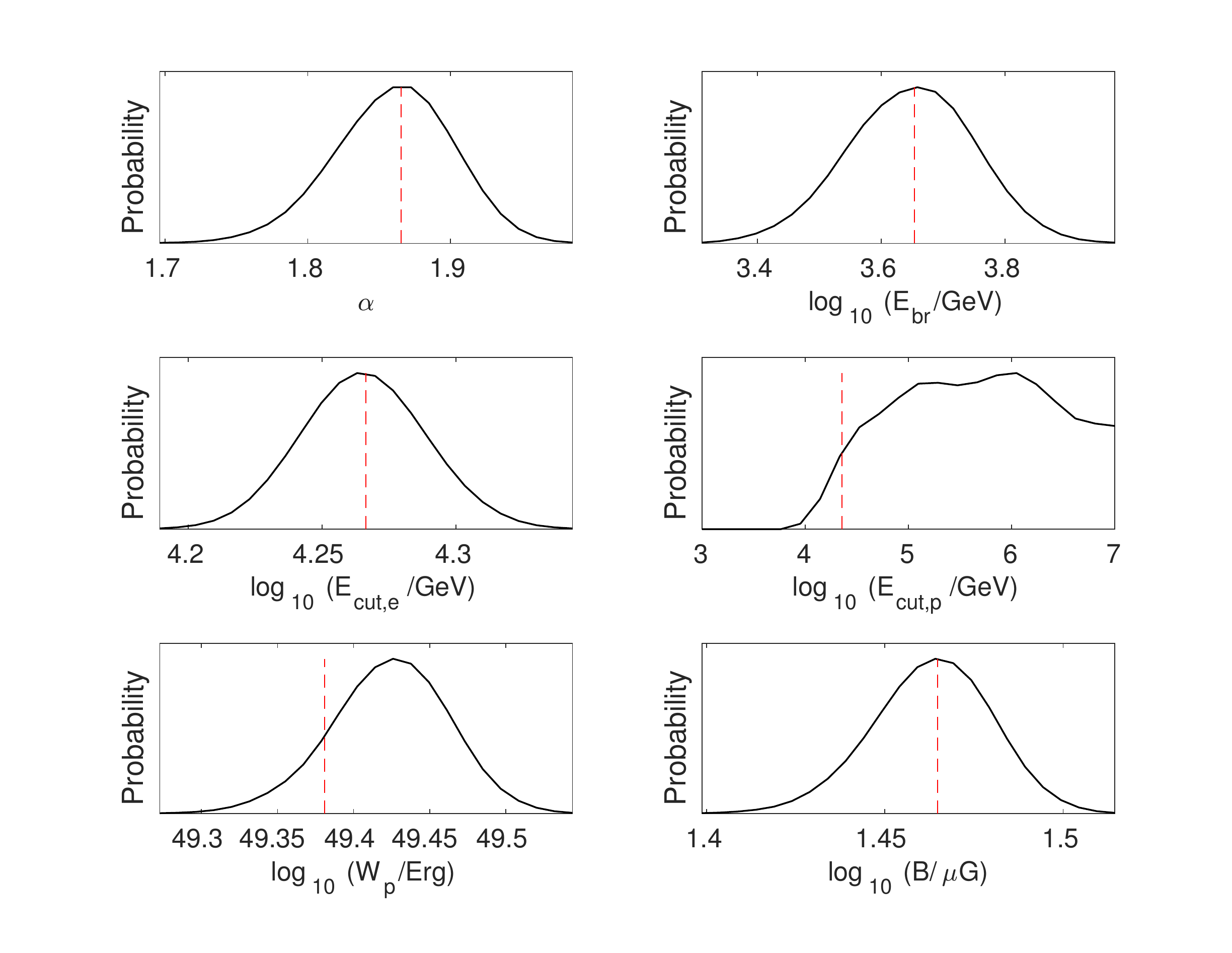}
\figsetgrpnote{ 1D probability distribution of the parameters for Hess J1731-347 }
\figsetgrpend

\figsetgrpstart
\figsetgrpnum{2.73}
\figsetgrptitle{G359.1-0.5 (a)
}
\figsetplot{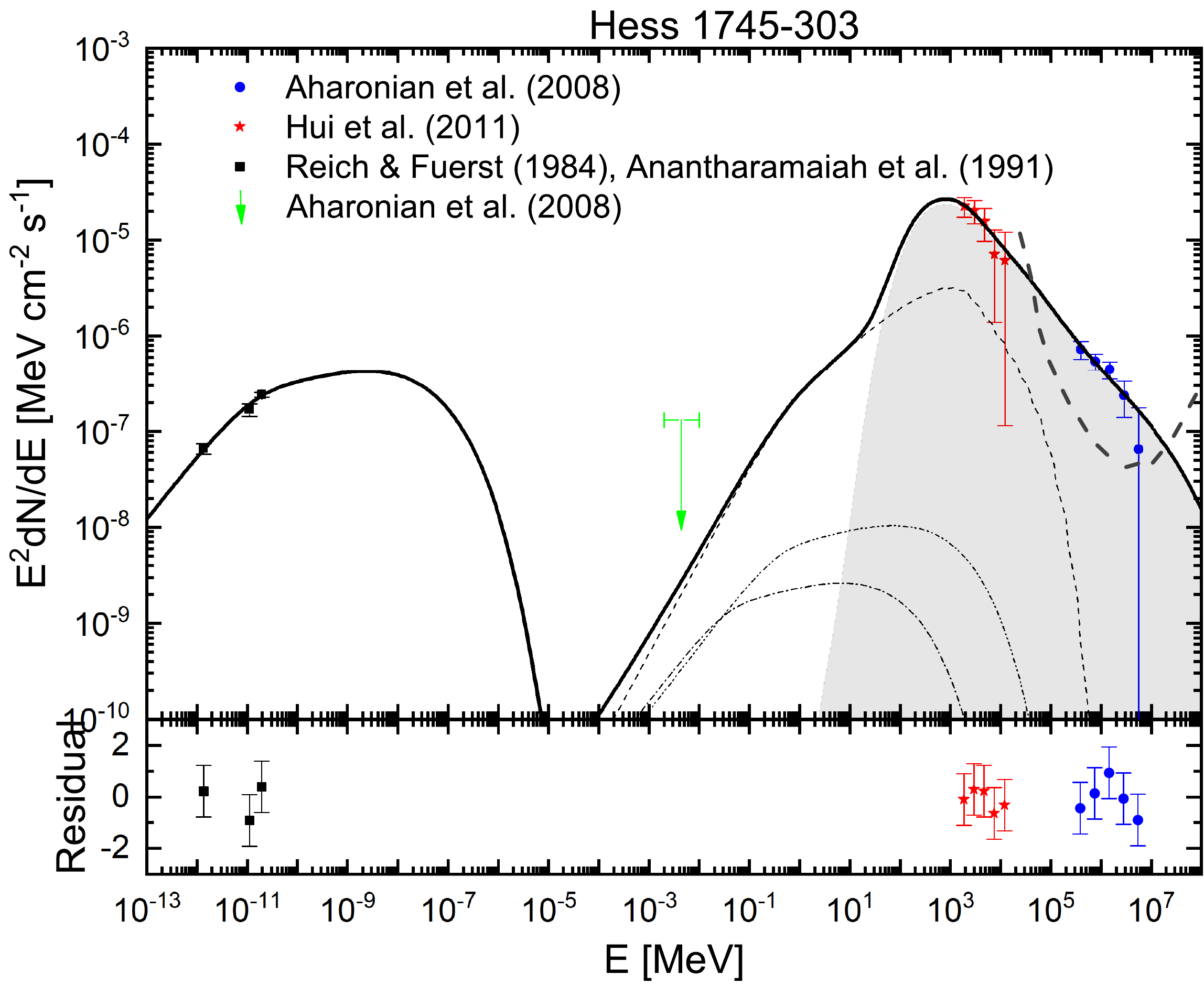}
\figsetgrpnote{The best fit to the spectral energy distribution (SED) for G359.1-0.5}
\figsetgrpend

\figsetgrpstart
\figsetgrpnum{2.74}
\figsetgrptitle{G359.1-0.5 (b)
}
\figsetplot{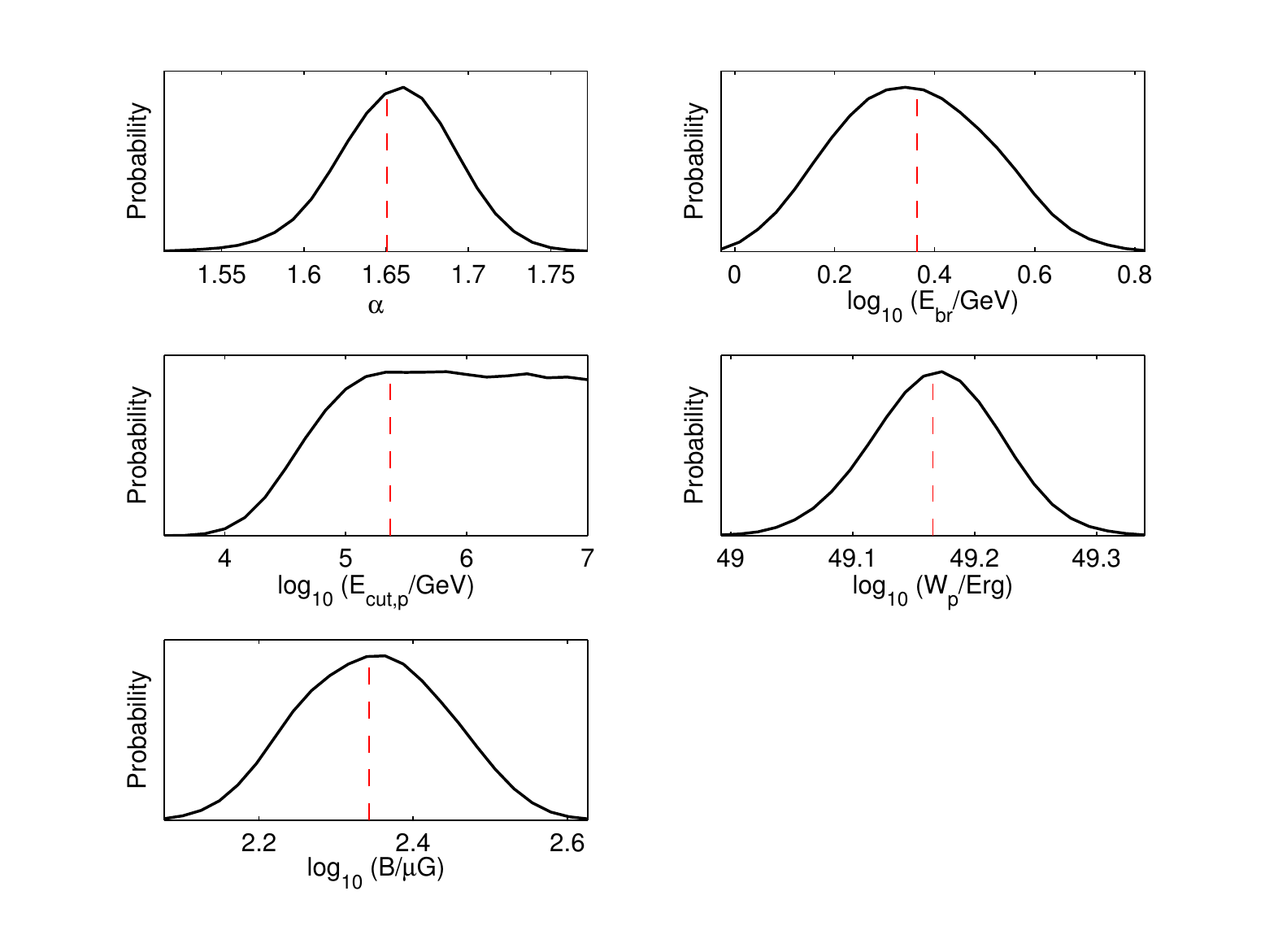}
\figsetgrpnote{1D probability distribution of the parameters for G359.1-0.5}
\figsetgrpend

\figsetend

\begin{figure}
\figurenum{2}
\gridline{\fig{W28a-eps-converted-to.pdf}{0.40\textwidth}{W28 (a) }
          \fig{W28b-eps-converted-to.pdf}{0.45\textwidth}{W28 (b)}
          }
\caption{
The best fit to the spectral energy distribution (SED) and 1D probability
distribution of the parameters for individual SNRs in our sample.
The thin solid lines and the dashed
lines represent the synchrotron and bremsstrahlung emissions, respectively.
The dash-dotted lines and the dash-dot-dot lines are for the IC component of the
CMB and the infrared photon fields, respectively. The dotted lines and shaded areas
show the hadronic component for those sources.
The data points
indicated with open boxes are not considered. And a 10\% error is assumed for
some of the radio data for G349.7+0.2, MSH 15-56, and RCW 103. The gray line is
the differential sensitivity (one year) of LHAASO (Decl. from $-10^{0}$ to $70^{0}$)
\citep{2016NPPP..279..166D}, and the dark gray and dashed dark gray lines represent
the differential sensitivities of North and South (50 hour) of CTA, respectively
(https://www.cta-observatory.org/). }
\end{figure}

\begin{figure}
\figurenum{2}
\gridline{\fig{W30a-eps-converted-to.pdf}{0.40\textwidth}{W30 (a)}
          \fig{W30b-eps-converted-to.pdf}{0.40\textwidth}{W30 (b)}
          }
\gridline{\fig{3C391a-eps-converted-to.pdf}{0.40\textwidth}{3C 391 (a) }
          \fig{3C391b-eps-converted-to.pdf}{0.40\textwidth}{3C 391 (b)}
          }
\gridline{\fig{Kes79a-eps-converted-to.pdf}{0.40\textwidth}{Kes 79 (a)}
          \fig{Kes79b-eps-converted-to.pdf}{0.40\textwidth}{Kes 79 (b)}
          }
\caption{continued}
\end{figure}

\begin{figure}
\figurenum{2}
\gridline{\fig{W44a-eps-converted-to.pdf}{0.40\textwidth}{W44 (a) }
          \fig{W44b-eps-converted-to.pdf}{0.40\textwidth}{W44 (b)}
          }
\gridline{\fig{W49Ba-eps-converted-to.pdf}{0.40\textwidth}{W49B (a)}
          \fig{W49Bb-eps-converted-to.pdf}{0.40\textwidth}{W49B (b)}
           }
\gridline{\fig{W51Ca-eps-converted-to.pdf}{0.40\textwidth}{W51C (a)}
          \fig{W51Cb-eps-converted-to.pdf}{0.40\textwidth}{W51C (b)}
          }
\caption{continued}
\end{figure}

\begin{figure}
\figurenum{2}
\gridline{\fig{W51Ca_single-eps-converted-to.pdf}{0.40\textwidth}{W51C ($\rm a^{'}$)}
          \fig{W51Cb_single-eps-converted-to.pdf}{0.40\textwidth}{W51C ($\rm b^{'}$)}
           }
\gridline{\fig{G73a-eps-converted-to.pdf}{0.40\textwidth}{G73.9+0.9 (a)}
          \fig{G73b-eps-converted-to.pdf}{0.40\textwidth}{G73.9+0.9 (b)}
          }
\gridline{\fig{loopa-eps-converted-to.pdf}{0.40\textwidth}{Cygnus Loop (a)}
          \fig{loopb-eps-converted-to.pdf}{0.40\textwidth}{Cygnus Loop (a)}
          }
\caption{continued}
\end{figure}

\begin{figure}
\figurenum{2}
\gridline{\fig{G78a-eps-converted-to.pdf}{0.40\textwidth}{$\gamma$ Cygni (a)}
          \fig{G78b-eps-converted-to.pdf}{0.40\textwidth}{$\gamma$ Cygni (b)}
          }
\gridline{\fig{HB21_4na-eps-converted-to.pdf}{0.40\textwidth}{HB 21 (a) }
          \fig{HB21_4nb-eps-converted-to.pdf}{0.40\textwidth}{HB 21 (b)}
           }
 \gridline{\fig{CTB109a-eps-converted-to.pdf}{0.40\textwidth}{CTB 109 (a) }
          \fig{CTB109b-eps-converted-to.pdf}{0.40\textwidth}{CTB 109 (b)}
          }
\caption{continued}
\end{figure}

\begin{figure}
\figurenum{2}
\gridline{\fig{Tycho_a-eps-converted-to.pdf}{0.40\textwidth}{Tycho (a) }
          \fig{Tycho_b-eps-converted-to.pdf}{0.40\textwidth}{Tycho (b)}
           }
 \gridline{\fig{HB3a-eps-converted-to.pdf}{0.40\textwidth}{HB 3 (a) }
          \fig{HB3b-eps-converted-to.pdf}{0.40\textwidth}{HB 3 (b)}
          }
\gridline{\fig{G150a-eps-converted-to.pdf}{0.40\textwidth}{G150.3+4.5 (a)}
          \fig{G150b-eps-converted-to.pdf}{0.40\textwidth}{G150.3+4.5 (b)}
           }
\caption{continued}
\end{figure}

\begin{figure}
\figurenum{2}
\gridline{\fig{HB9a-eps-converted-to.pdf}{0.40\textwidth}{HB 9 (a) }
          \fig{HB9b-eps-converted-to.pdf}{0.40\textwidth}{HB 9 (b)}
          }
\gridline{\fig{G166a-eps-converted-to.pdf}{0.40\textwidth}{G166.0+4.3 (a)}
          \fig{G166b-eps-converted-to.pdf}{0.40\textwidth}{G166.0+4.3 (b)}
           }
 \gridline{\fig{S147a_n=250_1-eps-converted-to.pdf}{0.40\textwidth}{S147 (a)}
          \fig{S147b_n=250_1-eps-converted-to.pdf}{0.40\textwidth}{S147 (b)}
          }
\caption{continued}
\end{figure}

\begin{figure}
\figurenum{2}
 \gridline{\fig{IC443a-eps-converted-to.pdf}{0.40\textwidth}{IC443 (a) }
          \fig{IC443b-eps-converted-to.pdf}{0.40\textwidth}{IC443 (b)}
          }
 \gridline{ \fig{G205a-eps-converted-to.pdf}{0.40\textwidth}{G205.5+0.5 (a) }
          \fig{G205b-eps-converted-to.pdf}{0.40\textwidth}{G205.5+0.5 (b)}
          }
 \gridline{\fig{PuppisAa-eps-converted-to.pdf}{0.40\textwidth}{Puppis A (a) }
          \fig{PuppisAb-eps-converted-to.pdf}{0.40\textwidth}{Puppis A (b)}
           }
 \caption{continued}
\end{figure}

\begin{figure}
\figurenum{2}
 \gridline{\fig{G296a-eps-converted-to.pdf}{0.40\textwidth}{G296.5+10.0 (a) }
          \fig{G296b-eps-converted-to.pdf}{0.40\textwidth}{G296.5+10.0 (b)}
          }
  \gridline{\fig{0852a-eps-converted-to.pdf}{0.40\textwidth}{RX J0852-4622 (a) }
          \fig{0852b-eps-converted-to.pdf}{0.40\textwidth}{RX J0852-4622 (b)}
           }
 \gridline{\fig{J0852a-eps-converted-to.pdf}{0.40\textwidth}{RX J0852-4622 ($\rm a^{'}$) }
          \fig{J0852b-eps-converted-to.pdf}{0.40\textwidth}{RX J0852-4622 ($\rm b^{'}$)}
          }
\caption{continued}
\end{figure}

\begin{figure}
\figurenum{2}
 \gridline{\fig{Kes17a-eps-converted-to.pdf}{0.40\textwidth}{Kes 17 (a) }
          \fig{Kes17b-eps-converted-to.pdf}{0.40\textwidth}{Kes 17 (b)}
           }
 \gridline{\fig{RCW86a-eps-converted-to.pdf}{0.40\textwidth}{RCW 86 (a)}
          \fig{RCW86b-eps-converted-to.pdf}{0.40\textwidth}{RCW 86 (b)}
          }
 \gridline{
          \fig{MSH15-56a-eps-converted-to.pdf}{0.40\textwidth}{MSH 15-56 (a)}
          \fig{MSH15-56b-eps-converted-to.pdf}{0.40\textwidth}{MSH 15-56 (b)}
           }
\caption{continued}
\end{figure}

\begin{figure}
\figurenum{2}
 \gridline{\fig{SN1006a-eps-converted-to.pdf}{0.40\textwidth}{SN 1006 (a)}
          \fig{SN1006b-eps-converted-to.pdf}{0.40\textwidth}{SN 1006 (b)}
          }
 \gridline{
          \fig{RCW103a-eps-converted-to.pdf}{0.40\textwidth}{RCW 103 (a)}
          \fig{RCW103b-eps-converted-to.pdf}{0.40\textwidth}{RCW 103 (b)}
           }
 \gridline{\fig{CTB33a-eps-converted-to.pdf}{0.40\textwidth}{CTB 33 (a) }
          \fig{CTB33b-eps-converted-to.pdf}{0.40\textwidth}{CTB 33 (b)}
          }
\caption{continued}
\end{figure}

\begin{figure}
\figurenum{2}
\gridline{
          \fig{J1713a-eps-converted-to.pdf}{0.40\textwidth}{RX J1713.7-3946 (a) }
          \fig{J1713b-eps-converted-to.pdf}{0.40\textwidth}{RX J1713.7-3946 (b)}
           }
 \gridline{\fig{CTB37Aa-eps-converted-to.pdf}{0.40\textwidth}{CTB 37A (a)}
          \fig{CTB37Ab-eps-converted-to.pdf}{0.40\textwidth}{CTB 37A (b)}
          }
 \gridline{\fig{CTB37Ba-eps-converted-to.pdf}{0.40\textwidth}{CTB 37B (a)}
          \fig{CTB37Bb-eps-converted-to.pdf}{0.40\textwidth}{CTB 37B (b)}
 }
\caption{continued}
\end{figure}

\begin{figure}
\figurenum{2}
 \gridline{\fig{G349a-eps-converted-to.pdf}{0.40\textwidth}{G349.7+0.2 (a) }
          \fig{G349b-eps-converted-to.pdf}{0.40\textwidth}{G349.7+0.2 (b)}
          }
 \gridline{\fig{J1731a-eps-converted-to.pdf}{0.40\textwidth}{Hess J1731-347 (a)}
          \fig{J1731b-eps-converted-to.pdf}{0.40\textwidth}{Hess J1731-347 (b)}
           }
 \gridline{\fig{G359a-eps-converted-to.pdf}{0.40\textwidth}{G359.1-0.5 (a)}
         \fig{G359b-eps-converted-to.pdf}{0.40\textwidth}{G359.1-0.5 (b)}
          }
\caption{continued}
\end{figure}

\begin{figure}
\figurenum{3}
\gridline{\fig{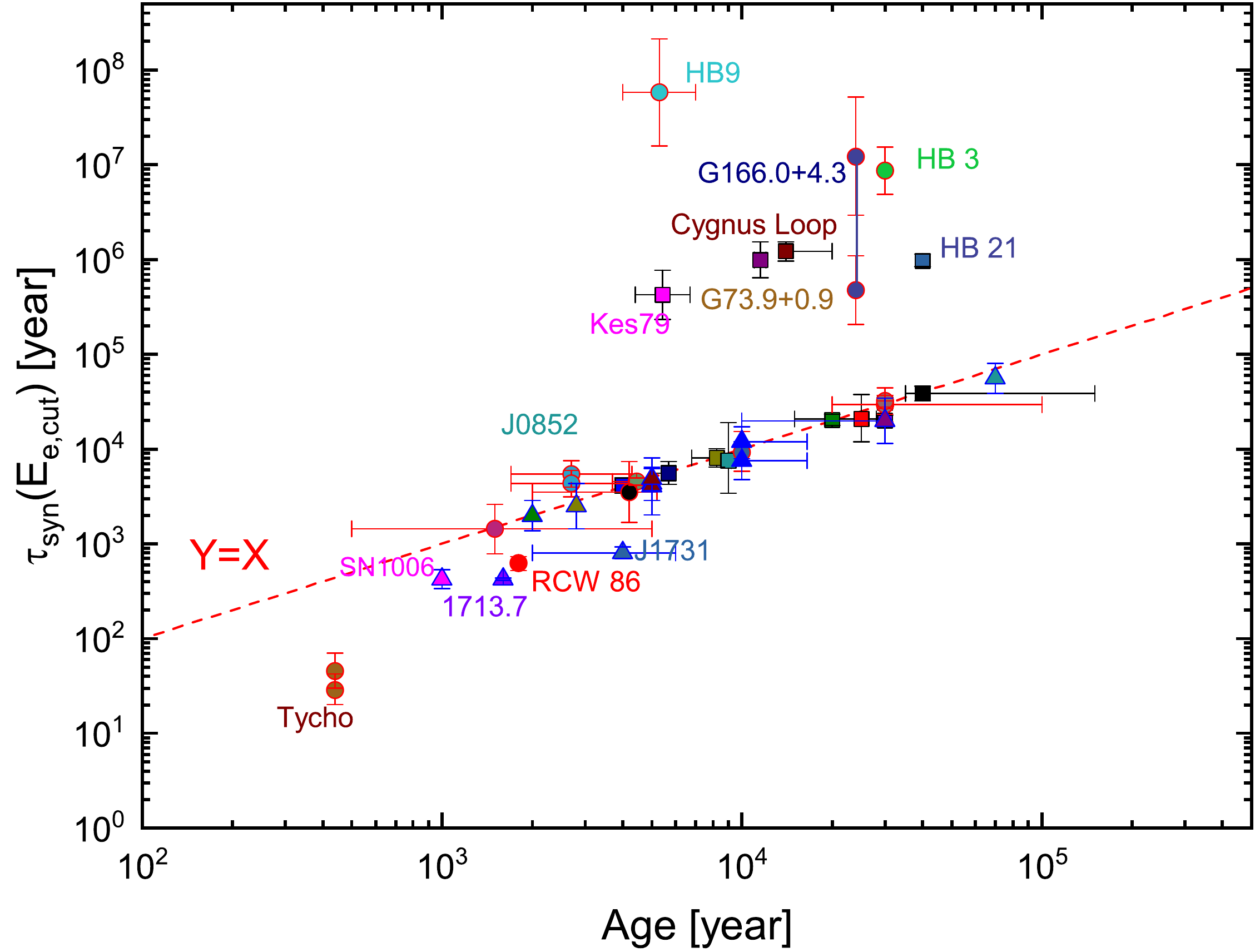}{0.40\textwidth}{ (a) }
          \fig{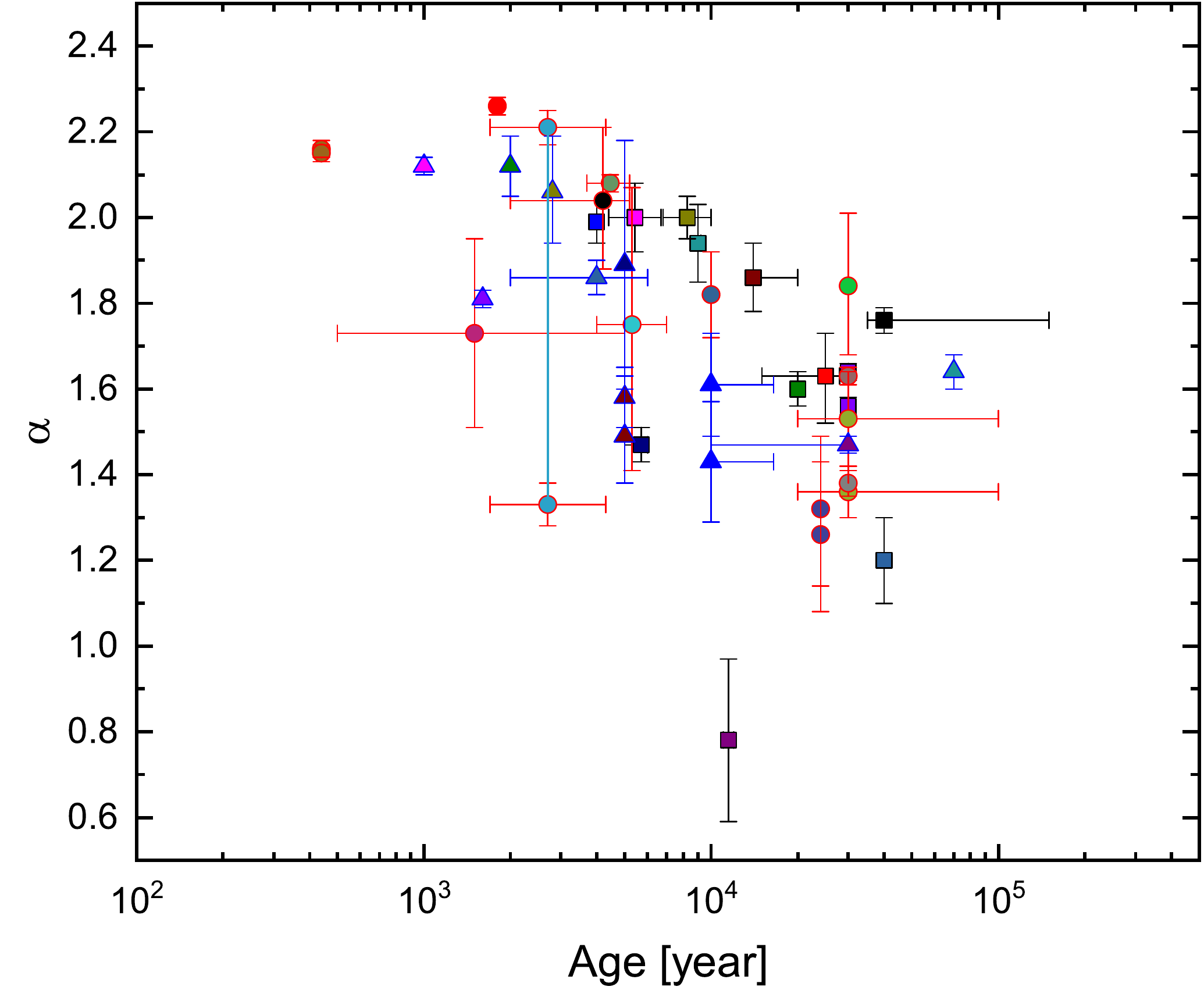}{0.40\textwidth}{ (b) }
          }
\gridline{\fig{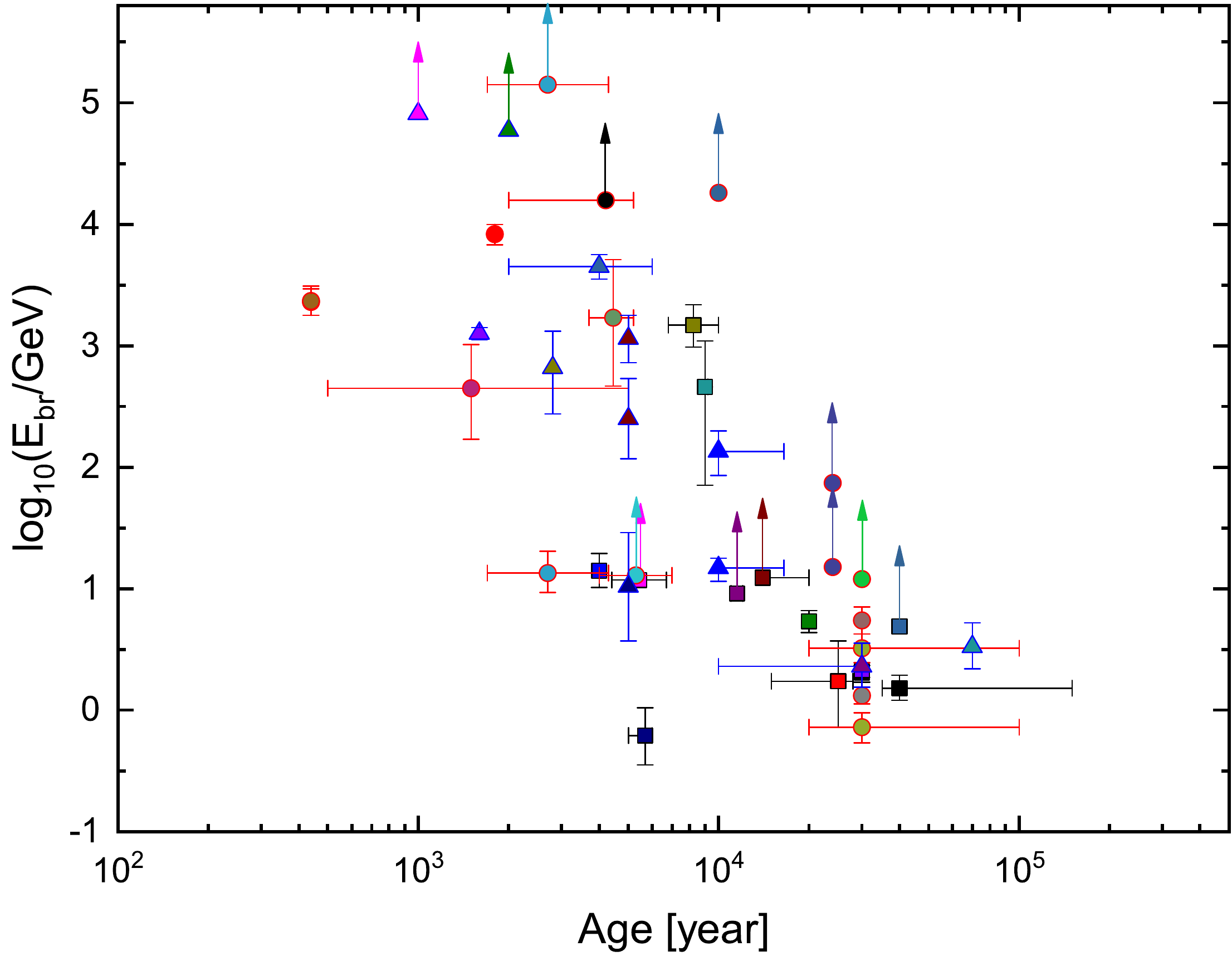}{0.40\textwidth}{ (c)} 
          \fig{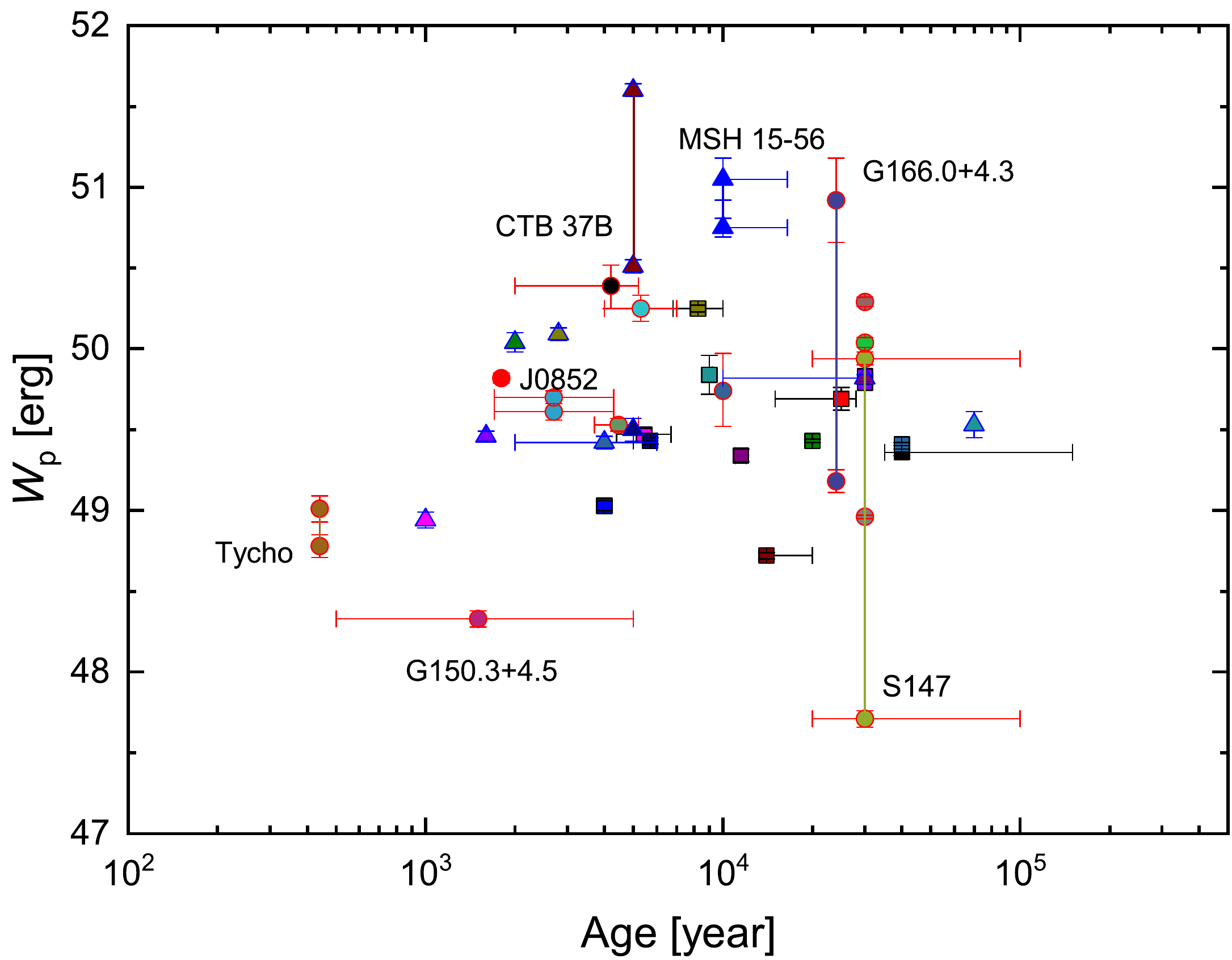}{0.40\textwidth}{ (d) }
          }
\gridline{\fig{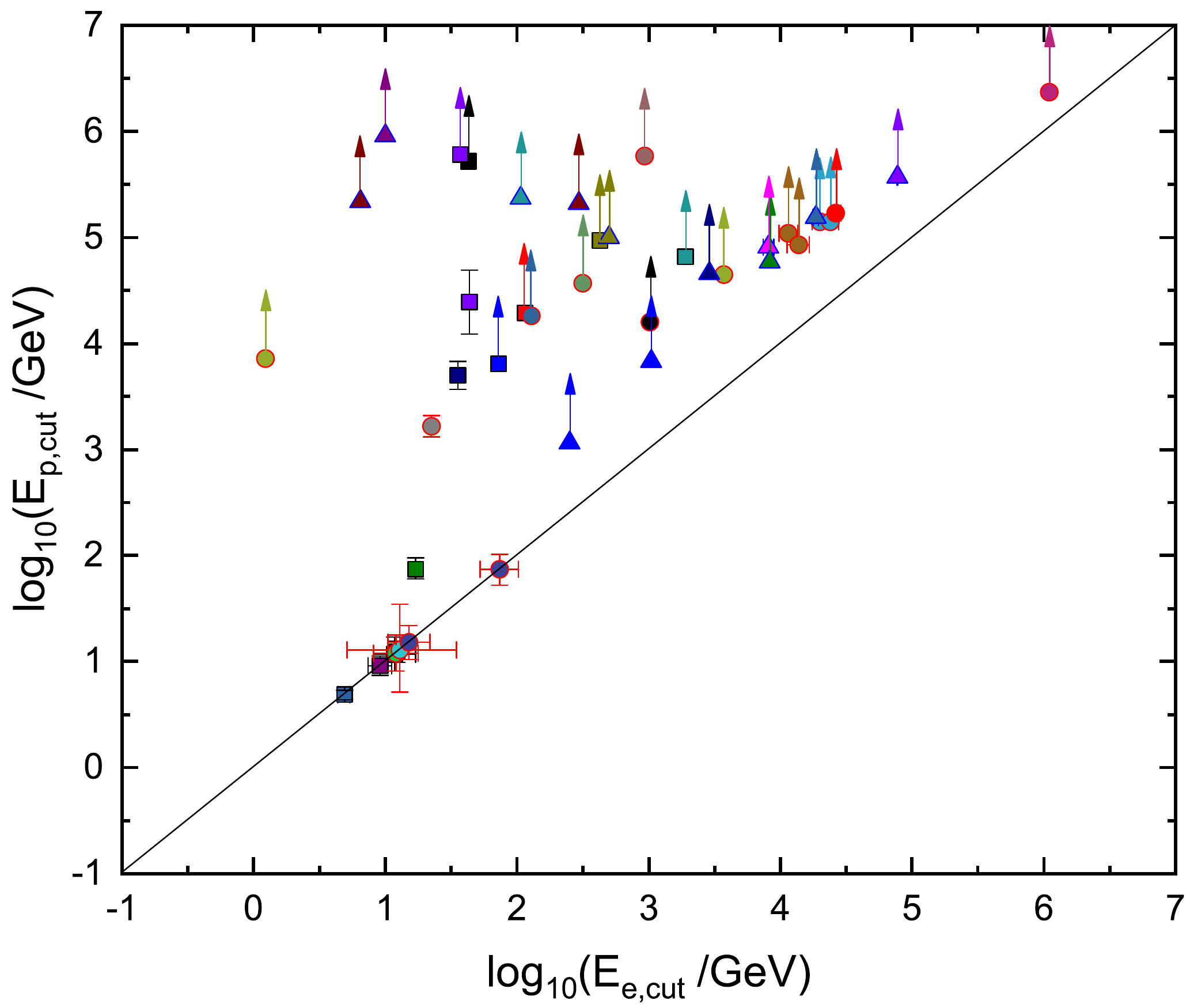}{0.40\textwidth}{ (e)} 
          \fig{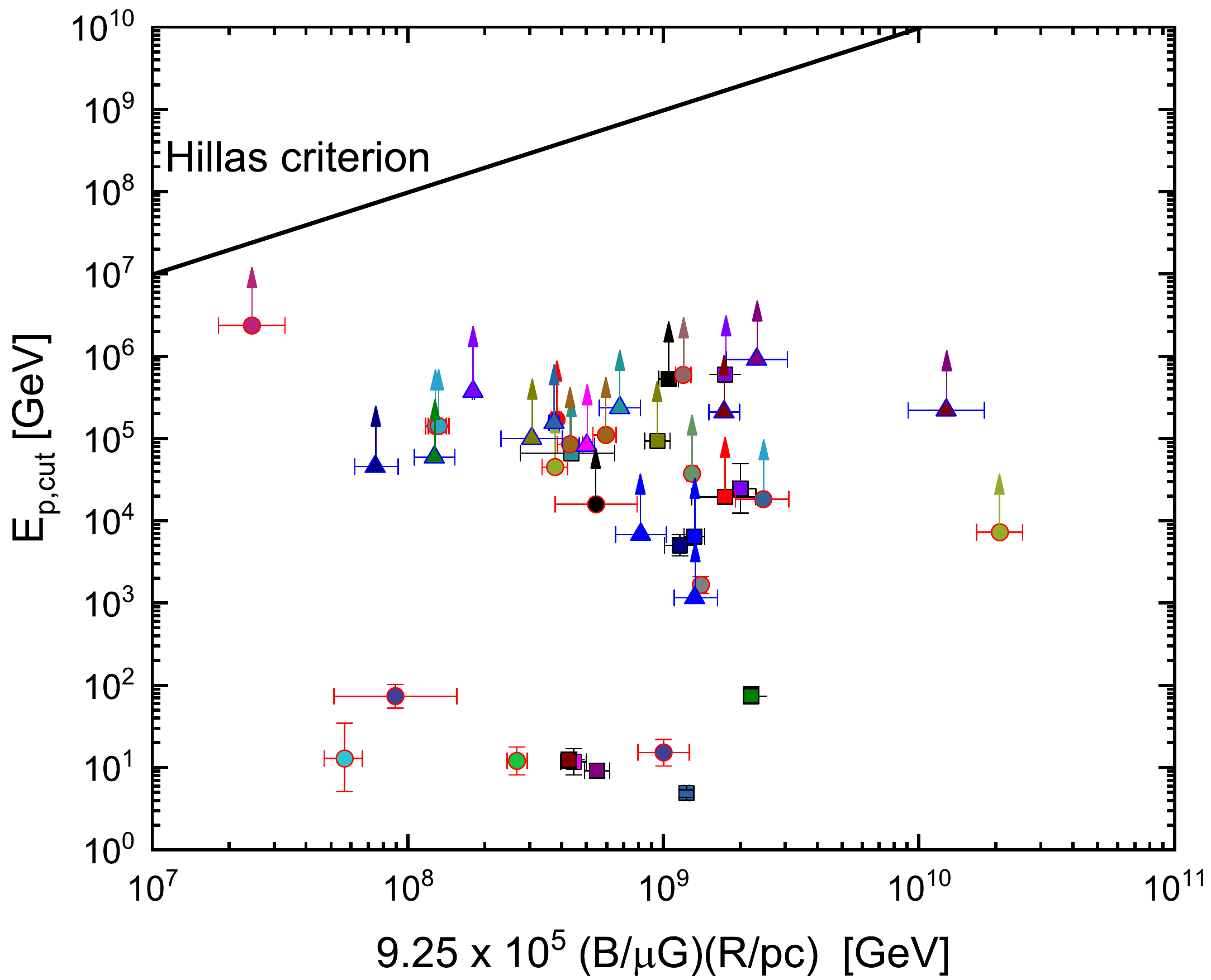}{0.40\textwidth}{ (f) }
          }
\end{figure}
\begin{figure}
\figurenum{3}
\gridline{\fig{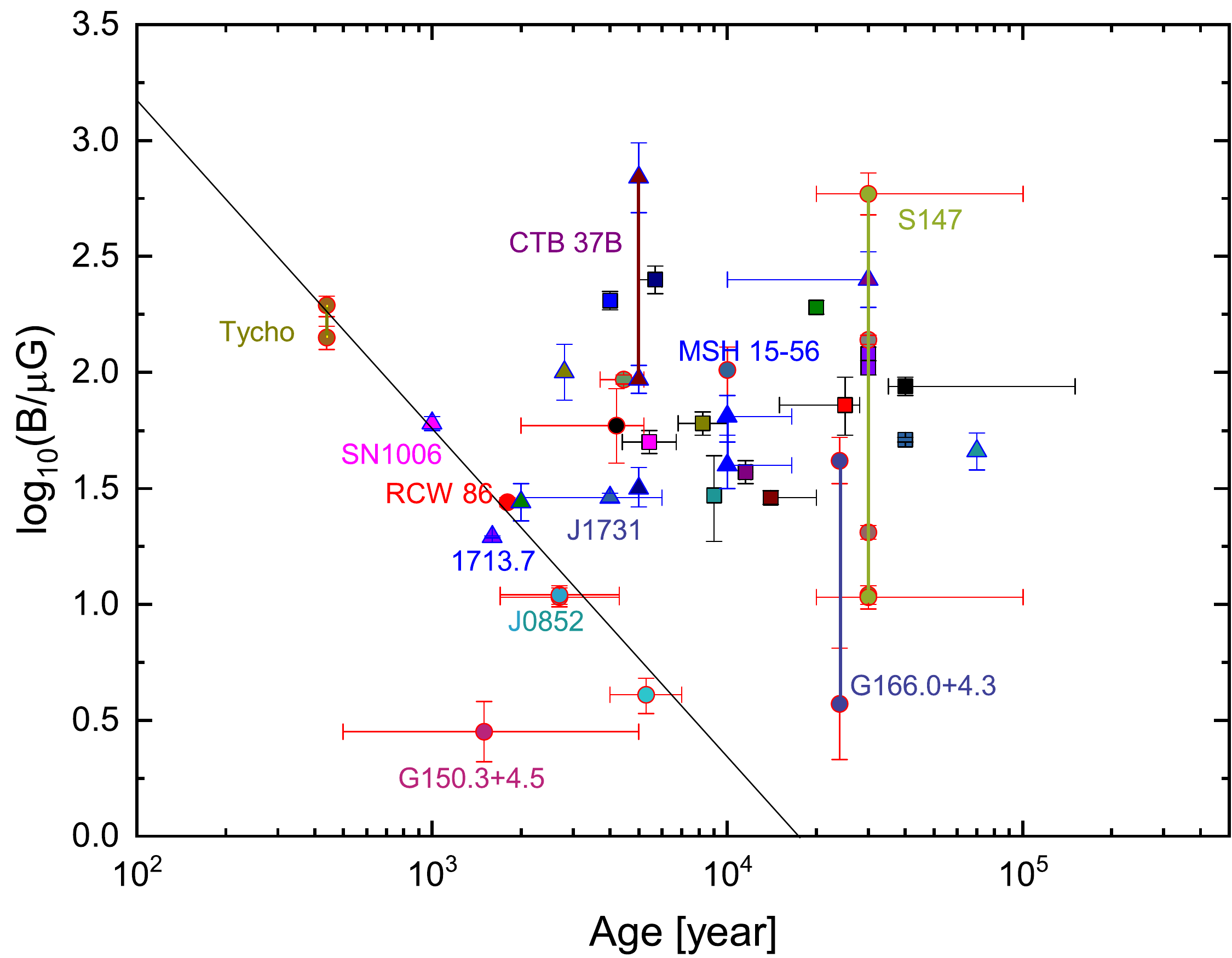}{0.40\textwidth}{ (g) }
          \fig{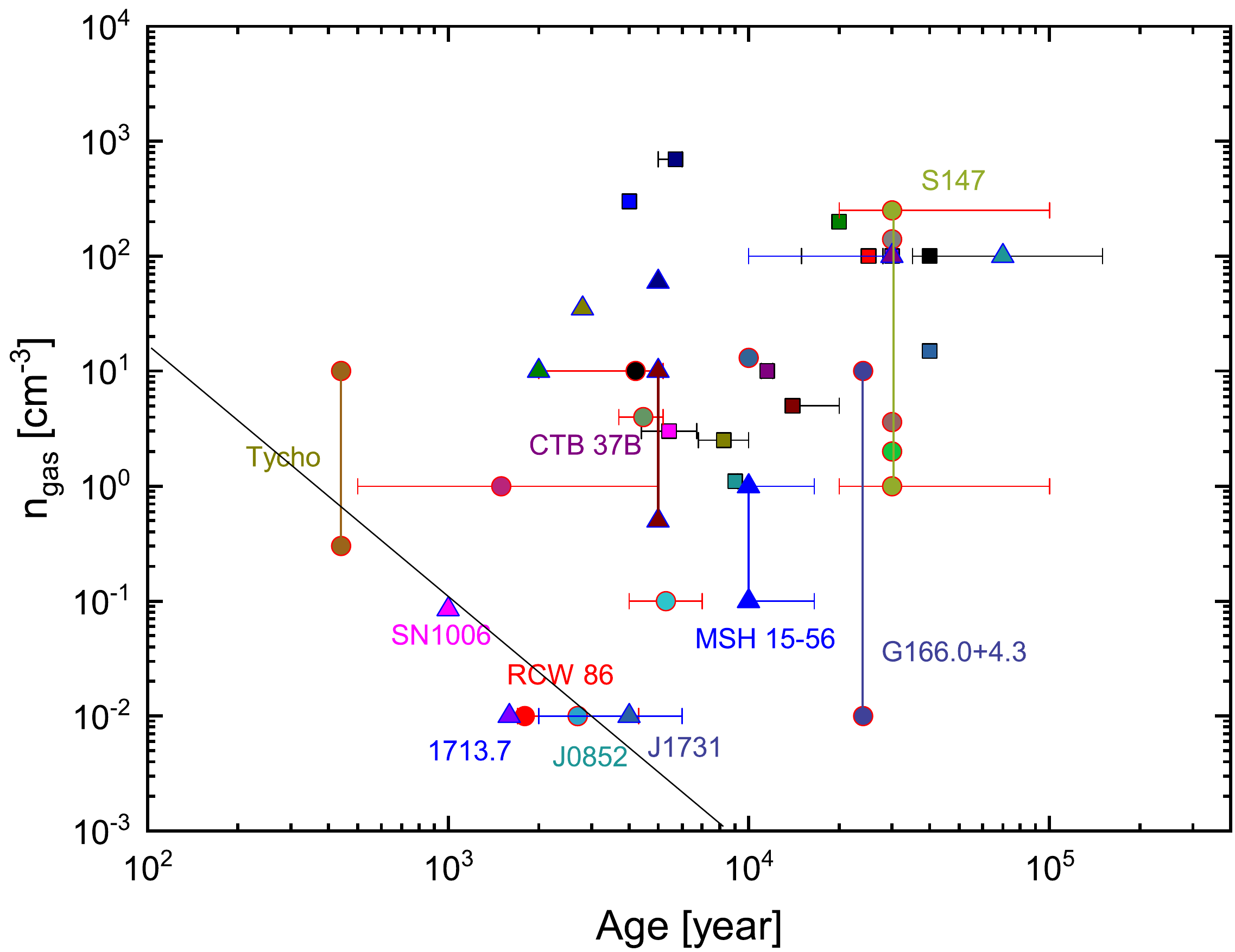}{0.40\textwidth}{ (h) }
          }
\gridline{\fig{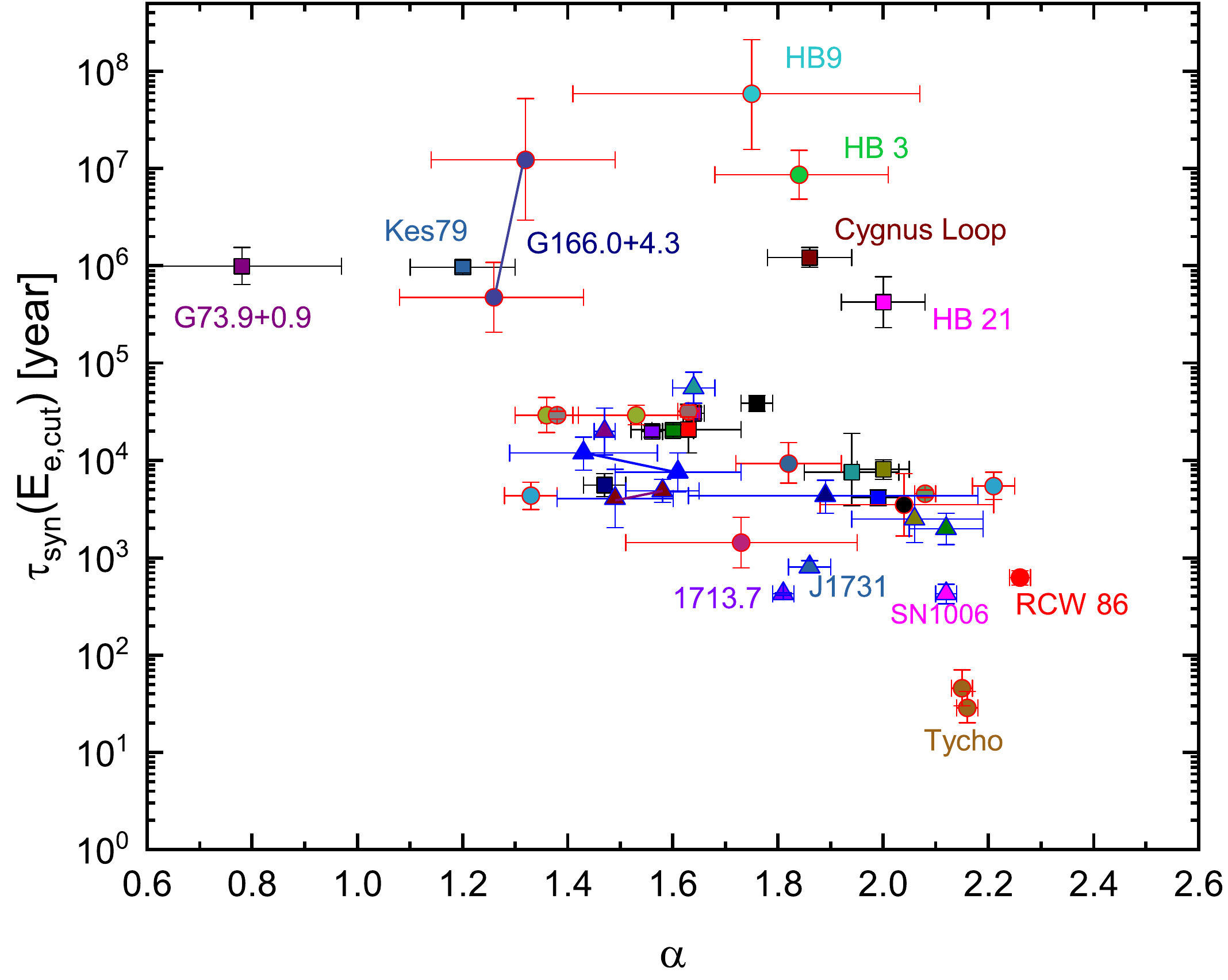}{0.40\textwidth}{ (i) }
          \fig{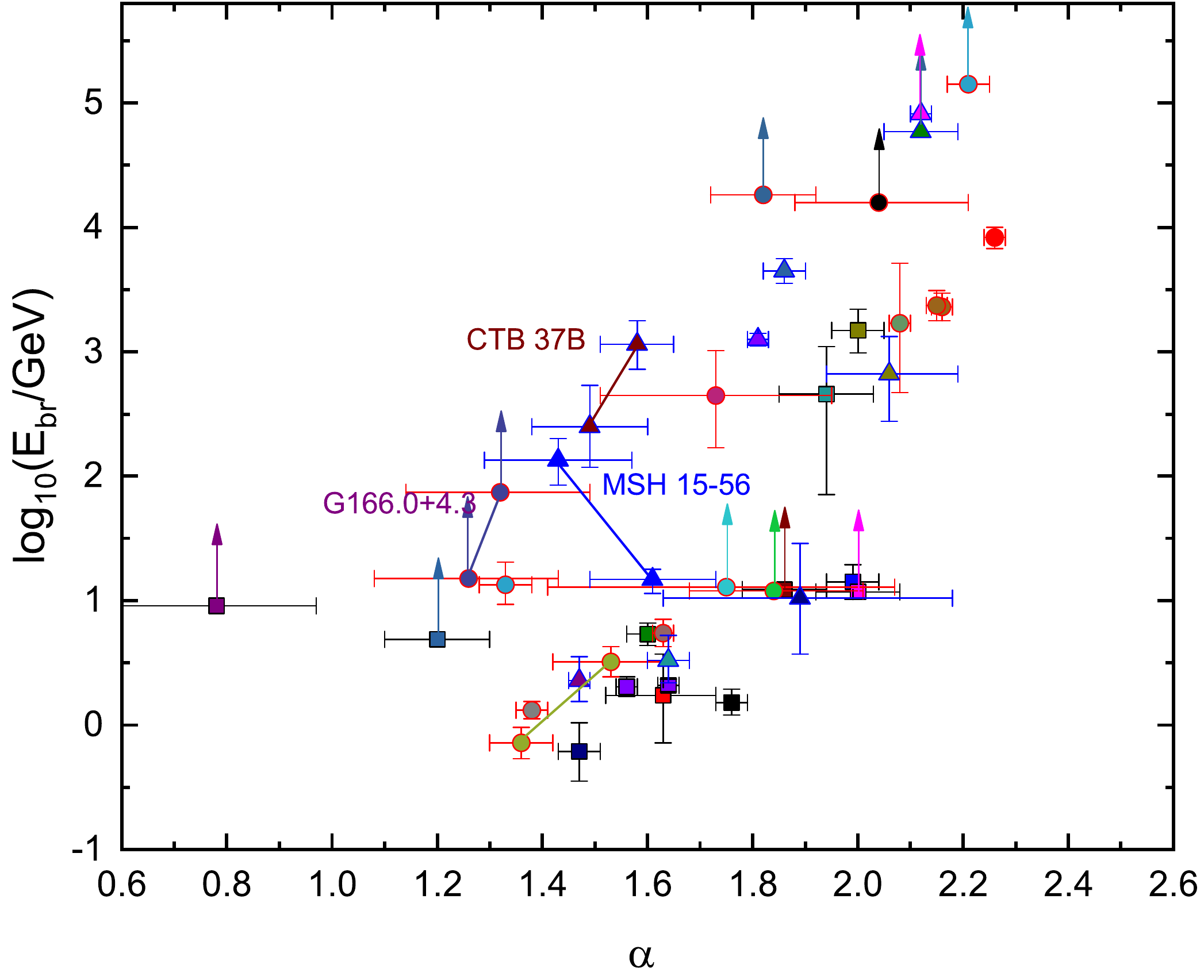}{0.40\textwidth}{ (j) }
          }
\gridline{\fig{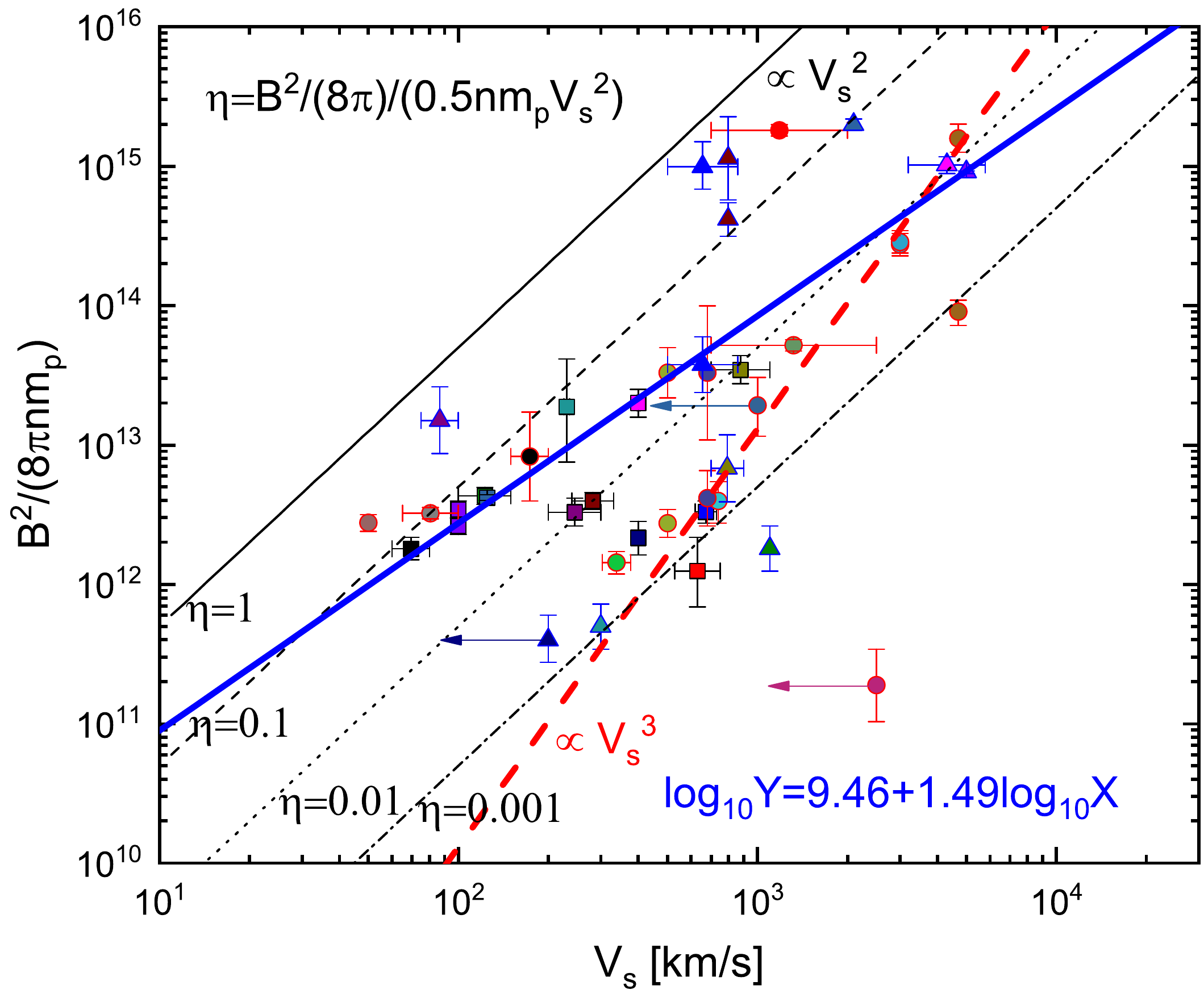}{0.40\textwidth}{ (k) }
          }
\caption{Correlation of model parameters. Lines connect sources with two different values of the background density adopted.
}
\end{figure}

\end{document}